\documentclass[12pt]{article}
\usepackage{epsfig,amssymb,amsmath,psfrag}
\usepackage{tikz}
\usepackage{graphicx}

\usepackage{bbm}
\usetikzlibrary{decorations.pathmorphing} % For photon propagator decoration
\def \bl  {\begin{align*}}
\def \el  {\end{align*}}

\def \be  {\begin{equation}}
\def \ee  {\end{equation}}
\def \ba  {\begin{eqnarray}}
\def \ea  {\end{eqnarray}}
\def \baa {\begin{eqnarray*}}
\def \eaa {\end{eqnarray*}}
\def \bb  {\begin {thebibliography} }
\def \eb  {\end{thebibliography}}
\def \lab #1 {\label{#1}}

\newcommand{\beq}{\begin{equation}}
\newcommand{\eeq}{\end{equation}}
\newcommand{\beqa}{\begin{eqnarray}}
\newcommand{\eeqa}{\end{eqnarray}}

\def\l<{\langle}
\def\r>{\rangle}

\def\XXint#1#2#3{{\setbox0=\hbox{$#1{#2#3}{\int}$}
     \vcenter{\hbox{$#2#3$}}\kern-.5\wd0}}

\renewcommand{\title}[1]{\vbox{\center\LARGE{#1}}\vspace{5mm}}
\renewcommand{\author}[1]{\vbox{\center#1}\vspace{5mm}}

\usepackage{pgfkeys}
\usepackage{pgffor}
\usepackage{tikz}
\usepackage{ifthen}
\usetikzlibrary{intersections}
\usetikzlibrary{decorations.pathmorphing}
\usetikzlibrary{calc}

\tikzset{%
  set style/.style={#1},
  expand style/.style={%completely expands the <value> of its argument (using \edef) before processing the result as a <key>
    set style/.expanded={#1}% needs braces around the argument
  },
  expand style once/.style={%expands the <value> of its argument once before processing the result as a <key>
    set style/.expand once={#1}% needs braces around the argument
  },
  expand style twice/.style={%expands the <value> of its argument twice before processing the result as a <key>
    set style/.expand twice={#1}% needs braces around the argument
  }
}

\pgfkeys{
  /ww/.is family,
  /ww/.unknown/.code = {} % Prevent "unknown key" errors globally if needed
}

\ExplSyntaxOn
\cs_new_eq:NN \ipLoop \int_step_inline:nn
\ExplSyntaxOff

\pgfkeys{
 /ww/defineline/.is family,/ww/defineline/.cd,
  /ww/defineline/.unknown/.code = {},/ww/defineline/linelength/.store in =\linelength, /ww/defineline/linelength/.default =1%
}

\newcommand{\defineline}[2]{%

  \pgfkeys{/ww/defineline/.cd,linelength}
  \pgfkeys{/ww/defineline/.cd,#2} 

  \pgfkeys{/ww/#1/.is family}%
  \pgfkeys{/ww/#1/.unknown/.code = \pgfkeyssetvalue{/ww/#1/\pgfkeyscurrentname}{##1}}%

  \pgfkeys{/ww/#1/.cd, #2}%

  \pgfkeysgetvalue{/ww/#1/pointcount}{\pointcount}%
  \def\pointsrange{(1.4*\linelength)}

  \pgfmathparse{\pointsrange/(\pointcount-1)}
  \def\pointgap{\pgfmathresult}
  
  \ipLoop{\pointcount}{

    \pgfkeysgetvalue{/ww/#1/originx}{\originx}
    \pgfkeysgetvalue{/ww/#1/originy}{\originy}
    \pgfkeysgetvalue{/ww/#1/angle}{\angle}

   \pgfmathsetmacro{\pointx}{\originx +(-\pointsrange/2 + (##1-1)*\pointgap)*cos(\angle)}   
   \pgfmathsetmacro{\pointy}{\originy +(-\pointsrange/2 + (##1-1)*\pointgap)*sin(\angle)}
   \pgfmathsetmacro{\linelengthtwo}{\linelength}

   \edef\keypx{/ww/#1/ipx##1}
   \edef\keypy{/ww/#1/ipy##1}
   \pgfkeyslet{\keypx}{\pointx}
   \pgfkeyslet{\keypy}{\pointy}

   \edef\keylinelength{/ww/#1/ilinelength}
   \pgfkeyslet{\keylinelength}{\linelengthtwo}

  }

}

\tikzset{gluon/.style={decorate, decoration={coil,amplitude=2pt, segment length=4pt},color=purple}}

\pgfkeys{
  /ww/prop/.cd,
  /ww/prop/.is family,
  /ww/prop/fromline/.initial = 1,
  /ww/prop/fromline/.store in = \fromline,
  /ww/prop/toline/.initial = 1,
  /ww/prop/toline/.store in=\toline,
  /ww/prop/fromip/.initial = 1,
  /ww/prop/fromip/.store in=\fromip,
  /ww/prop/toip/.initial = 1,
  /ww/prop/toip/.store in=\toip,
  /ww/prop/stublength/.store in=\stublength,
  /ww/prop/stublength/.default = 0.5,
  /ww/prop/markendpoints/.store in = \markendpoints,
  /ww/prop/markendpoints/.default = false,
  /ww/prop/startpointlabel/.store in=\startpointlabel,
  /ww/prop/startpointlabel/.default = \phantom{.},  
  /ww/prop/endpointlabel/.store in=\endpointlabel,
  /ww/prop/endpointlabel/.default = \phantom{.},  /ww/prop/labeldistone/.store in = \labeldistone,
  /ww/prop/labeldistone/.default = 0.5,
  /ww/prop/labeldisttwo/.store in = \labeldisttwo,
  /ww/prop/labeldisttwo/.default = 0.5,
  /ww/prop/looseval/.store in = \looseval,
  /ww/prop/looseval/.default = 1.4,
  /ww/prop/color/.store in = \propcolor,
  /ww/prop/color/.default = purple,
}%

\pgfkeys{
  /ww/drawline/.is family,/ww/drawline/.cd,
  /ww/drawline/.unknown/.code = {},%
  /ww/drawline/customstyle/.initial = {line width=1},
  /ww/drawline/customstyle/.default = {line width=1},
  /ww/drawline/showtwistors/.store in = \showtwistors,
  /ww/drawline/showtwistors/.default = false,
  /ww/drawline/showtwistorsout/.store in = \showtwistorsout,
  /ww/drawline/showtwistorsout/.default = false,
  /ww/drawline/showwings/.store in = \showwings,
  /ww/drawline/showwings/.default = false,
  /ww/drawline/showvertex/.store in = \showvertex,
  /ww/drawline/showvertex/.default = false,
  /ww/drawline/labeldist/.store in = \labeldist,
  /ww/drawline/labeldist/.default = 0.5,
  /ww/drawline/wingorientation/.store in = \wingorientation,
  /ww/drawline/wingorientation/.default = inner,
  /ww/drawline/tolabel/.store in = \tolabel,
  /ww/drawline/tolabel/.default = ,
  /ww/drawline/fromlabel/.store in = \fromlabel,
  /ww/drawline/fromlabel/.default = ,
  /ww/drawline/linelength/.store in = \linelength,
}

\newcommand{\drawline}[2][]{%
  \pgfkeys{/ww/drawline/.cd,tolabel,fromlabel,customstyle,showtwistors,showtwistorsout,showwings,wingorientation,showvertex,labeldist}
  \pgfkeys{/ww/drawline/.cd,#1}
  
  \pgfkeysgetvalue{/ww/#2/originx}{\ox}%
  \pgfkeysgetvalue{/ww/#2/originy}{\oy}%
  \pgfkeysgetvalue{/ww/#2/angle}{\ang}%
  \pgfkeysgetvalue{/ww/#2/ilinelength}{\linelength}%

  \pgfmathsetmacro{\oxf}{\ox}%
  \pgfmathsetmacro{\oyf}{\oy}%
  \pgfmathsetmacro{\angf}{\ang}%
  \pgfmathsetmacro{\linelengthf}{\linelength}

  \pgfkeysgetvalue{/ww/drawline/customstyle}{\customstyle}
  
  \draw[expand style = \customstyle, name path=#2]
    ($(\oxf,\oyf) - (\angf:\linelengthf+0.3)$) 
    -- 
    ($(\oxf,\oyf) + (\angf:\linelengthf+0.3)$);

    \ifthenelse{\equal{\showtwistors}{true}}
           {
           
           \ifthenelse{\equal{\showtwistorsout}{false}}{\path ($(\oxf,\oyf) + (\angf:\linelengthf+0.0) + (\angf+90:\labeldist)$) node {\fromlabel};
    \path ($(\oxf,\oyf) - (\angf:\linelengthf+0.0) + (\angf+90:\labeldist)$) node {\tolabel}}{\path ($(\oxf,\oyf) + (\angf:\linelengthf+0.3) + (\angf+90:\labeldist)$) node {\fromlabel};
    \path ($(\oxf,\oyf) - (\angf:\linelengthf+0.3) + (\angf+90:\labeldist)$) node {\tolabel}};

    \ifthenelse{\equal{\showvertex}{true}}{
    \draw[fill=black] ($(\oxf,\oyf) - (\angf:\linelengthf)$) circle (0.05);
    \draw[fill=black] ($(\oxf,\oyf) + (\angf:\linelengthf)$) circle (0.05);
    }
    }{}

    \ifthenelse{\equal{\showwings}{true}}
    {
      \ifthenelse{\equal{\wingorientation}{inner}}
      {
        \draw[expand style = \customstyle] ($(\oxf,\oyf) - (\angf:\linelengthf) - (\angf-30:-0.2) $) -- ($(\oxf,\oyf) - (\angf:\linelengthf) - (\angf-30:0.5)$);
        \draw[expand style = \customstyle] ($(\oxf,\oyf) + (\angf:\linelengthf) - (\angf-150:-0.2)$) -- ($(\oxf,\oyf) + (\angf:\linelengthf) - (\angf-150:0.5)$);
      }{
        \draw[expand style = \customstyle] ($(\oxf,\oyf) - (\angf:\linelengthf) - (\angf+30:-0.2) $) -- ($(\oxf,\oyf) - (\angf:\linelengthf) - (\angf+30:0.5)$);
        \draw[expand style = \customstyle] ($(\oxf,\oyf) + (\angf:\linelengthf) - (\angf+150:-0.2)$) -- ($(\oxf,\oyf) + (\angf:\linelengthf) - (\angf+150:0.5)$);
        }
    }{}

}%

\newcommand{\drawpropagator}[1]{

  \pgfkeys{/ww/prop/.cd,markendpoints,startpointlabel,endpointlabel,labeldistone,labeldisttwo,looseval,color}
  \pgfkeys{/ww/prop/.cd,#1}
  \pgfkeysgetvalue{/ww/\fromline/ipx\fromip}{\ix}%
  \pgfkeysgetvalue{/ww/\fromline/ipy\fromip}{\iy}%
  \pgfkeysgetvalue{/ww/\toline/ipx\toip}{\fx}%
  \pgfkeysgetvalue{/ww/\toline/ipy\toip}{\fy}%

  \pgfkeysgetvalue{/ww/\fromline/angle}{\angi}%
  \pgfkeysgetvalue{/ww/\fromline/pointcount}{\pointcount}%
  \pgfkeysgetvalue{/ww/\toline/angle}{\angt}%

  \pgfmathsetmacro{\angout}{\angi - 90}
  \pgfmathsetmacro{\angin}{\angt - 90}

  \draw[/tikz/gluon, color=\propcolor] (\ix,\iy) to[out=\angout, in = \angin,looseness=\looseval] (\fx,\fy);

  \ifthenelse{\equal{\markendpoints}{true}}
  {
    \path ($(\fx,\fy) + (\angin:-\labeldistone) $) node {\endpointlabel};
    \path ($(\ix,\iy) + (\angout:-\labeldisttwo) $) node {\startpointlabel};
  }

}%

\newcommand{\drawpropagatorstub}[1]{
  \pgfkeys{/ww/prop/.cd,markendpoints,startpointlabel,labeldistone,stublength}
  \pgfkeys{/ww/prop/.cd,#1}
  
  \pgfkeysgetvalue{/ww/\fromline/ipx\fromip}{\ix}%
  \pgfkeysgetvalue{/ww/\fromline/ipy\fromip}{\iy}%

  \pgfkeysgetvalue{/ww/\fromline/angle}{\angi}%
  \pgfkeysgetvalue{/ww/\fromline/pointcount}{\pointcount}%

  \pgfmathsetmacro{\angout}{\angi - 90}

  \draw[/tikz/gluon] (\ix,\iy) -- ++ (\angout:\stublength);

  \ifthenelse{\equal{\markendpoints}{true}}
  {
    \path ($(\ix,\iy) + (\angout:-\labeldistone) $) node {\startpointlabel};
  }
  
}%

\numberwithin{equation}{section}
\begin{document}

% Front page here
\thispagestyle{empty}

\begin{flushright}

\end{flushright}

\vskip2.2truecm
\begin{center}
\vskip 0.2truecm {\Large\bf
%\titleline
{\Large Leading singularities of Wilson loop correlators from twistor Wilson loop diagrams}
}\\
\vskip 1truecm
\vskip 1truecm
{\bf J.M. Drummond\footnote{J.M.Drummond@soton.ac.uk}, M. Rochford\footnote{M.J.Rochford@soton.ac.uk}, R. Wright\footnote{Rowan.Wright@soton.ac.uk}
}
\vskip 0.4truecm
{\emph{School of Physics and Astronomy, University of Southampton, Southampton, SO17 1BJ, UK
}}

\vskip 0.4truecm

\begingroup\bf\large

\endgroup
\vspace{0mm}

\begingroup
\textit{
 }\\
\par
%\texttt{drummond@lapp.in2p3.fr\phantom{\ldots}}
\endgroup

\end{center}

\vskip 1truecm %\Large
%\noindent
\centerline{\bf Abstract} % \normalsize
The leading singularities of one-loop scattering amplitudes in planar \(\mathcal{N}=4\) super Yang-Mills theory are known to factorise into products of tree-level amplitudes, and this can be seen from a number of different perspectives e.g. generalised unitarity or on-shell diagrams. Here we investigate the leading singularities from the perspective of the Wilson loop expectation values to which these amplitudes are dual, in particular making use of the twistor Wilson loop formalism. We show that the factorisation of one-loop leading singularities of a null Wilson loop's expectation value into a product of tree-level objects is manifest at the level of twistor Wilson loop diagrams, and is a simple consequence of planarity, without appeal to e.g. unitarity on the amplitude side of the duality. We then use the same approach to derive compact formulae for the one-loop leading singularities of correlators of multiple light-like Wilson loop operators in terms of tree-level objects. Via the chiral box expansion, these formulae provide a simple route to writing down the $O(g^2)$ correlation function of any number of Wilson loops at any MHV degree.

\medskip

 \noindent

\newpage
\tableofcontents
\newpage

\newpage
\setcounter{page}{1}\setcounter{footnote}{0}
%\tableofcontents
%\newpage

\section{Introduction}
The leading singularities of one-loop planar scattering amplitudes in maximally supersymmetric Yang-Mills theory have been studied from a number of perspectives, including via unitarity cuts \cite{Bern:1994zx}, generalised unitarity \cite{Britto:2004nc}, on-shell diagrams \cite{Arkani-Hamed:2012zlh}, and Grassmannian integral formulae \cite{Arkani-Hamed:2009ljj,Mason:2009qx,Arkani-Hamed:2009nll}.
As has been studied extensively, these amplitudes are dual to the expectation values of null polygonal Wilson loops \cite{Alday:2007hr,Drummond:2007aua,Brandhuber:2007yx,Drummond:2007cf,Bern:2008ap,Drummond:2008aq}, which can be supersymmetrically extended, \cite{Mason:2010yk,Caron-Huot:2010ryg}, in order to extend the duality beyond the MHV sector. While much of the rich structure of scattering amplitudes is made most manifest on the Wilson loop side of the amplitude-Wilson loop duality, the simple structure of one-loop leading singularities (i.e. the residues on maximal cuts), and especially the factorisation of these leading singularities into a product of tree-level objects, has typically been studied as a feature native to amplitudes, e.g. as a consequence of unitarity. 

In a series of recent and forthcoming papers \cite{Drummond:2025ulh, Drummond:2026lvq, multiBCFW}, the correlators of \emph{multiple} light-like Wilson loops in $\mathcal{N}=4$ super Yang-Mills theory have been studied. These objects are not immediately dual to scattering amplitudes, however they have been shown to enjoy much of the same rich structure, such as a BCFW-like recursion relation and a natural generalisation of the \(\bar{Q}\)-equation. A study of leading singularities \emph{from the Wilson loop side of the duality} is therefore timely. Understanding the leading singularities is sufficient to be able to write down the correlator of loop operators in the chiral box basis of local one-loop integrals \cite{Bourjaily:2013mma} as reviewed in \cite{Drummond:2026lvq}. The explicit expressions for the chiral box integrals are known and therefore the problem of writing down the integrated $O(g^2)$ contribution to the correlator of any number of Wilson loops in a compact form may be reduced to the problem of finding formulae for these leading singularities. The direct diagrammatic computation of the integrand and taking residues becomes cumbersome for high Grassmann degree and high multiplicity and therefore it is desirable to have to hand explicit formulae which give the leading singularities directly in terms of tree-level data. Here we will derive such formulae which one can think of as a Wilson loop version of the generalised unitarity formulae which we can apply to correlators of an arbitrary number of light-like loop operators.

This paper is set out as follows. In Section \ref{Sec-Background}, we provide an overview of some basic background material, including the twistor Wilson loop formalism. In Section \ref{Sec-singleWL}, we restrict our attention to the case of the expectation of a single Wilson loop, making use of the twistor Wilson loop formalism of Mason and Skinner \cite{Mason:2010yk} to re-derive expressions for four-mass leading singularities of these objects from the perspective of twistor Wilson loop diagrams. We also sketch out examples for how the same procedure can be applied to lower mass leading singularities. Then, in Section \ref{Sec-multiWL} we extend the analysis to the correlators of multiple Wilson loops and outline the derivation for all of the types of four-mass leading singularities which can arise. In Section \ref{Sec-allLeadingSing}, we present explicit, general formulae for all four-mass leading singularities of Wilson loop correlators, with all lower mass leading singularities included in appendix \ref{lowermassleading}. This amounts to the general solution to the \(O(g^2)\) problem for Wilson loop correlators (for any number of polygonal Wilson loops with any number of sides at any MHV degree).

\section{Review of background material}
\label{Sec-Background}
\subsection{The twistor Wilson loop}
The duality between the expectation values of light-like loop operators and planar MHV scattering amplitudes is well known, and by introducing a supersymmetric extension of the Wilson loop this duality may be extended beyond the MHV sector \cite{Mason:2010yk,Caron-Huot:2010ryg}. Here we work within the twistor Wilson loop formalism of Mason and Skinner \cite{Mason:2010yk} to compute the expectation values of light-like Wilson loops (whose planar limit coincides with planar partial-ordered scattering amplitudes divided by the equivalent MHV tree-level amplitude), as well as the correlation functions of multiple light-like Wilson loops. Note that the latter objects are not (at least in an obvious way) dual to scattering amplitudes.

We will not review the twistor Wilson loop from first principles here, and instead refer the reader to the original references \cite{Boels:2006ir,Mason:2010yk} or to \cite{Drummond:2025ulh} and \cite{Drummond:2026lvq} for a recent overview in the context of multiple Wilson loop correlators. However, let us collect the key operational details of this formalism, and in particular how one writes down and evaluates the twistor Wilson loop diagrams which correspond to a given tree-level object (or loop integrand). Note that here we will be taking the large $N$ limit, working in the $SU(N)$ theory.

\subsubsection{Tree level calculations}
We parametrise the nodal curve of a Wilson loop in twistor space by a sequence of intersecting \(\mathbb{CP}_1\) lines \(X_i\); the intersection point of \(X_{i-1}\) and \(X_i\) is given by the twistor \(Z_i\), which is the momentum twistor of the $i$'th particle in the dual scattering amplitude (in the case of a single Wilson loop's expectation value). Each twistor $Z_i$ has an associated Grassmann odd counterpart $\chi_i$ and the pair form a supertwistor $\mathcal{Z}_i = (Z_i | \chi_i) \in \mathbb{P}^{3|4}$. We parametrise the lines as
\be
\mathcal{Z}_i(s) = s \mathcal{Z}_{i-1} + \mathcal{Z}_i
\ee
so that \(s = \infty\) corresponds to \( \mathcal{Z}_{i-1}\) and \(s=0\) corresponds to \( \mathcal{Z}_i\).

At \(O(g^0)\) in the N$^k$MHV sector, a correlator of \(m\) Wilson loops (or expectation value, in the case \(m=1\)) receives contributions from twistor diagrams with \(k\) propagators, since each propagator,
\be
\langle \mathcal{A}_a(\mathcal{Z}_i(s)) \mathcal{A}_b(\mathcal{Z}_j(t)) \rangle = -\frac{4 \pi^2}{C_F} \delta_{ab} \int \frac{D^2c}{c_1 c_2 c_3} \bar{\delta}^{4|4}\bigl(c_1 \mathcal{Z}_* + c_2 \mathcal{Z}_i(s) + c_3 \mathcal{Z}_j(t)\bigr)
\label{prop}
\ee
carries Grassmann degree \(4\). Note that here \(\mathcal{Z}_*\) is an arbitrary \emph{reference twistor} which arises as a result of having fixed an axial gauge. Individually, twistor Wilson loop diagrams will retain dependence on this twistor, but this dependence will cancel out in the overall answer for  well-defined observables. The colour factor in (\ref{prop}) is given by $C_F = \frac{N^2-1}{2N}$.

Twistor lines may have multiple propagators ending on them (and if the order of the insertions differs between two diagrams, they are distinct diagrams), but diagrams with multiple propagators between the same pair of twistor lines evaluate to zero and are thus discarded. As described in \cite{Drummond:2025ulh}, we also use a prescription such that diagrams where a propagator runs between \emph{adjacent} twistor lines will evaluate to zero and are thus also discarded. 

\begin{figure}
\begin{center}

\begin{tikzpicture}[scale=0.9]
    \path[use as bounding box] (-3.5,-3) rectangle (3.5,2.0);
  \defineline{lin1m}{originx=-3, originy=1.75, angle=120, pointcount=5}
  \defineline{lin2m}{originx=-3, originy=-1.75, angle=60, pointcount=5}
  \defineline{lin3m}{originx=1, originy=1.75, angle=240, pointcount=3}
  \defineline{lin4m}{originx=1, originy=-1.75, angle=300, pointcount=3}

  \drawline[showwings=true, showtwistors=true, fromlabel={\small $$ \hspace{2mm}},tolabel={\small $$}]{lin1m}
  \drawline[showwings=true, wingorientation=inner, showtwistors=true, fromlabel={\small $$ \, },tolabel={\small $$}]{lin2m}
  \drawline[showwings=true, showtwistors=true, fromlabel={\small \hspace{2mm} $$},tolabel={\small $$}]{lin3m}
  \drawline[showwings=true, showtwistors=true, fromlabel={\small $$},tolabel={\small $$}]{lin4m}

  \drawpropagator{fromline=lin1m, toline=lin3m, fromip=4, toip=2,looseval=0.8}

  \drawpropagator{fromline=lin1m, toline=lin2m, fromip=2, toip=4,looseval=2}

  \drawpropagator{fromline=lin4m, toline=lin2m, fromip=2, toip=2,looseval=0.8}
 
\end{tikzpicture}
\end{center}
\caption{An example of a planar diagram which contributes to the correlator of two Wilson loops at N\(^3\)MHV.}
\label{planarExample1}
\end{figure}
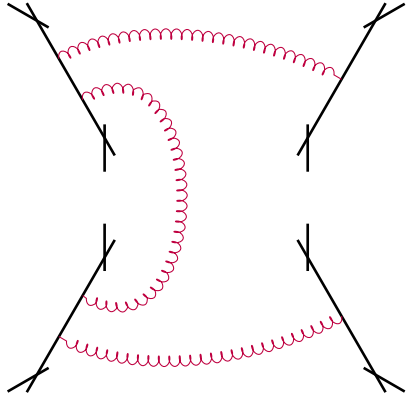

For a single Wilson loop, the heuristic to determine planar diagrams is straightforward: one depicts propagators as running \emph{inside} the Wilson loop, and planar diagrams are those which can be drawn without any propagators crossing. Equivalently, we may draw the propagators \emph{outside} the Wilson loop(s), and we prefer this heuristic here since it generalises to the connected contributions to correlators of multiple Wilson loops (provided we draw the colour trace going consistently clockwise or anticlockwise for all of the depicted Wilson loops; in this work we always depict the colour trace running clockwise). An example of a planar twistor diagram which contributes to the connected part of a correlation function of two Wilson loops is given in Fig. \ref{planarExample1}. In the case of connected contributions to correlators of multiple Wilson loops, non-crossing propagators are necessary but not sufficient for a diagram to be colour-dominant: we have the additional requirement that each Wilson loop must have at least two propagators which run from it to different Wilson loops (or the Lagrangian line(s) in the case of loop-level diagrams). Otherwise, the diagrams come with vanishing colour factor due to the tracelessness of the $SU(N)$ generators.

An equivalent heuristic which works for connected parts of correlators of multiple Wilson loops is that one of the Wilson loops may be depicted outside all of the others but with its colour trace running in the opposite direction, e.g. \emph{anticlockwise} instead of clockwise. Then, if all of the propagators are drawn within this Wilson loop (but outside the others), planar diagrams are again those which can be drawn without crossing and with at least two propagators on each Wilson loop running to different loops or the Lagrangian line. 

For a single Wilson loop, in the large \(N\) limit planar diagrams come with an overall colour factor of \(1\). More generally, for the correlator of \(m\) Wilson loops, the colour factor of the planar, fully-connected diagrams is \(\frac{1}{N^{2m-2}}\) in the large \(N\) limit (and therefore in fact colour suppressed relative to disconnected contributions). 

After generating all of the appropriate connected leading-$N$ twistor diagrams, the contribution (to be dressed with the appropriate colour factor, \(\frac{1}{N^{2m-2}}\)) from a given diagram may be written down using the following rules:
\begin{itemize}
\item For an external twistor line \(X_i\) on which \(n \geq 1\) propagators end,
\be
\int \frac{ds_{i,1}}{s_{i,1}} \int \frac{ds_{i,2}}{s_{i,2}-s_{i,1}} \int  \frac{ds_{i,3}}{s_{i,3}-s_{i,2}}\, ...  \int  \frac{ds_{i,n}}{s_{i,n}-s_{i,n-1}}.
\ee
where integration variables \(s_{i,k}\) are associated to the ends of the propagators which end on the twistor line \(X_i\), with \(s_{i,1}\) closest to \(Z_{i}\) and \(s_{i,n}\) closest to \(Z_{i-1}\).

\item For a propagator which runs from \((X_{i-1} X_i)\) to \((X_{j-1} X_j)\), with integration variables \(s_{i,m_1}\) and \(s_{j,m_2}\) associated to the insertion points on the twistor lines,
\be
\int \frac{D^2a_{(i,j)}}{a_{(i,j),1}a_{(i,j),2}a_{(i,j),3}} \bar{\delta}^{4|4}\bigl(a_{(i,j),1}\mathcal{Z}_* + a_{(i,j),2} s_{i,m_1} \mathcal{Z}_{i-1} + a_{(i,j),2}\mathcal{Z}_i + a_{(i,j),3} s_{j,m_2} \mathcal{Z}_{j-1} + a_{(i,j),3}\mathcal{Z}_j\bigr)
\ee
Note that we don't have a numerical prefactor in front of the propagator because we have absorbed such terms into the overall `colour factor' which evalutes to \(\frac{1}{N^{2m-2}}\) (where \(m\) is the number of Wilson loops) for colour-dominant diagrams in the large \(N\) limit. 
\end{itemize}
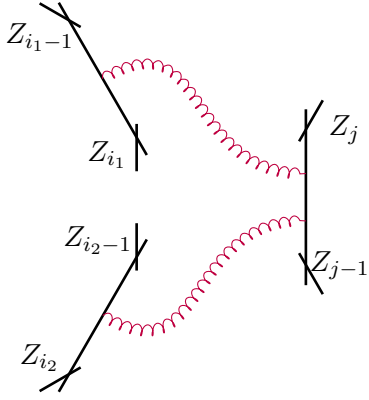
\begin{figure}
\begin{center}
\begin{tikzpicture}[scale=0.9]
   \path[use as bounding box] (-3.5,-2.5) rectangle (22.5,2.5);
  \defineline{lin1m}{originx=4.5, originy=2, angle=120, pointcount=3}
  \defineline{lin2m}{originx=4.5, originy=-1.5, angle=60, pointcount=3}
  \defineline{lin3m}{originx=7.5, originy=0.25, angle=270, pointcount=5}

  \drawline[showwings=true, showtwistors=true, fromlabel={\small $Z_{i_1-1}$},tolabel={\small $Z_{i_1}$}]{lin1m}
  \drawline[showwings=true, wingorientation=inner, showtwistors=true, fromlabel={\small $Z_{i_2-1}\,\,\,\,$},tolabel={\small $Z_{i_2}$}]{lin2m}
  \drawline[showwings=true, showtwistors=true, fromlabel={\small $Z_{j-1}$},tolabel={\small $\,\,Z_{j}$}]{lin3m}

  \drawpropagator{fromline=lin1m, toline=lin3m, fromip=2, toip=2}
  \drawpropagator{fromline=lin2m, toline=lin3m, fromip=2, toip=4}

\end{tikzpicture}
\end{center}
\caption{An example of a planar diagram which contributes at N\(^2\)MHV to the connected part of the correlation function of two Wilson loops. Here \(i_1\) and \(i_2\) label cusps on one Wilson loop, and \(j\) labels a cusp on another Wilson loop}
\label{planarExample2}
\end{figure}
For instance, for the diagram depicted in Fig. \ref{planarExample2} we have contribution (here for notational brevity we relabel some integration variables)
\begin{align}
\frac{1}{N^2}\int \frac{D^2a}{a_1a_2a_3}\frac{D^2b}{b_1b_2b_3} \frac{ds_1}{s_1} \frac{ds_2}{s_2-s_1} &\bar{\delta}^{4|4}(a_1 \mathcal{Z}_* + a_2s \mathcal{Z}_{i_1-1} + a_2\mathcal{Z}_{i_1} + a_3u_1\mathcal{Z}_{j-1} + a_3\mathcal{Z}_j ) \notag \\
&\times \bar{\delta}^{4|4}(b_1 \mathcal{Z}_* + b_2t\mathcal{Z}_{i_2-1}+b_2\mathcal{Z}_{i_2} + b_3u_2\mathcal{Z}_{j-1}+b_3\mathcal{Z}_{j})
\end{align}
Either by explicitly integrating out over the support of the delta functions, or by using the Feynman rules spelled out in \cite{Drummond:2025ulh}, it is straightforward to obtain expressions for $O(g^0)$ N\(^k\)MHV diagrams as a product of \(k\) $R$-invariants, sometimes involving shifted arguments such as 
\be
(i_1-1 \, i_1) \cap (* \, j-1 \, j) \equiv \mathcal{Z}_{i_1-1} \langle i_1 \,*\, j-1\, j \rangle - \mathcal{Z}_{i_1} \langle i_1-1\, *\, j-1\, j\rangle.
\ee
The diagram illustrated in Fig. \ref{planarExample2} then evaluates to
\be
\frac{1}{N^2}[*,i_1-1,i_1,j-1,j][*,i_2-1,i_2,(j-1 j) \cap (* \, i_1-1 \, i_1), j]\,,
\ee
where we recall the definition of the dual-superconformal `$R$-invariant'
\be
[a,b,c,d,e] = \frac{\bar{\delta}^{0|4}(\chi_a \langle bcde \rangle + \rm{cyc} )}{\langle abcd \rangle \langle bcde \rangle \langle cdea \rangle \langle deab \rangle \langle eabc \rangle}.
\ee

For a single Wilson loop, the result of this calculation is precisely the planar tree-level \(\mathcal{N}=4\) amplitude at the corresponding MHV degree, divided by the MHV tree-level amplitude. For multiple Wilson loops, one obtains the connected part of the correlation function, and such computations were presented in detail at tree-level in \cite{Drummond:2025ulh}.

\subsubsection{Calculations at O($g^2$)}
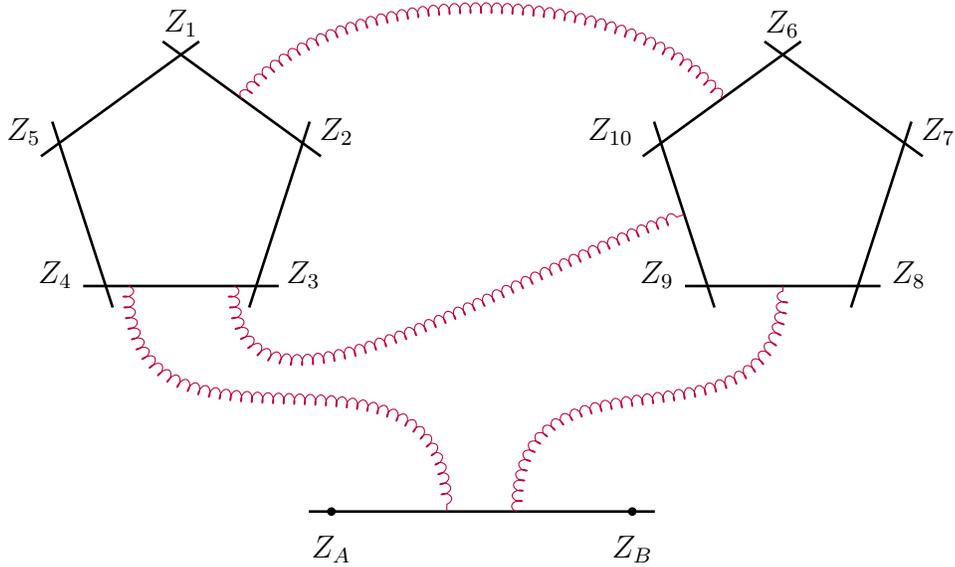
\begin{figure}
\begin{center}
\   \begin{tikzpicture}[baseline={([yshift=-.5ex]current bounding box.center)}]

  \clip (-3.25,-3.75) rectangle (10.75, 4);
  \defineline{lin1}{originx=0.809, originy=2.490, angle=144, pointcount=3}
  \defineline{lin2}{originx=1+0.309, originy=0.951, angle=72, pointcount=3}
  \defineline{lin3}{originx=0, originy=0, angle=0, pointcount=3}
  \defineline{lin4}{originx=-1-0.309, originy=0.951, angle=-72, pointcount=3}
  \defineline{lin5}{originx=-0.809, originy=2.490, angle=216, pointcount=3}
  
  \drawline[wingorientation=outer,showtwistors=true]{lin1}
  \drawline[wingorientation=outer,showtwistors=true,fromlabel=$\hspace{18.5mm}Z_{2}$,tolabel=$\hspace{22mm}Z_{3}$]{lin2}
  \drawline[showtwistors=true,wingorientation=outer]{lin3}
  \drawline[wingorientation=outer,showtwistors=true,fromlabel=$\hspace{-23mm}Z_{4}$,tolabel=$\hspace{-19mm}Z_{5}$]{lin4}
  \drawline[wingorientation=outer,showtwistors=true]{lin5}

  \defineline{linA}{originx=4, originy=-3, angle=180, pointcount=4,linelength=2}
  \drawline[showtwistors=true,showvertex=true,wingorientation=outer,tolabel=$Z_B$,fromlabel=$Z_A$]{linA}

  \defineline{lin6}{originx=8.809, originy=2.490, angle=-36, pointcount=3}
  \defineline{lin7}{originx=9+0.309, originy=0.951, angle=72, pointcount=3}
  \defineline{lin8}{originx=8, originy=0, angle=0, pointcount=3}
  \defineline{lin9}{originx=7-0.309, originy=0.951, angle=-72, pointcount=3}
  \defineline{lin10}{originx=8-0.809, originy=2.490, angle=216, pointcount=3}

    \drawline[showtwistors=true,wingorientation=outer,tolabel=$Z_{6} \hspace{6mm}$]{lin6}
      \drawline[showtwistors=true,wingorientation=outer]{lin8}
\drawline[showtwistors=true,wingorientation=outer,fromlabel=$\hspace{-23mm}Z_{9}$,tolabel=$\hspace{-23mm}Z_{10}$]{lin9}
  \drawline[showtwistors=true,wingorientation=outer]{lin10}
\drawline[showtwistors=true,wingorientation=outer,tolabel=$\hspace{23mm}Z_{8}$,fromlabel=$\hspace{18.5mm}Z_{7}$]{lin7}

  \drawpropagator{fromline=lin1, toline=lin10, fromip=2, toip=2,looseval=0.8}
  \drawpropagator{fromline=lin3, toline=lin9, fromip=3, toip=2,looseval=1}
  \drawpropagator{fromline=lin3, toline=linA, fromip=1, toip=3}
  \drawpropagator{fromline=linA, toline=lin8, fromip=2, toip=2}

\node[] at (0,3.5) {$Z_1$};

       \end{tikzpicture}

    \end{center}
    \caption{A planar twistor diagram which contributes at N\({}^2\)MHV to a pentagon-pentagon correlator at O($g^2$).}
\label{oneLoopExample}
\end{figure}

Computations at $O(g^2)$ may be performed using the method of \emph{Lagrangian insertions}, as reviewed in \cite{Drummond:2026lvq}. In terms of an expansion in twistor diagrams, one includes diagrams with a \emph{Lagrangian line} \(X_{AB}\) (beneath the Wilson loops, with propagators attaching to this line from above and with the colour trace running from left to right to preserve our `non-crossing' planarity heuristics) and only includes diagrams where at least two propagators end on this Lagrangian line. The rules for tree-level diagrams given above are then supplemented with the rule for the Lagrangian line: if \(r\) propagators end on it, we include 
\be
g^2\int \frac{d^{4|8}x_{AB}}{\pi^2} \biggl(\prod_{p=1}^n \int ds_{x,p} \biggr) \frac{1}{(s_{x,1}-s_{x,2})(s_{x,2}-s_{x,3})...(s_{x,n}-s_{x,1})}.
\ee
Here,
\be
g^2 = \frac{g^2_{\rm YM}N}{16\pi^2}
\ee
where we continue to work in the conventions and normalisations of \cite{Drummond:2026lvq}. The variables \(s_{x,m}\) are associated to the propagators which end on the Lagrangian line and the integral over \(x_{AB}\) is simply the loop integration. For instance, an $O(g^2)$ contribution to the connected part of the correlator of two Wilson loops at N$^2$MHV is depicted in Fig. \ref{oneLoopExample}. This diagram has associated expression
\begin{align}
\frac{g^2}{N^2} &\int \frac{d^{4|8}x_{AB}}{\pi^2} \frac{D^2a}{a_1a_2a_3} \frac{D^2b}{b_1b_2b_3} \frac{D^2c}{c_1c_2c_3} \frac{D^2d}{d_1d_2d_3} \notag \\
& \times \frac{d\rho_1 d\rho_2}{(\rho_1-\rho_2)(\rho_2-\rho_1)} \frac{ds_1}{s_1}\frac{ds_2}{s_2-s_1} \frac{dt}{t} \frac{du}{u} \frac{dv}{v} \frac{dw}{w} \notag \\
&\times \bar{\delta}^{4|4}(a_1\mathcal{Z}_* + a_2s_1Z_{3} + a_2\mathcal{Z}_{4} + a_3\rho_2\mathcal{Z}_{A} + a_3\mathcal{Z}_{B} ) \notag \\
&\times \bar{\delta}^{4|4}(b_1\mathcal{Z}_* + b_2 s_2 \mathcal{Z}_3 + b_2 \mathcal{Z}_4 + b_3 u \mathcal{Z}_{9} + b_3 \mathcal{Z}_{10}) \notag \\
&\times \bar{\delta}^{4|4}(c_1 \mathcal{Z}_* + c_2 t \mathcal{Z}_{1} + c_2 \mathcal{Z}_{2} + w c_3 \mathcal{Z}_{10} + c_3 \mathcal{Z}_6) \notag \\
&\times \bar{\delta}^{4|4}(d_1 \mathcal{Z}_* + d_2v \mathcal{Z}_{8} + d_2 \mathcal{Z}_{9} + d_3\rho_2 \mathcal{Z}_{A} + d_3 \mathcal{Z}_{B}) \,,
\end{align}
which evaluates (either by manually integrating out the delta functions, or by using the Feynman rules spelled out in \cite{Drummond:2025ulh}) to
\begin{equation}
 \frac{g^2}{N^2} \int \frac{d^{4|8}}{\pi^2} [*,1,2,10,6][*,3,4,A,\widehat{B}_9][*,3,\widehat{4}_B,9,10][*,8,9,A,\widehat{B}_4].
\end{equation}
Here we introduce the notation
\be
\widehat{\mathcal{Z}}_{i,j} \equiv (i-1 \,i) \cap (j-1 \, j \, *)
\ee
where, for the Lagrangian line \((AB)\), we identify \(A-1=B\) and \(B-1=A\).

Although loop integrands are therefore easy to generate, the loop \emph{integration} remains to be performed. In \cite{Drummond:2026lvq}, we performed the loop integration for correlators of two Wilson loops at \(O(g^2)\) by using the \emph{chiral box basis} for one-loop integrals originally introduced in \cite{Bourjaily:2013mma}. Since all of these integrals are known, the only obstacle to solving the general $O(g^2)$ problem is therefore determining an analytic form of the leading singularities of the loop integrand, which supply the coefficients of the chiral box integrals. The derivation of these leading singularities is the focus of the present paper. 

\subsection{Schubert problems and Schubert solutions}
\label{schubertproblem}
\begin{figure}
\begin{center}
  \begin{tikzpicture}
% The square
\draw (0,0) -- (2,0) -- (2,2) -- (0,2) -- cycle;

% The legs
\draw (0,2) -- (-0.4,2);
\draw (0,2) -- (0,2.4);
\draw (0,0) -- (-0.4,0);
\draw (0,0) -- (0,-0.4);
\draw (2,2) -- (2.283,2.283);
\draw (2,0) -- (2.4,0);
\draw (2,0) -- (2,-0.4);

% The labels 
\node at (-0.4,1.7) {$7$};
\node at (0.2,2.5) {$1$};
\node at (2.4,2.4) {$2$};
\node at (2.3,0.3) {$3$};
\node at (1.8,-0.4) {$4$};
\node at (0.2,-0.4) {$5$};
\node at (-0.4,0.3) {$6$};
\end{tikzpicture}
\end{center}
\caption{A box diagram drawn for the three-mass cut \((12)(23)(45)(67)\).}
\label{boxDiagram}
\end{figure}
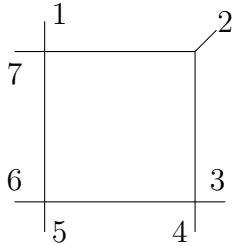

The loop integrands of Wilson loop expectation values, and correlators of multiple Wilson loops, have physical poles of the form \(\langle A \, B \, i-1 \, i\rangle\), and maximal cuts correspond to setting four such twistor brackets simultaneously to zero. For reasons native to the dual amplitude, it is common to refer to the different types of singular configuration as 'zero mass', 'one mass', 'two mass easy', 'two mass hard', 'three mass' and 'four mass'. For instance an example of a three-mass singular configuration would be 
\be
\langle AB12 \rangle = \langle AB23 \rangle = \langle AB45 \rangle = \langle AB67 \rangle = 0
\ee
and the nomenclature becomes clear when drawing the `box' picture which is natural on the amplitude side of the duality, as depicted in Fig. \ref{boxDiagram}; namely, there are three massive corners! Although this terminology is much less natural for correlators of multiple Wilson loops, we will continue to use it here. \bigskip

The twistor bracket \(\langle A B i-1 \, i\rangle\) will equal zero precisely when the twistor line \(AB\) intersects the twistor line \((i-1 \, i)\), and so finding a configuration of the Lagrangian line which simultaneously sets all four poles equal to zero is equivalent to finding a line which simultaneously intersects four other specified lines in \(\mathbb{PT} \subset \mathbb{CP}^3\). This a classical problem from enumerative geometry usually called a \emph{Schubert problem}, and a key result is that, for four sufficiently generic lines, there are \emph{two} discrete solutions, which we will call \emph{Schubert solutions}. Even on the same cut, an observable therefore has \emph{two} residues (i.e. leading singularities), with one for each Schubert solution. For a detailed review of the various Schubert problems and their solutions in the context of planar scattering amplitudes, we refer the reader to \cite{Arkani-Hamed:2010pyv}.

In \cite{Drummond:2026lvq}, we gave a full taxonomy of the Schubert solutions to each class of Schubert problem, and we adopt precisely the same conventions here. 

\section{Leading singularities for a single Wilson loop}
\label{Sec-singleWL}
Let us specialise for now to the case of the expectation value of a single Wilson loop. Although compact expressions for the leading singularities are known via the dual amplitude, this will provide a useful introduction to the Wilson loop-based derivations which we will then apply to correlators of multiple Wilson loops in the next section. 

\subsection{Four mass leading singularities}
Let us restrict our attention to leading singularities of four-mass type, by which we mean that we consider the residues associated to the maximal cut 
\begin{equation}
\langle A \, B \, i-1 \, i \rangle = \langle A \, B \, j-1 \, j \rangle = \langle A \, B \, k-1 \, k \rangle = \langle A \, B \, l-1 \, l \rangle = 0
\label{quadruplePole}
\end{equation}
where all of \(i\), \(j\), \(k\) and \(l\) are separated by at least two (so that, in the 'box diagram', all four corners are massive), and we take \(i < j < k < l\) modulo \(n\), where \(n\) is the multiplicity of the Wilson loop/planar scattering amplitude. These leading singularities will receive contributions from the fewest diagrams (e.g. just one, at N\(^2\)MHV, and none at lower MHV degree), and moreover the other types of leading singularity can be obtained from these by taking an appropriate collinear limit as discussed in \cite{Bourjaily:2013mma}.

\begin{figure}
\begin{tikzpicture}[scale=0.9]

  \path[use as bounding box] (-3.5,-3) rectangle (3.5,2.5);

  \defineline{lin1m}{originx=7.75, originy=-1, angle=15, pointcount=3}
  \defineline{lin4m}{originx=9.75, originy=1.5, angle=80, pointcount=3}
  \defineline{lin5m}{originx=2, originy=1.5, angle=280, pointcount=3}
  \defineline{lin6m}{originx=4.25, originy=-1, angle=345, pointcount=3}

  \defineline{linA}{originx=6, originy=-2.5, angle=180, pointcount=4,linelength=6}
  
\drawline[showtwistors=true,showvertex=true,wingorientation=outer,tolabel=$Z_B$,fromlabel=$Z_A$]{linA}
  
    \drawline[showwings=true, showtwistors=true, fromlabel={\small $Z_{j-1}$},tolabel={\small $Z_{j}$}]{lin1m}
  \drawline[showwings=true, showtwistors=true, fromlabel={\small $Z_{i-1}$},tolabel={\small $Z_{i}$}]{lin4m}
  \drawline[showwings=true, showtwistors=true, fromlabel={\small \,  $Z_{l-1}$},tolabel={\small $\,\,\,Z_{l}$}]{lin5m}
  \drawline[showwings=true, showtwistors=true, fromlabel={\small $Z_{k-1}$},tolabel={\small $\,\,Z_{k}$}]{lin6m}

   \drawpropagator{fromline=lin5m, toline=linA, fromip=2, toip=4}
   \drawpropagator{fromline=lin6m, toline=linA, fromip=2, toip=3}
   \drawpropagator{fromline=lin1m, toline=linA, fromip=2, toip=2}
   \drawpropagator{fromline=lin4m, toline=linA, fromip=2, toip=1}

\end{tikzpicture}
    \caption{The only planar twistor diagram for a single Wilson loop expectation value which has non-zero residue on the quadruple pole in Eq. \ref{quadruplePole}.}
\label{planarFourMassDiagram}

\end{figure}
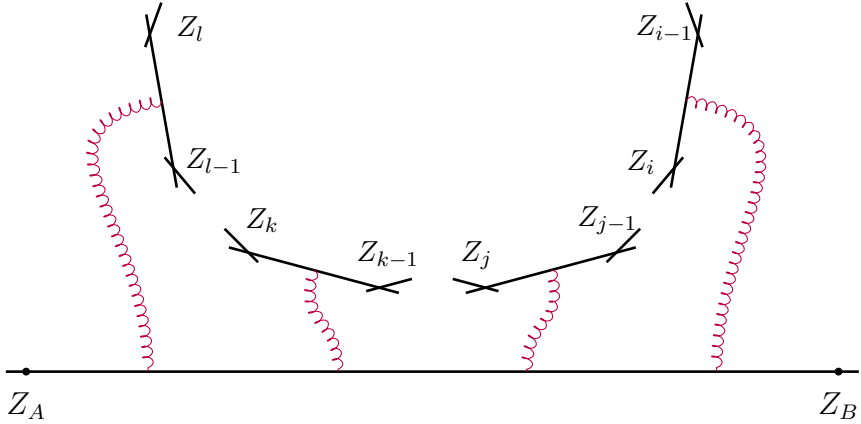

In the expansion of an N\({}^2\)MHV Wilson loop expectation value in terms of twistor Wilson loop diagrams, there is only a single planar diagram which has a non-zero residue on this cut, shown in Fig. \ref{planarFourMassDiagram}; note that the same diagram is relevant for both Schubert solutions. 

This diagram evaluates to (here we take the planar limit so that all planar diagrams come with factor \(1\))
\be
g^2 \int \frac{d^{4|8}x_{AB}}{\pi^2} [*, A, \widehat{B}_i,j-1,j][*,A,\widehat{B}_j,k-1,k][*,A,\widehat{B}_k,l-1,l][*,A,\widehat{B}_l,i-1,i].
\ee

Now recall that the four-mass Schubert problem under consideration has two solutions; that is, there are two different configurations of the Lagrangian line \((AB)\) which will simultaneously intersect \((i-1 \ i)\), \((j-1 \ j)\), \((k-1 \ k)\) and \((l-1 \ l)\). Let us adopt the notation of \cite{Bourjaily:2013mma} and write \(\alpha_1,\beta_1,\gamma_1,\delta_1\) for the intersection of the first Schubert solution with each of \((i-1,i)\), \((j-1,j)\), \((k-1,k)\) and \((l-1,l)\) respectively, and similarly \(\alpha_2,\beta_2,\gamma_2,\delta_2\) for the equivalent intersections for the second Schubert solution. Explicit expressions for these intersection twistors are given in that reference and we also supply expressions in Section \ref{Sec-allLeadingSing} when we collect together expressions for all leading singularities. 

Computing the residue of this diagram on this quadruple cut\footnote{Strictly speaking, for the integrand as defined each of these should come with a pre-factor of \(\frac{1}{\pi^2}\), which we omit here and throughout as is standard practice in the literature.} (on each Schubert solution), one finds residues (after performing the fermionic integration) 
\begin{equation}
C_1(i,j,k,l) = -\phi_1(i,j,k,l)[\delta_1, i-1, i, j-1,j ][\beta_1, k-1, k, l-1, l]
\label{C1}
\end{equation}
and
\begin{equation}
C_2(i,j,k,l)=\phi_2(i,j,k,l)[\alpha_2, j-1, j, k-1,k ][\gamma_2, l-1, l, i-1, i]
\label{C2}
\end{equation}
where we have defined the conformally-invariant factors
\be
\phi_1(i,j,k,l) = \biggl(1 - \frac{\langle\beta_1 \, l \, i-1 \, i\rangle \langle \delta_1 \, j \, k-1 \, k \rangle}{\langle\beta_1 \, l \, k-1 \, k\rangle\langle \delta_1 j \, i-1 \, i \rangle} \biggr)^{-1}
\ee
and
\be
\phi_2(i,j,k,l) = \biggl(1 - \frac{\langle \alpha_2 \, k \, l-1 \, l \rangle \langle \gamma_2 \, i \, j-1 \, j \rangle}{\langle \alpha_2 \, k \, j-1 \, j \rangle \langle \gamma_2 \, i \, l-1 \, l\rangle} \biggr)^{-1}.
\ee
which we will continue to make use of later when discussing four mass cuts for correlators of multiple Wilson loops.

Note that this way of writing the leading singularities obscures the fact that they are simply related by interchanging the two Schubert solutions (which in this case amounts to flipping the sign choice on a discriminant); however, as described in \cite{Bourjaily:2013mma}, this choice is advantageous since each solution then separately reduces smoothly to the three-mass, two-mass (easy and hard), one-mass and zero-mass leading singularities under collinear limits.  Note also that these equations are for the leading singularities, \emph{not} the `box coefficients' which may differ by a factor of \(-1\) on either Schubert solution depending on the convention for the chiral box integrands/integrals.

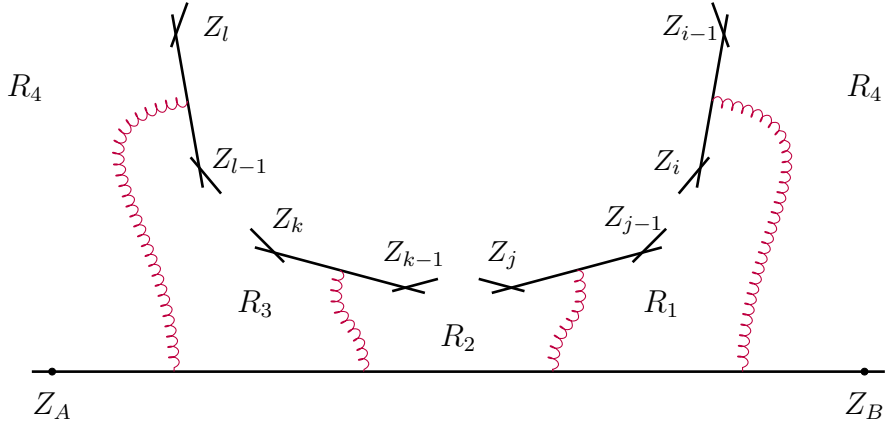
\begin{figure}
\begin{tikzpicture}[scale=0.9]

  \path[use as bounding box] (-3.5,-3) rectangle (3.5,2.5);

  \defineline{lin1m}{originx=7.75, originy=-1, angle=15, pointcount=3}
  \defineline{lin4m}{originx=9.75, originy=1.5, angle=80, pointcount=3}
  \defineline{lin5m}{originx=2, originy=1.5, angle=280, pointcount=3}
  \defineline{lin6m}{originx=4.25, originy=-1, angle=345, pointcount=3}

  \defineline{linA}{originx=6, originy=-2.5, angle=180, pointcount=4,linelength=6}
  
\drawline[showtwistors=true,showvertex=true,wingorientation=outer,tolabel=$Z_B$,fromlabel=$Z_A$]{linA}
  
    \drawline[showwings=true, showtwistors=true, fromlabel={\small $Z_{j-1}$},tolabel={\small $Z_{j}$}]{lin1m}
  \drawline[showwings=true, showtwistors=true, fromlabel={\small $Z_{i-1}$},tolabel={\small $Z_{i}$}]{lin4m}
  \drawline[showwings=true, showtwistors=true, fromlabel={\small \,  $Z_{l-1}$},tolabel={\small $\,\,\,Z_{l}$}]{lin5m}
  \drawline[showwings=true, showtwistors=true, fromlabel={\small $Z_{k-1}$},tolabel={\small $\,\,Z_{k}$}]{lin6m}

   \drawpropagator{fromline=lin5m, toline=linA, fromip=2, toip=4}
   \drawpropagator{fromline=lin6m, toline=linA, fromip=2, toip=3}
   \drawpropagator{fromline=lin1m, toline=linA, fromip=2, toip=2}
   \drawpropagator{fromline=lin4m, toline=linA, fromip=2, toip=1}

\node at (-0.4,1.7) {$R_4$};
\node at (12,1.7) {$R_4$};
\node at (9,-1.5) {$R_1$};
\node at (6,-2) {$R_2$};
\node at (3,-1.5) {$R_3$};

\end{tikzpicture}
    \caption{The four propagators which were required to ensure a non-zero residue on the four mass cut in (\ref{quadruplePole}) split the twistor Wilson loop into four regions, which we label as \(R_i\) for \(i=1,2,3,4\). Additional propagators which are added to increase the MHV degree are forbidden by planarity from crossing between these regions.}
\label{partitions}
\end{figure}

To inspect N\(^k\)MHV contributions to this box coefficient for \(k \geq 3\), one needs to consider copies of this N\({}^2\)MHV diagram with the appropriate number of additional propagators added (e.g. one more for N\(^3\)MHV, two more for N\(^4\)MHV, and so on). We will thus refer to this as the `seed diagram' which generates the relevant diagrams at higher MHV degree in this way. Diagrams in which propagators cross are colour-suppressed and so in the planar theory we essentially partition the Wilson loop into sub-regions in which further propagators can be added, as shown in Fig \ref{partitions}.

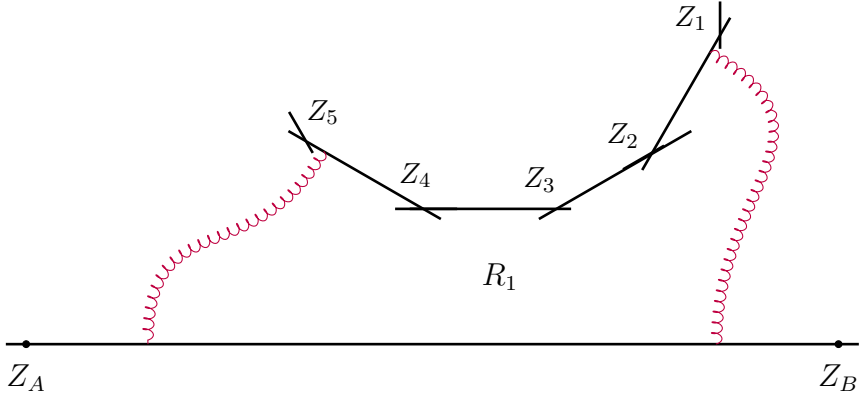
\begin{figure}
\begin{tikzpicture}[scale=0.9]

  \path[use as bounding box] (-3.5,-3) rectangle (3.5,2.5);

  \defineline{lin1m}{originx=6.75, originy=-0.5, angle=0, pointcount=3}
  \defineline{lin2m}{originx=8.7, originy=0, angle=30, pointcount=3}
  \defineline{lin4m}{originx=9.75, originy=1.2, angle=60, pointcount=3}
  \defineline{lin5m}{originx=5, originy=-0, angle=330, pointcount=3}
  \defineline{linA}{originx=6, originy=-2.5, angle=180, pointcount=4,linelength=6}
  
\drawline[showtwistors=true,showvertex=true,wingorientation=outer,tolabel=$Z_B$,fromlabel=$Z_A$]{linA}
  
    \drawline[showwings=false, showtwistors=true, fromlabel={\small $$},tolabel={\small $Z_{4}$}]{lin1m}
  \drawline[showwings=true, showtwistors=true, fromlabel={\small $Z_{1}$},tolabel={\small $Z_{2}$}]{lin4m}
    \drawline[showwings=false, showtwistors=true, fromlabel={\small $$},tolabel={\small $Z_{3}$}]{lin2m}
    \drawline[showwings=true, showtwistors=true, fromlabel={\small $$},tolabel={\small $Z_5$}]{lin5m}

   \drawpropagator{fromline=lin4m, toline=linA, fromip=3, toip=1}
   \drawpropagator{fromline=lin5m, toline=linA, fromip=1, toip=4}

\node at (7,-1.5) {$R_1$};

\end{tikzpicture}
    \caption{The region \(R_1\) from Fig. \ref{partitions}, restricting for simplicity to the case \(i=2\), \(j=5\).}
\label{R1}
\end{figure}

\begin{figure}
\centering
\begin{tabular}{c@{\hspace{3cm}}c}
\begin{tikzpicture}[scale=0.5]
  \path[use as bounding box] (-4,-3) rectangle (4,3);

  \defineline{lin1m}{originx=-0.25, originy=-0.5, angle=0, pointcount=3}
  \defineline{lin2m}{originx=1.7, originy=0, angle=30, pointcount=3}
  \defineline{lin4m}{originx=2.75, originy=1.2, angle=60, pointcount=3}
  \defineline{lin5m}{originx=-2, originy=-0, angle=330, pointcount=3}
  \defineline{linA}{originx=-1, originy=-2.5, angle=180, pointcount=4,linelength=6}
  
\drawline[showtwistors=true,showvertex=true,wingorientation=outer,tolabel={},fromlabel={}]{linA}
  
    \drawline[showwings=false, showtwistors=true, fromlabel={},tolabel={}]{lin1m}
  \drawline[showwings=true, showtwistors=true, fromlabel={},tolabel={}]{lin4m}
    \drawline[showwings=false, showtwistors=true, fromlabel={},tolabel={}]{lin2m}
    \drawline[showwings=true, showtwistors=true, fromlabel={},tolabel={}]{lin5m}

   \drawpropagator{fromline=lin4m, toline=linA, fromip=3, toip=1}
   \drawpropagator{fromline=lin5m, toline=linA, fromip=1, toip=4}
   \drawpropagator{fromline=lin5m, toline=lin2m, fromip=2, toip=2, color=blue}

\end{tikzpicture} &
\begin{tikzpicture}[scale=0.5]
  \path[use as bounding box] (-4,-3) rectangle (4,3);

  \defineline{lin1m}{originx=-0.25, originy=-0.5, angle=0, pointcount=3}
  \defineline{lin2m}{originx=1.7, originy=0, angle=30, pointcount=3}
  \defineline{lin4m}{originx=2.75, originy=1.2, angle=60, pointcount=3}
  \defineline{lin5m}{originx=-2, originy=-0, angle=330, pointcount=3}
  \defineline{linA}{originx=-1, originy=-2.5, angle=180, pointcount=4,linelength=6}
  
\drawline[showtwistors=true,showvertex=true,wingorientation=outer,tolabel={},fromlabel={}]{linA}
  
    \drawline[showwings=false, showtwistors=true, fromlabel={},tolabel={}]{lin1m}
  \drawline[showwings=true, showtwistors=true, fromlabel={},tolabel={}]{lin4m}
    \drawline[showwings=false, showtwistors=true, fromlabel={},tolabel={}]{lin2m}
    \drawline[showwings=true, showtwistors=true, fromlabel={},tolabel={}]{lin5m}

   \drawpropagator{fromline=lin4m, toline=linA, fromip=3, toip=1}
   \drawpropagator{fromline=lin5m, toline=linA, fromip=1, toip=4}
   \drawpropagator{fromline=lin1m, toline=lin4m, fromip=2, toip=2, color=blue}

\end{tikzpicture}\\
\begin{tikzpicture}[scale=0.5]
  \path[use as bounding box] (-4,-3) rectangle (4,3);

  \defineline{lin1m}{originx=-0.25, originy=-0.5, angle=0, pointcount=3}
  \defineline{lin2m}{originx=1.7, originy=0, angle=30, pointcount=3}
  \defineline{lin4m}{originx=2.75, originy=1.2, angle=60, pointcount=3}
  \defineline{lin5m}{originx=-2, originy=-0, angle=330, pointcount=3}
  \defineline{linA}{originx=-1, originy=-2.5, angle=180, pointcount=4,linelength=6}
  
\drawline[showtwistors=true,showvertex=true,wingorientation=outer,tolabel={},fromlabel={}]{linA}
  
    \drawline[showwings=false, showtwistors=true, fromlabel={},tolabel={}]{lin1m}
  \drawline[showwings=true, showtwistors=true, fromlabel={},tolabel={}]{lin4m}
    \drawline[showwings=false, showtwistors=true, fromlabel={},tolabel={}]{lin2m}
    \drawline[showwings=true, showtwistors=true, fromlabel={},tolabel={}]{lin5m}

   \drawpropagator{fromline=lin4m, toline=linA, fromip=3, toip=1}
   \drawpropagator{fromline=lin5m, toline=linA, fromip=1, toip=4}
   \drawpropagator{fromline=lin5m, toline=lin4m, fromip=2, toip=2, color=blue, looseval=2}

\end{tikzpicture} &
\begin{tikzpicture}[scale=0.5]
  \path[use as bounding box] (-4,-3) rectangle (4,3);

  \defineline{lin1m}{originx=-0.25, originy=-0.5, angle=0, pointcount=3}
  \defineline{lin2m}{originx=1.7, originy=0, angle=30, pointcount=3}
  \defineline{lin4m}{originx=2.75, originy=1.2, angle=60, pointcount=3}
  \defineline{lin5m}{originx=-2, originy=-0, angle=330, pointcount=3}
  \defineline{linA}{originx=-1, originy=-2.5, angle=180, pointcount=4,linelength=6}
  
\drawline[showtwistors=true,showvertex=true,wingorientation=outer,tolabel={},fromlabel={}]{linA}
  
    \drawline[showwings=false, showtwistors=true, fromlabel={},tolabel={}]{lin1m}
  \drawline[showwings=true, showtwistors=true, fromlabel={},tolabel={}]{lin4m}
    \drawline[showwings=false, showtwistors=true, fromlabel={},tolabel={}]{lin2m}
    \drawline[showwings=true, showtwistors=true, fromlabel={},tolabel={}]{lin5m}

   \drawpropagator{fromline=lin4m, toline=linA, fromip=3, toip=1}
   \drawpropagator{fromline=lin5m, toline=linA, fromip=1, toip=4}
   \drawpropagator{fromline=lin1m, toline=linA, fromip=2, toip=2, color=blue}

\end{tikzpicture}\\
\multicolumn{2}{c}{
\begin{tikzpicture}[scale=0.5]
  \path[use as bounding box] (-4,-3) rectangle (4,3);

  \defineline{lin1m}{originx=-0.25, originy=-0.5, angle=0, pointcount=3}
  \defineline{lin2m}{originx=1.7, originy=0, angle=30, pointcount=3}
  \defineline{lin4m}{originx=2.75, originy=1.2, angle=60, pointcount=3}
  \defineline{lin5m}{originx=-2, originy=-0, angle=330, pointcount=3}
  \defineline{linA}{originx=-1, originy=-2.5, angle=180, pointcount=4,linelength=6}
  
\drawline[showtwistors=true,showvertex=true,wingorientation=outer,tolabel={},fromlabel={}]{linA}
  
    \drawline[showwings=false, showtwistors=true, fromlabel={},tolabel={}]{lin1m}
  \drawline[showwings=true, showtwistors=true, fromlabel={},tolabel={}]{lin4m}
    \drawline[showwings=false, showtwistors=true, fromlabel={},tolabel={}]{lin2m}
    \drawline[showwings=true, showtwistors=true, fromlabel={},tolabel={}]{lin5m}

   \drawpropagator{fromline=lin4m, toline=linA, fromip=3, toip=1}
   \drawpropagator{fromline=lin5m, toline=linA, fromip=1, toip=4}
   \drawpropagator{fromline=lin2m, toline=linA, fromip=2, toip=2, color=blue}

\end{tikzpicture}
}
\end{tabular}
\caption{The five possibilities for the extra propagator in region $R_1$ as shown in Fig. \ref{R1}. }
\label{fiveChoices}
\end{figure}
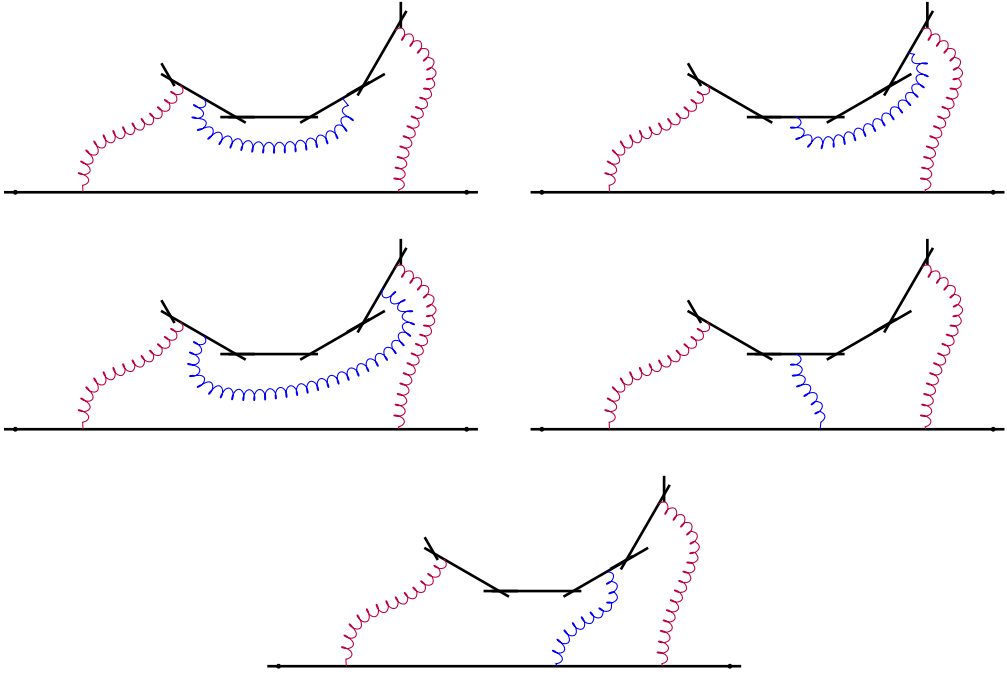

Now consider the contribution to this leading singularity at N\(^3\)MHV, for which we must add one extra propagator, which must stay entirely within one of the four regions. Let us focus our attention on the region \(R_1\). For illustrative purposes, consider the case where \(i=2\) and \(j=5\); we depict this region in Fig. \ref{R1}. Clearly there are five possibilities for the extra propagator one could draw in this region, and we depict all five of these possibilities in Fig. \ref{fiveChoices}. 

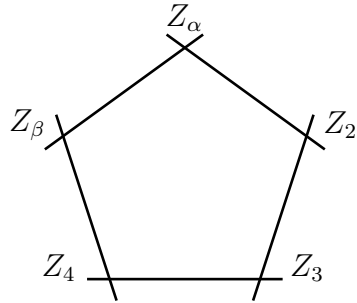
\begin{figure}
\begin{center}
\   \begin{tikzpicture}[baseline={([yshift=-.5ex]current bounding box.center)}]

  \clip (-3.25,-0.5) rectangle (10.75, 4);
  \defineline{lin1}{originx=4.809, originy=2.490, angle=-36, pointcount=3}
  \defineline{lin2}{originx=5+0.309, originy=0.951, angle=72, pointcount=3}
  \defineline{lin3}{originx=4, originy=0, angle=0, pointcount=3}
  \defineline{lin4}{originx=4-1-0.309, originy=0.951, angle=-72, pointcount=3}
  \defineline{lin5}{originx=4-0.809, originy=2.490, angle=216, pointcount=3}
  
  \drawline[wingorientation=outer,showtwistors=true,tolabel=$\hspace{-6mm}Z_{\alpha}$]{lin1}
  \drawline[wingorientation=outer,showtwistors=true,fromlabel=$\hspace{18.5mm}Z_{2}$,tolabel=$\hspace{22mm}Z_{3}$]{lin2}
  \drawline[showtwistors=true,wingorientation=outer]{lin3}
  \drawline[wingorientation=outer,showtwistors=true,fromlabel=$\hspace{-23mm}Z_{4}$,tolabel=$\hspace{-19mm}Z_{\beta}$]{lin4}
  \drawline[wingorientation=outer,showtwistors=true]{lin5}

       \end{tikzpicture}

    \end{center}
    \caption{After sending \((AB)\) to the appropriate Schubert solution, the region \(R_1\) depicted in Fig \ref{R1} takes the depicted geometry, and the cuts of each of the five one-loop diagrams depicted in Fig \ref{fiveChoices} are in one-to-one correspondence with the tree-level diagrams of this pentagonal Wilson loop (at tree level, NMHV).}
\label{treeVersion}
\end{figure}

Recall that, when we compute the residue, we send the Lagrangian line \((AB)\) to the configuration corresponding to the appropriate Schubert solution. Since we defined \(\alpha_{1,2}\) and \(\beta_{1,2}\) (with the subscript depending on the Schubert solution) to be the intersections of those Schubert solutions with the lines \((12)\) and \((45)\) respectively, this means that the region \(R_1\) effectively takes, upon cutting, the configuration of a tree-level, pentagonal Wilson loop with vertices \((\alpha_{1,2}, 2,3,4,\beta_{1,2})\), as shown in Figure \ref{treeVersion}. 

After computing the associated residue, each of the five one-loop diagrams shown in Fig. \ref{fiveChoices} are then in one-to-one correspondence with the five NMHV tree-level diagrams associated to this pentagonal Wilson loop - precisely, the cut is the original N\(^2\)MHV leading singularity multiplied by the value of that tree-level NMHV diagram. For instance, the bottom diagram in Fig \ref{fiveChoices} is identified with the tree-level diagram where the propagator runs from \((\alpha  \beta)\) to \((23)\), and the top-left diagram is identified with the tree-level diagram where the propagator runs from \((4 \beta)\) to \((23)\).  

Since we could instead have chosen one of the other three regions to add the extra propagator to, it is immediately apparent that, at N\(^3\)MHV, the leading singularities are given by (here we relax the specification that \(i=2\) and \(j=5\))
\begin{align}
C_1(i,j,k,l)&\bigl(\langle\mathcal{L}(\alpha_1, i,i+1, ..., j-1, \beta_1)\rangle^{(1,0)} + \langle\mathcal{L}(\beta_1, j,j+1, ..., k-1, \gamma_1)\rangle^{(1,0)} \notag \\
+&\langle\mathcal{L}(\gamma_1, k,k+1, ..., l-1, \delta_1)\rangle^{(1,0)} + \langle\mathcal{L}(\delta_1, l,l+1, ..., i-1, \alpha_1)\rangle^{(1,0)}\bigr) 
\end{align}
\label{cut1N3}
and
\begin{align}
C_2(i,j,k,l) &\bigl(\langle\mathcal{L}(\alpha_2, i,i+1, ..., j-1, \beta_2)\rangle^{(1,0)} + \langle\mathcal{L}(\beta_2, j,j+1, ..., k-1, \gamma_2)\rangle^{(1,0)} \notag \\
+&\langle\mathcal{L}(\gamma_2, k,k+1, ..., l-1, \delta_2)\rangle^{(1,0)} + \langle\mathcal{L}(\delta_2, l,l+1, ..., i-1, \alpha_2)\rangle^{(1,0)}\bigr) 
\end{align}
\label{cut2N3}
and where we have used superscript \((k,l)\) to denote the contribution at N\(^k\)MHV, \(O(g^{2l})\).

If we now consider the leading singularity beyond N\(^3\)MHV, we have to add more propagators. This can mean adding propagators in multiple regions, adding multiple propagators to the same region, or both. For instance, at N\(^4\)MHV we could add two propagators to one region (which would correspond, after cutting, to a tree-level N\(^2\)MHV diagram for that region), or we could add one propagator each to two different regions (which would correspond to a product of two NMHV diagrams from each region). In this way the factorisation of the four mass box coefficient into a product of four tree level amplitudes/Wilson loops is completely manifest at the level of the twistor Wilson loop diagrams; each amplitude is simply associated to one of the four regions into which we may add extra propagators to the original N\(^2\)MHV 'seed diagram'. 

Purely from the perspective of the twistor Wilson loop and without any appeal to e.g. the unitary properties of amplitudes, we thus arrive at the all-MHV degree expressions for the four-mass leading singularities which are familiar from the literature (e.g. given in twistor notation in \cite{Bourjaily:2013mma}), 
\begin{align}
C_1(i,j,k,l) &\langle\mathcal{L}(\alpha_1, i,i+1, ..., j-1, \beta_1)\rangle \times \langle\mathcal{L}(\beta_1, j,j+1, ..., k-1, \gamma_1)\rangle \notag \\
\times&\langle\mathcal{L}(\gamma_1, k,k+1, ..., l-1, \delta_1)\rangle \times \langle\mathcal{L}(\delta_1, l,l+1, ..., i-1, \alpha_1)\rangle
\end{align}
\label{cut1general}
and
\begin{align}
C_2(i,j,k,l) &\langle\mathcal{L}(\alpha_2, i,i+1, ..., j-1, \beta_2)\rangle \times \langle\mathcal{L}(\beta_2, j,j+1, ..., k-1, \gamma_2)\rangle \notag \\
\times&\langle\mathcal{L}(\gamma_2, k,k+1, ..., l-1, \delta_2)\rangle \times \langle\mathcal{L}(\delta_2, l,l+1, ..., i-1, \alpha_2)\rangle.
\end{align}
\label{cut2general}

\subsection{Lower mass example: two mass easy}

All lower-mass leading singularities follow smoothly from collinear limits on the four-mass case, as explained in \cite{Bourjaily:2013mma}. This means that obtaining the four-mass leading singularities is enough to determine all of the (planar) leading singularities for a one-loop single Wilson loop expectation value (or, equivalently, planar amplitude). However, for the sake of completeness, let us now sketch out how the same logic can be applied to lower mass coefficients. 

The derivation is exactly the same as for the four-mass case, except that now the first 'seed' diagram(s) may appear at MHV or NMHV, and there may be different seed diagram(s) for the two Schubert solutions. For instance, consider the two-mass easy cut 
\be
\langle A \, B \, i-1 \, i \rangle = \langle A \, B \, i \, i+1 \rangle = \langle A \, B \, j-1 \, j \rangle = \langle A \, B \, j \, j+1 \rangle = 0 
\label{2mecut}
\ee
where \(j > i+1\). Here we write \(\alpha_{1,2}\), \(\beta_{1,2}\), \(\gamma_{1,2}\) and \(\delta_{1,2}\) for the intersections of each Schubert solution with \((i-1 \, i)\), \((i \, i+1)\), \((j-1 \, j)\) and \((j \, j+1)\) respectively. The first solution to the Schubert problem is given by 
\be
(AB) = (ij)
\ee
and the second is given by 
\be
(AB) = (i-1 \, i \, i+1) \cap (j-1 \, j \, j+1).
\ee

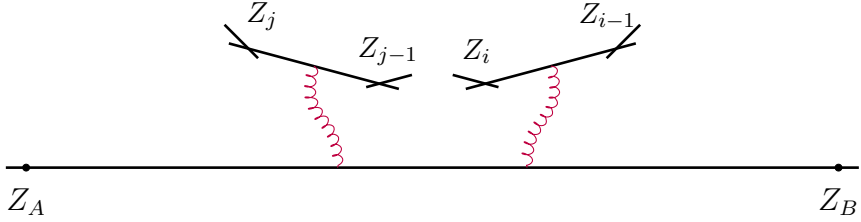
\begin{figure}
\begin{tikzpicture}[scale=0.9]

  \path[use as bounding box] (-3.5,-3) rectangle (3.5,2.5);

  \defineline{lin1m}{originx=7.75, originy=-1, angle=15, pointcount=3}
  \defineline{lin6m}{originx=4.25, originy=-1, angle=345, pointcount=3}

  \defineline{linA}{originx=6, originy=-2.5, angle=180, pointcount=4,linelength=6}
  
\drawline[showtwistors=true,showvertex=true,wingorientation=outer,tolabel=$Z_B$,fromlabel=$Z_A$]{linA}
  
    \drawline[showwings=true, showtwistors=true, fromlabel={\small $Z_{i-1}$},tolabel={\small $Z_{i}$}]{lin1m}
  \drawline[showwings=true, showtwistors=true, fromlabel={\small $Z_{j-1}$},tolabel={\small $\,\,Z_{j}$}]{lin6m}

   \drawpropagator{fromline=lin6m, toline=linA, fromip=2, toip=3}
   \drawpropagator{fromline=lin1m, toline=linA, fromip=2, toip=2}

\end{tikzpicture}
    \caption{A planar twistor diagram for a single Wilson loop expectation value at MHV which has non-zero residue on the quadruple pole in Eq. \ref{2mecut} for the first Schubert solution. The other three diagrams which contribute to this cut are related by \(i \to i+1\) and/or \(j \to j+1\).}
\label{kermit}
\end{figure}

On the first Schubert solution, there are \emph{four} contributing diagrams at MHV: the one depicted in Fig. \ref{kermit}, and those related by \(i \to i+1\) and/or \(j \to j+1\). Crucially, each of these four 'seed' diagrams provide the same tree-level correlators upon dressing with further propagators, because there is no difference between e.g. 
\(\langle \mathcal{L} (\gamma_1, j, j+1, \, ... \, i-1, \alpha_1) \rangle\) and \(\langle \mathcal{L}( \gamma_1, j, j+1, \, ... \, i, \alpha_1) \rangle\) upon using that on this Schubert solution we have \(\alpha_1 = i\). Note that just as the 'dual tree correlators' which one writes down for the other seed diagrams are the same, the extra diagrams which one can get by making use of the extra twistor line e.g. \((i \, i+1)\) in a region have zero residue and so (up to the bosonic factor which comes from the cut of the MHV seed diagram) the contributions from dressing the four seed diagrams with additional propagators are all identical.

Since the sum of the four residues of the MHV seed diagrams is easily seen by direct calculation to be \(-1\), we see by considering the partitioning into regions that the contribution will be 
\be
-\langle \mathcal{L}(\gamma_1,j,j+1,...,i-1,\alpha_1) \rangle \langle \mathcal{L}(\alpha_1,i,i+1,...,j-1,\gamma_1) \rangle.
\ee
However, clearly we have
\be
\alpha_1 = \beta_1 = \mathcal{Z}_i
\ee
and
\be
\gamma_1 = \delta_1 = \mathcal{Z}_j.
\ee
and so we can rewrite this as 
\be
-\langle \mathcal{L}(\delta_1,j,j+1,...,i-1,\alpha_1) \rangle \langle \mathcal{L}(\beta,i,i+1,...,j-1,\gamma_1) \rangle.
\ee
to make more manifest the descent of this expression from the four mass leading singularity on the first Schubert solution, and for a perfect match with the standard result in the literature.

\begin{figure}
\begin{tikzpicture}[scale=0.9]

  \path[use as bounding box] (-3.5,-3) rectangle (3.5,2.5);

  \defineline{lin1m}{originx=7.75, originy=-1, angle=15, pointcount=3}
  \defineline{lin4m}{originx=9, originy=0.4, angle=80, pointcount=3}
  \defineline{lin5m}{originx=3, originy=0.4, angle=280, pointcount=3}
  \defineline{lin6m}{originx=4.25, originy=-1, angle=345, pointcount=3}

  \defineline{linA}{originx=6, originy=-2.5, angle=180, pointcount=4,linelength=6}
  
\drawline[showtwistors=true,showvertex=true,wingorientation=outer,tolabel=$Z_B$,fromlabel=$Z_A$]{linA}
  
    \drawline[showwings=false, showtwistors=false, fromlabel={\small $Z_{j-1}$},tolabel={\small $Z_{j}$}]{lin1m}
  \drawline[showwings=false, showtwistors=true, fromlabel={\small $Z_{i-1}$},tolabel={\small $Z_{i}$}]{lin4m}
  \drawline[showwings=false, showtwistors=false, fromlabel={\small \,  $Z_{l-1}$},tolabel={\small $\,\,\,Z_{l}$}]{lin5m}
  \drawline[showwings=false, showtwistors=true, fromlabel={\small $Z_{j-1}$},tolabel={\small $\,\,Z_{j}$}]{lin6m}

   \drawpropagator{fromline=lin5m, toline=linA, fromip=2, toip=4}
   \drawpropagator{fromline=lin6m, toline=linA, fromip=2, toip=3}
   \drawpropagator{fromline=lin1m, toline=linA, fromip=2, toip=2}
   \drawpropagator{fromline=lin4m, toline=linA, fromip=2, toip=1}

   \node at (3.3,1.6) {$Z_{j+1}$};
   \node at (6.5,-0.75) {$Z_{i+1}$};

\end{tikzpicture}
    \caption{The only N\(^2\)MHV planar twistor diagram for a single Wilson loop expectation value which has non-zero residue on the quadruple pole in Eq. \ref{2mecut} for the second Schubert solution.}
\label{twoMassEasyN2diagram}
\end{figure}
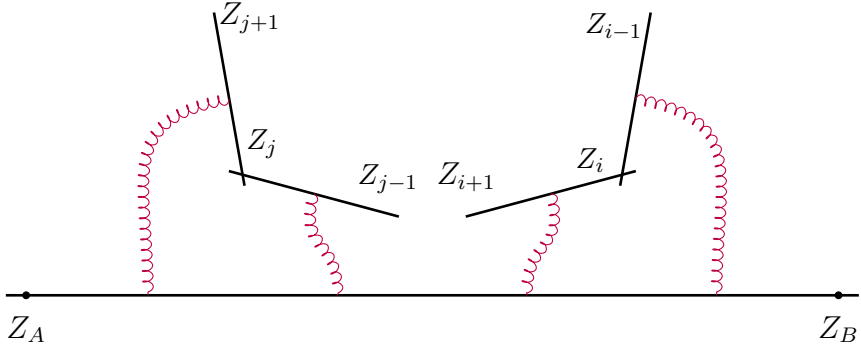

On the second Schubert solution, it is only at N$^{2}$MHV that we have a first seed diagram, depicted in Fig. \ref{twoMassEasyN2diagram} and with cut equal to 
\be
[\alpha_2, i, i+1, j-1, j][\gamma_2, j, j+1, i-1, i ].
\ee
To extend beyond N\(^2\)MHV we can, as usual, consider adding propagators in a way which respects planarity. We cannot add propagators to the regions spanned by \((\alpha_2,i,\beta_2)\) or \((\gamma_2,j,\delta_2)\) as these would correspond to tree-level triangular Wilson loops. Thus we can only dress the other two regions with additional propagators and as such arrive at the answer
\begin{align}
&[\alpha_2 , i , i+1 , j-1 , j][\gamma_2 , j , j+1 , i-1 , i ] \notag \\
\times& \langle\mathcal{L}(\beta_2,i+1,.,,,j-1,\gamma_2)\rangle \langle\mathcal{L}(\delta_2,j+1,.,,,i-1,\alpha_2)\rangle.
\end{align}
This is exactly the same as the expression given in e.g. \cite{Bourjaily:2013mma} and moreover clearly follows from the appropriate collinear limits on the four-mass expression. 

\section{Leading singularities for multiple Wilson Loop correlators}
\label{Sec-multiWL}
Let us now turn our attention to (the connected part of) correlators of multiple Wilson loops. While the leading-\(N\) contributions are \(\frac{1}{N^2}\) suppressed compared to the disconnected contributions, here we will suppress the \(\frac{1}{N^2}\) factor while still restricting our attention to planar connected diagrams.

We can apply exactly the same logic as applied to a single Wilson loop in the last section, in order to obtain compact, tree-level expressions for all one-loop leading singularities for these objects. As for a single Wilson loop, we will restrict our attention to the four-mass case since we would expect all others to thus descend via collinear limits. However, we will explicitly give all the lower mass coefficients in the next section and have verified that the formulae which arise from collinear limits are equivalent to those which follow from a direct derivation.

Even within the category of four-mass cuts, we must now split into cases according to which Wilson loops the physical poles being set to zero belong to. For example, a four-mass leading singularity where two of the poles belong to each of two different Wilson loops will be referred to as a `2-2 split'; if there are two on one Wilson loop and one on each of two others, we will refer to it as a `2-1-1 split', etc. In total in addition to these examples we also have the case of a `3-1' split and a  `1-1-1-1' split. 

In the case where there is a Wilson loop in the correlator which does not actually supply any of the poles which enter into the cut, we will refer to such Wilson loops as 'spectators' and will set out how to update our formulae accordingly. 

\subsection{2-2 split}
Let us focus first on the case of the correlator of two Wilson loops where two of the cut propagators lie on each of two Wilson loops. We consider the four-mass cut associated to 
\be
\langle A \, B \, i_1-1 \, i_1 \rangle = \langle A \, B \, i_2-1 \, i_2 \rangle  = \langle A \, B \, j_1-1 \, j_1 \rangle  = \langle A \, B \, j_2-1 \, j_2 \rangle  = 0.
\ee 
\label{2-2FourMassCut}
where we use \(i\)-labels for the first Wilson loop and \(j\)-labels for the second Wilson loop. Here we will write \(\alpha_{1,2}\), \(\beta_{1,2}\), \(\gamma_{1,2}\), \(\delta_{1,2}\) for the intersection of the Schubert solutions with the lines \((i_1-1 \, i_1)\), \((i_2-1 \, i_2)\), \((j_1-1 \, j_1)\), and \((j_2-1 \, j_2)\) respectively. 

\begin{figure}
\centering
\begin{tabular}{c@{\hspace{2cm}}c}
\begin{tikzpicture}[scale=0.6]
  \path[use as bounding box] (-5.5,-3) rectangle (5.5,2.5);
  \defineline{lin1m}{originx=1.25, originy=1.5, angle=280, pointcount=3}
  \defineline{lin4m}{originx=4.5, originy=1.5, angle=80, pointcount=3}
  \defineline{lin5m}{originx=-4.5, originy=1.5, angle=280, pointcount=3}
  \defineline{lin6m}{originx=-1.25, originy=1.5, angle=80, pointcount=3}
  \defineline{linA}{originx=0, originy=-2.5, angle=180, pointcount=4,linelength=6}
 
\drawline[showtwistors=true,showvertex=true,wingorientation=outer,tolabel=$Z_B$,fromlabel=$Z_A$]{linA}
 
    \drawline[showwings=true, showtwistors=true, fromlabel={\tiny $ \, \, \, Z_{i_2-1}$},tolabel={\tiny $Z_{i_2}$}]{lin1m}
  \drawline[showwings=true, showtwistors=true, fromlabel={\tiny $Z_{i_1-1} \, \, \, \, \, \, \, $},tolabel={\tiny $Z_{i_1} \, $}]{lin4m}
  \drawline[showwings=true, showtwistors=true, fromlabel={\tiny \, $Z_{j_2-1}$},tolabel={\tiny $\,\,\,Z_{j_2}$}]{lin5m}
  \drawline[showwings=true, showtwistors=true, fromlabel={\tiny $Z_{j_1-1} \, \, \, \, \, \, \,$},tolabel={\tiny $\,\,Z_{j_1}$}]{lin6m}
   \drawpropagator{fromline=lin5m, toline=linA, fromip=2, toip=4}
   \drawpropagator{fromline=lin6m, toline=linA, fromip=2, toip=3}
   \drawpropagator{fromline=lin1m, toline=linA, fromip=2, toip=2}
   \drawpropagator{fromline=lin4m, toline=linA, fromip=2, toip=1}
\end{tikzpicture} &
\begin{tikzpicture}[scale=0.6]
  \path[use as bounding box] (-5.5,-3) rectangle (5.5,2.5);
  \defineline{lin1m}{originx=1.25, originy=1.5, angle=280, pointcount=3}
  \defineline{lin4m}{originx=4.5, originy=1.5, angle=80, pointcount=3}
  \defineline{lin5m}{originx=-4.5, originy=1.5, angle=280, pointcount=3}
  \defineline{lin6m}{originx=-1.25, originy=1.5, angle=80, pointcount=3}
  \defineline{linA}{originx=0, originy=-2.5, angle=180, pointcount=4,linelength=6}
 
\drawline[showtwistors=true,showvertex=true,wingorientation=outer,tolabel=$Z_B$,fromlabel=$Z_A$]{linA}
 
    \drawline[showwings=true, showtwistors=true, fromlabel={\tiny $ \, \, \, Z_{i_2-1}$},tolabel={\tiny $Z_{i_2}$}]{lin1m}
  \drawline[showwings=true, showtwistors=true, fromlabel={\tiny $Z_{i_1-1} \, \, \, \, \, \, \, $},tolabel={\tiny $Z_{i_1} \, $}]{lin4m}
  \drawline[showwings=true, showtwistors=true, fromlabel={\tiny \, $Z_{j_1-1}$},tolabel={\tiny $\,\,\,Z_{j_1}$}]{lin5m}
  \drawline[showwings=true, showtwistors=true, fromlabel={\tiny $Z_{j_2-1} \, \, \, \,$},tolabel={\tiny $\,\,Z_{j_2}$}]{lin6m}
   \drawpropagator{fromline=lin5m, toline=linA, fromip=2, toip=4}
   \drawpropagator{fromline=lin6m, toline=linA, fromip=2, toip=3}
   \drawpropagator{fromline=lin1m, toline=linA, fromip=2, toip=2}
   \drawpropagator{fromline=lin4m, toline=linA, fromip=2, toip=1}
\end{tikzpicture}\\[1cm]
\begin{tikzpicture}[scale=0.6]
  \path[use as bounding box] (-5.5,-3) rectangle (5.5,2.5);
  \defineline{lin1m}{originx=1.25, originy=1.5, angle=280, pointcount=3}
  \defineline{lin4m}{originx=4.5, originy=1.5, angle=80, pointcount=3}
  \defineline{lin5m}{originx=-4.5, originy=1.5, angle=280, pointcount=3}
  \defineline{lin6m}{originx=-1.25, originy=1.5, angle=80, pointcount=3}
  \defineline{linA}{originx=0, originy=-2.5, angle=180, pointcount=4,linelength=6}
 
\drawline[showtwistors=true,showvertex=true,wingorientation=outer,tolabel=$Z_B$,fromlabel=$Z_A$]{linA}
 
    \drawline[showwings=true, showtwistors=true, fromlabel={\tiny $ \, \, \, Z_{i_1-1}$},tolabel={\tiny $Z_{i_1}$}]{lin1m}
  \drawline[showwings=true, showtwistors=true, fromlabel={\tiny $Z_{i_2-1} \, \, \, \, \, \, \, $},tolabel={\tiny $Z_{i_2} \, $}]{lin4m}
  \drawline[showwings=true, showtwistors=true, fromlabel={\tiny \, $Z_{j_2-1}$},tolabel={\tiny $\,\,\,Z_{j_2}$}]{lin5m}
  \drawline[showwings=true, showtwistors=true, fromlabel={\tiny $Z_{j_1-1} \, \, \,  \, \, \, \,$},tolabel={\tiny $\,\,Z_{j_1}$}]{lin6m}
   \drawpropagator{fromline=lin5m, toline=linA, fromip=2, toip=4}
   \drawpropagator{fromline=lin6m, toline=linA, fromip=2, toip=3}
   \drawpropagator{fromline=lin1m, toline=linA, fromip=2, toip=2}
   \drawpropagator{fromline=lin4m, toline=linA, fromip=2, toip=1}
\end{tikzpicture} &
\begin{tikzpicture}[scale=0.6]
  \path[use as bounding box] (-5.5,-3) rectangle (5.5,2.5);
  \defineline{lin1m}{originx=1.25, originy=1.5, angle=280, pointcount=3}
  \defineline{lin4m}{originx=4.5, originy=1.5, angle=80, pointcount=3}
  \defineline{lin5m}{originx=-4.5, originy=1.5, angle=280, pointcount=3}
  \defineline{lin6m}{originx=-1.25, originy=1.5, angle=80, pointcount=3}
  \defineline{linA}{originx=0, originy=-2.5, angle=180, pointcount=4,linelength=6}
 
\drawline[showtwistors=true,showvertex=true,wingorientation=outer,tolabel=$Z_B$,fromlabel=$Z_A$]{linA}
 
    \drawline[showwings=true, showtwistors=true, fromlabel={\tiny $ \, \, \, Z_{i_1-1}$},tolabel={\tiny $Z_{i_1}$}]{lin1m}
  \drawline[showwings=true, showtwistors=true, fromlabel={\tiny $Z_{i_2-1} \, \, \, \, \, \, \, $},tolabel={\tiny $Z_{i_2} \, $}]{lin4m}
  \drawline[showwings=true, showtwistors=true, fromlabel={\tiny \, $Z_{j_1-1}$},tolabel={\tiny $\,\,\,Z_{j_1}$}]{lin5m}
  \drawline[showwings=true, showtwistors=true, fromlabel={\tiny $Z_{j_2-1} \, \, \, \, \, \, \,$},tolabel={\tiny $\,\,Z_{j_2}$}]{lin6m}
   \drawpropagator{fromline=lin5m, toline=linA, fromip=2, toip=4}
   \drawpropagator{fromline=lin6m, toline=linA, fromip=2, toip=3}
   \drawpropagator{fromline=lin1m, toline=linA, fromip=2, toip=2}
   \drawpropagator{fromline=lin4m, toline=linA, fromip=2, toip=1}
\end{tikzpicture}\\
\end{tabular}
    \caption{Due to the different colour structure of a correlator of two Wilson loops, there are \emph{four} distinct planar diagrams which contribute to the four-mass cut (\ref{2-2FourMassCut}), unlike the case of a single Wilson loop where there was only one. The diagrams which contribute to the leading singularity at higher MHV degree are then those which arise by dressing any of these four 'seed' diagrams with further propagators. Note that in each instance we have rotated the Wilson loops to keep the overall shape of the diagram  the same.}
\label{2-2splitseeds}
\end{figure}
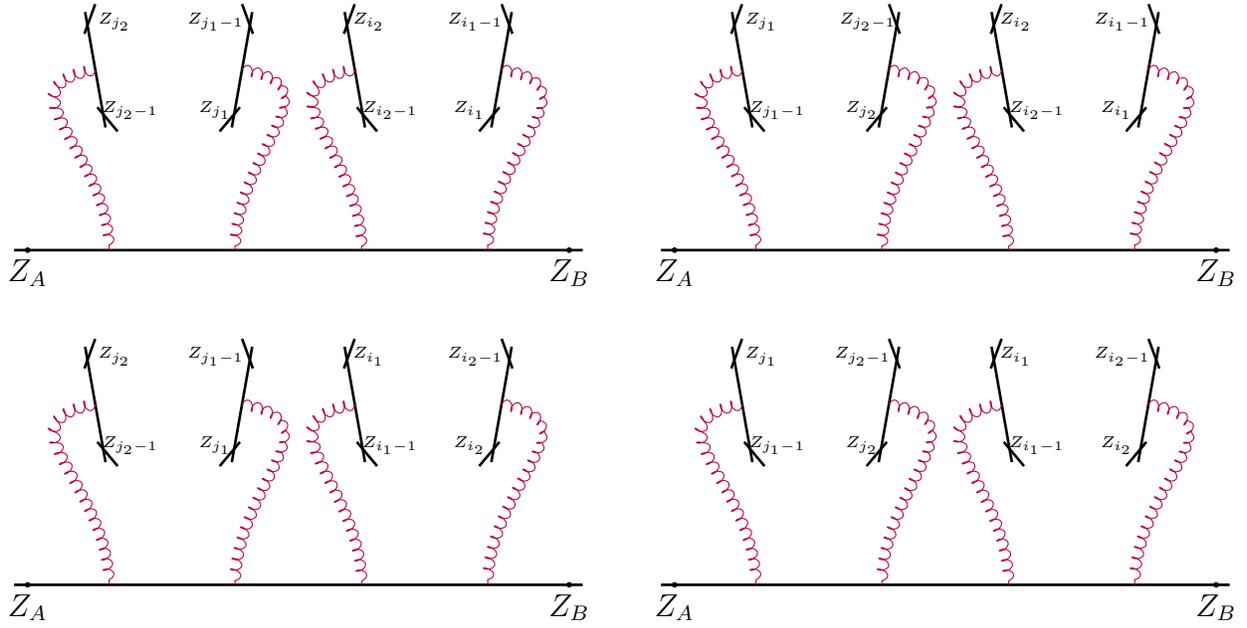

At N\(^2\)MHV for two Wilson loops, there are now \emph{four} planar diagrams which contribute a non-zero residue, in contrast to the case previously outlined of a single Wilson loop. These are depicted in Fig. \ref{2-2splitseeds}. Notably, whereas the propagators for a single Wilson loop divided the geometry into four regions, now each of these is divided into \emph{three} regions. This immediately suggests that we are expecting a formula involving products of three, rather than four, tree-level objects. 

Let us label the four diagrams in Fig. \ref{2-2splitseeds} as \(D_1\), \(D_2\), \(D_3\) and \(D_4\), clockwise from the top-left. Even prior to cutting we have \(D_1 = D_4\) and \(D_2=D_3\), but this is not a redundancy; they are genuinely distinct diagrams arising for the O($g^2$) N\(^2\)MHV integrand (as the cyclic order of the insertions of the propagators on the Lagrangian is different), and they will give rise to different contributions once we consider populating them with more propagators to calculate the contribution at higher MHV degree.

We have by simple calculation that (here we omit the loop integral to be performed and just write the loop integrand, pre-fermionic integration) 
\begin{align}
D_1 &= D_4 =  \frac{g^2}{\pi^2}[*,A,\widehat{B}_{i_1},i_{2}-1,i_2][*,A,\widehat{B}_{i_2},j_1-1,j_1][*,A,\widehat{B}_{j_1},j_2-1,j_2][*,A,\widehat{B}_{j_2},i_1-1,i_1] \notag \\
D_2 &= D_3 = \frac{g^2}{\pi^2}[*,A,\widehat{B}_{i_1},i_{2}-1,i_2][*,A,\widehat{B}_{i_2},j_2-1,j_2][*,A,\widehat{B}_{j_2},j_1-1,j_1][*,A,\widehat{B}_{j_1},i_1-1,i_1] 
\end{align}
Taking the cut\footnote{As discussed in \cite{Drummond:2026lvq}, one must take care to specify the ordering of the four poles when referring to the residue on a given cut. Here we take the ordering \((i_1-1, i_1)\), \((i_2-1,i_2)\), \((j_1-1,j_1)\), \((j_2-1,j_2)\); orderings which differ from this by an odd permutation will be out by a factor of \(-1\) relative to our given expressions. The ordering is similarly implied throughout the paper whenever a cut is given in this form.} on the first Schubert solution and performing the fermionic integration, we then have for \(D_1\) and \(D_4\) (again omitting the \(\frac{g^2}{\pi^2}\)) 
\be
C_1(i_1,i_2,j_1,j_2)
\label{2-2cutsol1}
\ee
which is as the equivalent contribution to a single Wilson loop, and for \(D_2\) and \(D_3\)
\be
 \Bigl(-1-\frac{1}{2}x\Bigr)C_1(i_1,i_2,j_1,j_2)
 \label{2-2cutsol2}
\ee
where we recall the definitions given in (\ref{C1}) and (\ref{C2}), with the intersection twistors \(\alpha_{1,2},\ldots,\delta_{1,2}\) now updated to reflect the cut being considered. We also define
\be
x(i_1,i_2,j_1,j_2) = \Delta-u+v-1
\label{xdef}
\ee
where we define conformal cross-ratios \(u(i_1,i_2,j_1,j_2)\) and \(v(i_1,i_2,j_1,j_2)\) and square root \(\Delta\) via 
\begin{equation}
u(i_1,i_2,j_1,j_2) = \frac{\langle i_1 -1 \, i_1 \, i_2-1 \, i_2 \rangle \langle j_1-1 \, j_1 \, j_2-1 \, j_2\rangle}{\langle i_1 -1 \, i_1 \, j_1-1 \, j_1\rangle \langle i_2-1 \, i_2 \, j_2-1 \, j_2\rangle},
\end{equation}
\begin{equation}
v(i_1,i_2,j_1,j_2) = \frac{\langle i_1 -1 \, i_1 \, j_2-1 \, j_2 \rangle \langle i_2-1 \, i_2 \, j_1-1 \, j_1\rangle}{\langle i_1-1 \, i_1 \, j_1-1 \, j_1\rangle \langle i_2 -1 \, i_2 \, j_2-1 \, j_2 \rangle},
\end{equation}
and
\begin{equation}
\Delta(i_1,i_2,j_1,j_2) = \sqrt{(1-u-v)^2-4uv}.
\end{equation}

The leading singularities associated to the second Schubert solution are given by taking (\ref{2-2cutsol1}) and (\ref{2-2cutsol2}) by replacing \(C_1 \to C_2\) and \(x \to \bar{x}\) where
\be
\bar{x}(i_1,i_2,j_1,j_2) = -\Delta-u+v-1
\label{xbardef}
\ee
Summing over the four contributions, we can see that at N\(^2\)MHV, the value of this leading singularity is the same (on each Schubert solution) as for a single Wilson loop with labels \(i_1 < i_2 < j_1 < j_2\), except for multiplication by the simple bosonic factor \(-x\) or \(-\bar{x}\) on the first and second Schubert solutions respectively. At higher MHV degree, we will clearly have a sum of four terms, each of which corresponds to dressing one of \(D_1\), \(D_2\), \(D_3\) or \(D_4\) with further propagators in a way which respects planarity.  

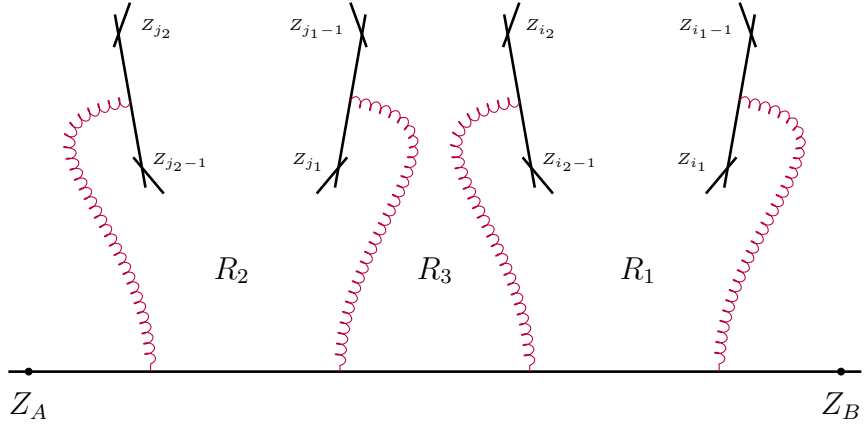
\begin{figure}
\begin{center}
\begin{tikzpicture}[scale=0.9]
  \path[use as bounding box] (-5.5,-3) rectangle (5.5,2.5);
  \defineline{lin1m}{originx=1.25, originy=1.5, angle=280, pointcount=3}
  \defineline{lin4m}{originx=4.5, originy=1.5, angle=80, pointcount=3}
  \defineline{lin5m}{originx=-4.5, originy=1.5, angle=280, pointcount=3}
  \defineline{lin6m}{originx=-1.25, originy=1.5, angle=80, pointcount=3}
  \defineline{linA}{originx=0, originy=-2.5, angle=180, pointcount=4,linelength=6}
 
\drawline[showtwistors=true,showvertex=true,wingorientation=outer,tolabel=$Z_B$,fromlabel=$Z_A$]{linA}
 
    \drawline[showwings=true, showtwistors=true, fromlabel={\tiny $ \, \, \, Z_{i_2-1}$},tolabel={\tiny $Z_{i_2}$}]{lin1m}
  \drawline[showwings=true, showtwistors=true, fromlabel={\tiny $Z_{i_1-1} \, \, \, \, \, \, \, $},tolabel={\tiny $Z_{i_1} \, $}]{lin4m}
  \drawline[showwings=true, showtwistors=true, fromlabel={\tiny \, $Z_{j_2-1}$},tolabel={\tiny $\,\,\,Z_{j_2}$}]{lin5m}
  \drawline[showwings=true, showtwistors=true, fromlabel={\tiny $Z_{j_1-1} \, \, \, \, \, \, \,$},tolabel={\tiny $\,\,Z_{j_1}$}]{lin6m}
   \drawpropagator{fromline=lin5m, toline=linA, fromip=2, toip=4}
   \drawpropagator{fromline=lin6m, toline=linA, fromip=2, toip=3}
   \drawpropagator{fromline=lin1m, toline=linA, fromip=2, toip=2}
   \drawpropagator{fromline=lin4m, toline=linA, fromip=2, toip=1}

   \node at (3,-1) {$R_1$};
   \node at (-3,-1) {$R_2$};
   \node at (0,-1) {$R_3$};

\end{tikzpicture} 
\end{center}
\caption{The partitioning of one of the N\(^2\)MHV 'seed' diagrams for a 2-2 split four mass cut into three regions.}
\label{2-2regions}
\end{figure}

Now let us focus our attention on \(D_1\) and consider the diagrams of higher MHV degree which can be generated by adding extra propagators. Planarity requires that additional propagators stay within one of \emph{three} regions, depicted in Fig. \ref{2-2regions}, and do not cross between them, which is analogous to the partitioning into four regions seen for a single Wilson loop.

By analogy with the case of a single Wilson loop we therefore see that the contributions from dressing this diagram arrange themselves as the N\(^2\)MHV cut for \(D_1\) multiplied by a product of three Wilson loops, each associated to one of the three regions. In particular, the three tree-level Wilson loops in our product should clearly be
\be
\langle\mathcal{L}(\alpha_1,i_1,i_1+1,...,i_2-1,\beta_1) \rangle
\ee
\be
\langle\mathcal{L}(\gamma_1,j_1,j_1+1,...,j_2-1,\delta_1) \rangle
\ee
and
\be
\langle\mathcal{L}(\delta_1,j_2,j_2+1,...,j_1-1,\gamma_1,\beta_1,i_2,i_2+1,...,i_1-1,\alpha_1) \rangle
\ee

for \(R_1\), \(R_2\) and \(R_3\) respectively on the first Schubert solution, and replacing \((\alpha_1,\beta_1,\gamma_1,\delta_1) \to (\alpha_2,\beta_2,\gamma_2,\delta_2)\) for the second Schubert solution. 

Let us now remark on an important subtlety for the term corresponding to the region \(R_3\), involving indices from both Wilson loops. If we simply take an ordinary Wilson loop correlator with the correct multiplicity and set the twistors equal to the given values, we would (diagrammatically) include contributions from diagrams where e.g. a propagator runs from the line \((\alpha_1\delta_1)\) to the line \((\gamma_1\beta_1)\), because those are two well-separated lines in the ordering of the loop labels on the Wilson loop. Of course such a diagram is degenerate, because in fact \((\alpha \delta)\) and \((\beta \gamma)\) both lie on the singular configuration of the Lagrangian line and as such are the same line! Of course, pre-cutting there is no dual one-loop diagram either because we don't have a one-loop diagram where a propagator runs from the Lagrangian to itself. 

What we formally mean by such a term in all of our expressions is therefore to take the Wilson loop correlator of the appropriate multiplicity, initially using e.g. \(\mathcal{Z}_{\alpha_1} + \epsilon \mathcal{Z}_r\) for an arbitrary twistor \(\mathcal{Z}_r\), and then take the \(\epsilon \to 0\) limit in a way that respects \(Q\)-supersymmetry i.e. which treats the bosonic and fermionic parts of the super-twistors equally. Doing so, the apparantly degenerate diagrams go to zero on the support of the pre-multiplying $R$-invariants (which in some cases are required to eliminate for Grassmann reasons apparent divergences in this limit). If one wishes to generate this term using twistor diagrams, it is an adequate prescription to simply plug in for \(\alpha_1\), \(\beta_1\), \(\gamma_1\), \(\delta_1\) and the values of the other twistors and then discard diagrams which appear naively divergent. 

Clearly overall we find a sum over four such three-fold products, each originating from one of the four original 'seed' diagrams at N\(^2\)MHV. Each will simply be the cut of that 'seed' diagram at N\(^2\) MHV (on the appropriate Schubert solution), multiplied by the appropriate product of three tree-level correlators. 

Explicitly we have for the first Schubert solution contributions
\begin{align}
 &C_1(i_1,i_2,j_1,j_2)\langle\mathcal{L}(\alpha_1,i_1,i_1+1,...,i_2-1,\beta_1) \rangle \langle\mathcal{L}(\gamma_1,j_1,j_1+1,...,j_2-1,\delta_1) \rangle  \notag \\
 &\times\langle\mathcal{L}(\delta_1,j_2,j_2+1,...,j_1-1,\gamma_1,\beta_1,i_2,i_2+1,...,i_1-1,\alpha_1) \rangle,  
\end{align}
 \begin{align}
 &\Bigl(-1-\frac{1}{2}x(i_1,i_2,j_1,j_2)\Bigr)C_1(i_1,i_2,j_1,j_2) \notag \\
  &\times \langle\mathcal{L}(\alpha_1,i_1,i_1+1,...,i_2-1,\beta_1) \rangle \langle\mathcal{L}(\delta_1,j_2,j_2+1,...,j_1-1,\gamma_1) \rangle  \notag \\
 &\times\langle\mathcal{L}(\gamma_1,j_1,j_1+1,...,j_2-1,\delta_1,\beta_1,i_2,i_2+1,...,i_1-1,\alpha_1) \rangle,  
\end{align}
 \begin{align}
 &\Bigl(-1-\frac{1}{2}x(i_1,i_2,j_1,j_2)\Bigr)C_1(i_1,i_2,j_1,j_2) \notag \\
 &\times \langle\mathcal{L}(\beta_1,i_2,i_2+1,...,i_1-1,\alpha_1) \rangle \langle\mathcal{L}(\gamma_1,j_1,j_1+1,...,j_2-1,\delta_1) \rangle  \notag \\
 &\times\langle\mathcal{L}(\delta_1,j_2,j_2+1,...,j_1-1,\gamma_1,\alpha_1,i_1,i_1+1,...,i_2-1,\beta_1) \rangle,  
\end{align}
and
\begin{align}
 &C_1(i_1,i_2,j_1,j_2)\langle\mathcal{L}(\beta_1,i_2,i_2+1,...,i_1-1,\alpha_1) \rangle \langle\mathcal{L}(\delta_1,j_2,j_2+1,...,j_1-1,\gamma_1) \rangle  \notag \\
 &\times\langle\mathcal{L}(\gamma_1,j_1,j_1+1,...,j_2-1,\delta_1,\alpha_1,i_1,i_1+1,...,i_2-1,\beta_1) \rangle,  
\end{align}
from \(D_1\), \(D_2\), \(D_3\) and \(D_4\) respectively. The leading singularity is then given by summing over these four contributions. The leading singularity on the second Schubert solution follows from the above under the replacements \(C_1 \to C_2\), \((\alpha_1, \beta_1, \gamma_1,\delta_1) \to (\alpha_2, \beta_2, \gamma_2,\delta_2)\), and \(x \to \bar{x}\).

\subsubsection{Adding spectator Wilson loops}
Now consider the analogous case for a correlator of \emph{three} Wilson loops. If we consider a four mass cut where two propagators associated to each of two Wilson loops are involved, then we can consider the third Wilson loop as a 'spectator'. Returning to the seed diagram \(D_1\) for the case without a spectator, the diagrams which are planar are now those where the spectator Wilson sits in one of the three regions and any propagators from this Wilson loop (note that we need at least one propagator ending on the spectator Wilson loop since we are considering the connected part of the correlator, and moreover since we work in $SU(N)$ we in fact need at least two to avoid a vanishing colour trace) cannot cross between regions. Thus with three Wilson loops we would effectively have three versions of this seed diagram \(D_1\) where the spectator Wilson loop is inserted in one of the three regions \(R_1\), \(R_2\) or \(R_3\).

The requirement to include the spectator Wilson loop in a given region, combined with the fact we are looking here at specifically the connected part of the correlator (and thus there must be propagators ending on every Wilson loop), means that the term in the product associated to that region gets replaced with the connected part of the correlator of the spectator Wilson loop with the Wilson loop originally associated to that region. 

For instance, if we compute the four-mass cut in (\ref{2-2FourMassCut}) but we are dealing with the correlator of \emph{three} Wilson loops, the third of which \(\mathcal{W}_3\) does not supply any of the propagators to be cut, then (ignoring for a moment the $R$-invariants and bosonic prefactor which are unchanged) \(D_1\)'s contribution on the first Schubert solution (the second Schubert solution is the same with \((\alpha_1,\beta_1,\gamma_1,\delta_1) \to (\alpha_2,\beta_2,\gamma_2,\delta_2)\)),
\begin{align}
 & \langle\mathcal{L}(\alpha_1,i_1,i_1+1,...,i_2-1,\beta_1) \rangle \langle\mathcal{L}(\gamma_1,j_1,j_1+1,...,j_2-1,\beta_1) \rangle \notag \\
 &\times \langle\mathcal{L}(\delta_1,j_2,j_2+1,...,j_1-1,\gamma_1,\beta_1,i_2,i_2+1,...,i_1-1,\alpha_1) \rangle,
\end{align}
is replaced with the sum over three terms
\begin{align}
  \langle\mathcal{L}(\alpha_1&,i_1,i_1+1,...,i_2-1,\beta_1) \mathcal{W}_3\rangle^c \langle\mathcal{L}(\gamma_1,j_1,j_1+1,...,j_2-1,\delta_1) \rangle \notag \\
 &\times \langle\mathcal{L}(\delta_1,j_2,j_2+1,...,j_1-1,\gamma_1,\beta_1,i_2,i_2+1,...,i_1-1,\alpha_1) \rangle \notag \\
 + \langle\mathcal{L}(\alpha_1&,i_1,i_1+1,...,i_2-1,\beta_1) \rangle \langle\mathcal{L}(\gamma_1,j_1,j_1+1,...,j_2-1,\delta_1) \mathcal{W}_3\rangle^c \notag \\
 &\times \langle\mathcal{L}(\delta_1,j_2,j_2+1,...,j_1-1,\gamma_1,\beta_1,i_2,i_2+1,...,i_1-1,\alpha_1) \rangle \notag \\
+   \langle\mathcal{L}(\alpha_1&,i_1,i_1+1,...,i_2-1,\beta_1) \rangle \langle\mathcal{L}(\gamma_1,j_1,j_1+1,...,j_2-1,\delta_1) \rangle \notag \\
 &\times \langle\mathcal{L}(\delta_1,j_2,j_2+1,...,j_1-1,\gamma_1,\beta_1,i_2,i_2+1,...,i_1-1,\alpha_1) \mathcal{W}_3\rangle^c 
\end{align}
where each corresponds to the spectator Wilson loop sitting in a different region. Here we use the superscript `c' to denote taking the \emph{connected} part of the corresponding tree-level correlator. The contributions associated to \(D_2\), \(D_3\) and \(D_4\) are similarly extended to a sum over three terms in the same way.

If we have \emph{multiple} spectator Wilson loops, we need to sum over all the possibilities for which regions they are added in; note that there is no issue with splitting them across different regions (if e.g. two spectators are added to the same region, that will of course be the connected part of a \emph{three} Wilson loop correlator). Note that, at a given multiplicity and MHV degree, many of these connected correlators may evaluate to zero.  

\subsection{4- split}
Having addressed the matter of `spectator' Wilson loops which do not participate in the cut, it is straightforward to handle the case where all four cut propagators lie on the same Wilson loop as in (\ref{quadruplePole}), but with additional spectator Wilson loops: we simply take the result for a single Wilson loop, (\ref{C1}) and (\ref{C2}) with \(i<j<k<l\) all on the same Wilson loop, and replace the product (here we write the version on the first Schubert solution)
\begin{align}
 \langle\mathcal{L}(\alpha_1&, i,i+1, ..., j-1, \beta_1)\rangle \times \langle\mathcal{L}(\beta_1, j,j+1, ..., k-1, \gamma_1)\rangle \notag \\
\times&\langle\mathcal{L}(\gamma_1, k,k+1, ..., l-1, \delta_1)\rangle \times \langle\mathcal{L}(\delta_1, l,l+1, ..., i-1, \alpha_1)\rangle
\end{align}
with a sum over all of the ways of inserting the extra Wilson loop(s) into one of the four Wilson loops to form a correlator. Here \(\alpha_{1,2}\), \(\beta_{1,2}\), \(\gamma_{1,2}\) and \(\delta_{1,2}\) are still the intersections of the Schubert solution with \((i-1\,i)\), \((j-1 \, j)\), \((k-1 \, k)\), and \((l-1 \, l)\) respectively. For instance, for a single spectator Wilson loop \(\mathcal{W}_2\) this term is replaced with the sum 
\begin{align}
 \langle\mathcal{L}(\alpha_1&, i,i+1, ..., j-1, \beta_1)\mathcal{W}_2\rangle^c \times \langle\mathcal{L}(\beta_1, j,j+1, ..., k-1, \gamma_1)\rangle \notag \\
\times&\langle\mathcal{L}(\gamma_1, k,k+1, ..., l-1, \delta_1)\rangle \times \langle\mathcal{L}(\delta_1, l,l+1, ..., i-1, \alpha_1)\rangle \notag \\
+ \langle\mathcal{L}(\alpha_1&, i,i+1, ..., j-1, \beta_1)\rangle \times \langle\mathcal{L}(\beta_1, j,j+1, ..., k-1, \gamma_1)\mathcal{W}_2\rangle^c \notag \\
\times&\langle\mathcal{L}(\gamma_1, k,k+1, ..., l-1, \delta_1)\rangle \times \langle\mathcal{L}(\delta_1, l,l+1, ..., i-1, \alpha_1)\rangle \notag \\
+\langle\mathcal{L}(\alpha_1&, i,i+1, ..., j-1, \beta_1)\rangle \times \langle\mathcal{L}(\beta_1, j,j+1, ..., k-1, \gamma_1)\rangle \notag \\
\times&\langle\mathcal{L}(\gamma_1, k,k+1, ..., l-1, \delta_1)\mathcal{W}_2\rangle^c \times \langle\mathcal{L}(\delta_1, l,l+1, ..., i-1, \alpha_1)\rangle \notag \\
+\langle\mathcal{L}(\alpha_1&, i,i+1, ..., j-1, \beta_1)\rangle \times \langle\mathcal{L}(\beta_1, j,j+1, ..., k-1, \gamma_1)\rangle \notag \\
\times&\langle\mathcal{L}(\gamma_1, k,k+1, ..., l-1, \delta_1)\rangle \times \langle\mathcal{L}(\delta_1, l,l+1, ..., i-1, \alpha_1)\mathcal{W}_2\rangle^c.
\end{align}

\subsection{3-1 split }

\begin{figure}
\centering
\begin{tabular}{c@{\hspace{2cm}}c}
\begin{tikzpicture}[scale=0.6]
  \defineline{lin1m}{originx=5.25, originy=1, angle=90, pointcount=3}
  \defineline{lin4m}{originx=8.75, originy=-1, angle=0, pointcount=3}
  \defineline{lin5m}{originx=1.0, originy=1, angle=270, pointcount=3}
  \defineline{lin6m}{originx=3.25, originy=-1, angle=0, pointcount=3}
  \defineline{linA}{originx=4.75, originy=-2.5, angle=180, pointcount=4,linelength=5.5}
  \drawline[showtwistors=true,showvertex=true,wingorientation=outer,tolabel=$Z_B$,fromlabel=$Z_A$]{linA}
  \drawline[showwings=true, showtwistors=true, fromlabel={\tiny $Z_{i_1-1} \, \, \, \, \, \, \, \, $},tolabel={\tiny $Z_{i_1}$}]{lin1m}
  \drawline[showwings=true, showtwistors=true, fromlabel={\tiny $Z_{j-1}$},tolabel={\tiny $Z_{j}$}]{lin4m}
  \drawline[showwings=true, showtwistors=true, fromlabel={\tiny$ \, \, \, \, \, Z_{i_3-1}$},tolabel={\tiny $\,\,\,Z_{i_3}$}]{lin5m}
  \drawline[showwings=true, showtwistors=true, fromlabel={\tiny $Z_{i_2-1}$},tolabel={\tiny $\,\,Z_{i_2}$}]{lin6m}
  \drawpropagator{fromline=lin5m, toline=linA, fromip=2, toip=4}
  \drawpropagator{fromline=lin6m, toline=linA, fromip=2, toip=3}
  \drawpropagator{fromline=lin1m, toline=linA, fromip=2, toip=2}
  \drawpropagator{fromline=lin4m, toline=linA, fromip=2, toip=1}
\end{tikzpicture} &
\begin{tikzpicture}[scale=0.6]
  \defineline{lin1m}{originx=5.25, originy=1, angle=90, pointcount=3}
  \defineline{lin4m}{originx=8.75, originy=-1, angle=0, pointcount=3}
  \defineline{lin5m}{originx=1.0, originy=1, angle=270, pointcount=3}
  \defineline{lin6m}{originx=3.25, originy=-1, angle=0, pointcount=3}
  \defineline{linA}{originx=4.75, originy=-2.5, angle=180, pointcount=4,linelength=5.5}
  \drawline[showtwistors=true,showvertex=true,wingorientation=outer,tolabel=$Z_B$,fromlabel=$Z_A$]{linA}
  \drawline[showwings=true, showtwistors=true, fromlabel={\tiny $Z_{i_2-1} \, \, \, \, \, \, \, $},tolabel={\tiny $Z_{i_2}$}]{lin1m}
  \drawline[showwings=true, showtwistors=true, fromlabel={\tiny $Z_{j-1}$},tolabel={\tiny $Z_{j}$}]{lin4m}
  \drawline[showwings=true, showtwistors=true, fromlabel={\tiny \, $\, \, \, \, Z_{i_1-1}$},tolabel={\tiny $\,\,\,Z_{i_1}$}]{lin5m}
  \drawline[showwings=true, showtwistors=true, fromlabel={\tiny $Z_{i_3-1}$},tolabel={\tiny $\,\,Z_{i_3}$}]{lin6m}
  \drawpropagator{fromline=lin5m, toline=linA, fromip=2, toip=4}
  \drawpropagator{fromline=lin6m, toline=linA, fromip=2, toip=3}
  \drawpropagator{fromline=lin1m, toline=linA, fromip=2, toip=2}
  \drawpropagator{fromline=lin4m, toline=linA, fromip=2, toip=1}
\end{tikzpicture}\\[0.8cm]
\multicolumn{2}{c}{
\begin{tikzpicture}[scale=0.6]
  \defineline{lin1m}{originx=5.25, originy=1, angle=90, pointcount=3}
  \defineline{lin4m}{originx=8.75, originy=-1, angle=0, pointcount=3}
  \defineline{lin5m}{originx=1.0, originy=1, angle=270, pointcount=3}
  \defineline{lin6m}{originx=3.25, originy=-1, angle=0, pointcount=3}
  \defineline{linA}{originx=4.75, originy=-2.5, angle=180, pointcount=4,linelength=5.5}
  \drawline[showtwistors=true,showvertex=true,wingorientation=outer,tolabel=$Z_B$,fromlabel=$Z_A$]{linA}
  \drawline[showwings=true, showtwistors=true, fromlabel={\tiny $Z_{i_3-1} \, \, \, \, \, \, \, $},tolabel={\tiny $Z_{i_3}$}]{lin1m}
  \drawline[showwings=true, showtwistors=true, fromlabel={\tiny $Z_{j-1}$},tolabel={\tiny $Z_{j}$}]{lin4m}
  \drawline[showwings=true, showtwistors=true, fromlabel={\tiny \, $\, \, \, \, Z_{i_2-1}$},tolabel={\tiny $\,\,\,Z_{i_2}$}]{lin5m}
  \drawline[showwings=true, showtwistors=true, fromlabel={\tiny $Z_{i_1-1}$},tolabel={\tiny $\,\,Z_{i_1}$}]{lin6m}
  \drawpropagator{fromline=lin5m, toline=linA, fromip=2, toip=4}
  \drawpropagator{fromline=lin6m, toline=linA, fromip=2, toip=3}
  \drawpropagator{fromline=lin1m, toline=linA, fromip=2, toip=2}
  \drawpropagator{fromline=lin4m, toline=linA, fromip=2, toip=1}
\end{tikzpicture}
}
\end{tabular}
\caption{The three N$^2$MHV diagrams which contribute to the cut (\ref{3-1cut}); note that planarity partitions each into three regions, and thus the resulting term in the expression will have a three-fold product of Wilson loops.}
\label{3-1seeds}
\end{figure}
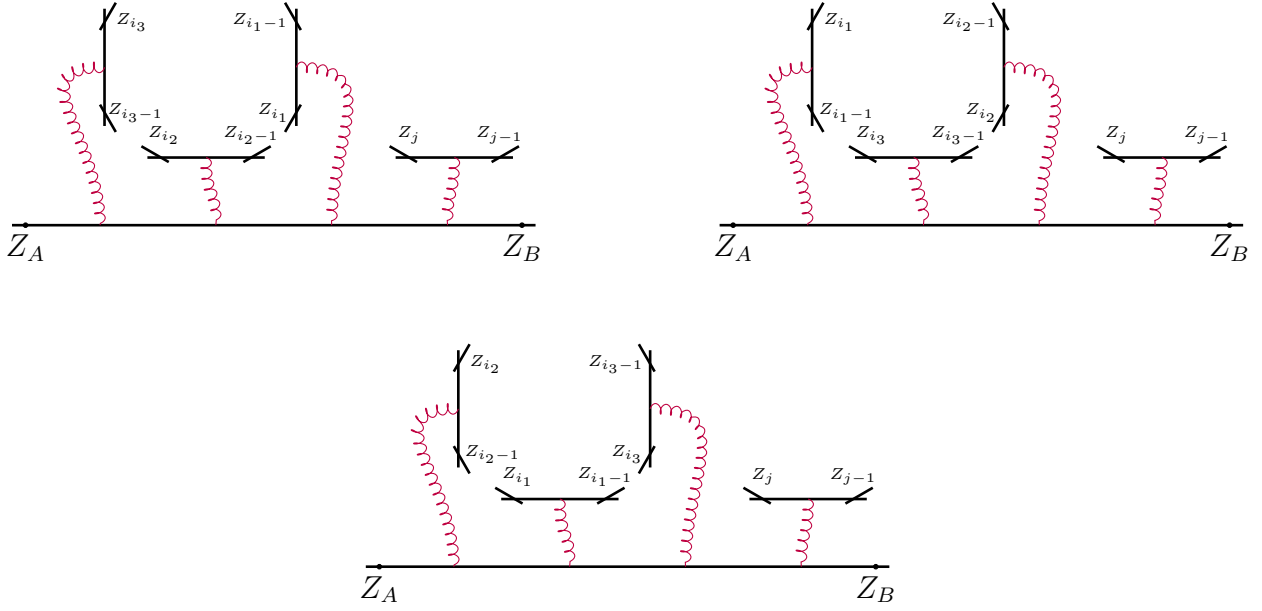

Now let us consider the case of a 3-1 split, initially with no spectators i.e. for a correlator of two Wilson loops. There are clearly three different seed diagrams with four propagators. These are depicted in Fig. \ref{3-1seeds} for the cut
\be 
\langle A \, B \, i_1-1 \, i_1 \rangle =  \langle A \, B \, i_2-1 \, i_2 \rangle =  \langle A \, B \, i_3-1 \, i_3 \rangle =  \langle A \, B \, j-1 \, j \rangle = 0
\ee
\label{3-1cut}
with \(i_1,i_2,i_3\) on one loop and \(j\) on another. Here we take \(i_1 < i_2 < i_3\) modulo the multiplicity of the first Wilson loop, and identify \(\alpha_{1,2}\), \(\beta_{1,2}\), \(\gamma_{1,2}\) and \(\delta_{1,2}\) as the intersections of the Lagrangian line on each Schubert solution with \((i_1-1 \, i_1)\), \((i_2-1 \, i_2)\), \((i_3-1 \, i_3)\), and \((j-1 \, j)\) respectively. 

Now, regardless of the fact that these N\({}^2\)MHV seed diagrams are vanishing for trace reasons in the $SU(N)$ theory, we can still compute their cut (e.g. taking the equivalent diagram in the Abelian theory) to identify the factors which will accompany the descendants of these diagrams at higher MHV diagree (i.e. a product of two $R$-invariants multiplied by a bosonic factor). These prefactors are then multiplied by the appropriate product of three Wilson loops, as can be seen from the fact that each 'seed' diagram is partitioned into three regions. 

The end result for the first Schubert solution is (note that the bosonic prefactors follow from simple calculation of the cut of the N\(^2\)MHV 'seed' diagrams in the Abelian theory) the sum over three contributions 
\begin{align}
C_1(i_1,i_2,i_3,j)\langle \mathcal{L}(\alpha_1&,i_1, i_1+1, ..., i_2-1, \beta_1)\rangle \langle \mathcal{L}(\beta_1,i_2,i_2+1,...,i_3-1,\gamma_1)\rangle  \notag \\ 
&\times \langle\mathcal{L}(\delta_1,\gamma_1,i_3,i_3+1,...,i_1-1,\alpha_1,\delta_1,j,j+1,...,j-1) \rangle,
\end{align}
\begin{align}
\frac{x(i_1,i_2,i_3,j)}{2}&C_1(i_1,i_2,i_3,j)  \notag \\
&\times\langle \mathcal{L}(\beta_1,i_2, i_2+1, ..., i_3-1, \gamma_1, )\rangle \langle \mathcal{L}(\gamma_1,i_3,i_3+1,...,i_1-1,\alpha_1)\rangle  \notag \\ 
&\times \langle\mathcal{L}(\delta_1,\alpha_1,i_1,i_1+1,...,i_2-1,\beta_1,\delta_1,j,j+1,...,j-1) \rangle,
\end{align}
and
\begin{align}
\Bigl(-1-&\frac{x(i_1,i_2,i_3,j)}{2}\Bigr)C_1(i_1,i_2,i_3,j) \notag \\
 &\times \langle \mathcal{L}(\gamma_1,i_3, i_3+1, ..., i_1-1, \alpha_1 )\rangle \langle \mathcal{L}(\alpha_1,i_1,i_1+1,...,i_2-1,\beta_1)\rangle   \notag \\
&\times \langle \mathcal{L}(\delta_1,\beta_1,i_2,i_2+1,...,i_3-1,\gamma_1,\delta_1,j,j+1,...,j-1) \rangle
\end{align}

Note that the tree-level correlators which we write down do include contributions from cuts of diagrams which have vanishing colour-factors: namely, those where the Wilson loop containing the twistor line \((j-1 \, j)\) has no additional propagators running to the other Wilson loop or to the Lagrangian. However, we recall from the discussion of the cancellation of all but the two-vertex in the Abelian theory in \cite{Drummond:2025ulh} that if one sums over a family of  twistor Wilson loop diagrams where one insertion on the Lagrangian is fixed and order of the other insertions on the Lagrangian are cycled over, the result is zero (provided all are dressed with the same colour factor). It is therefore straightforward to see that the sum of the one-loop diagrams whose cuts we spuriously include will be precisely zero, and thus the cuts of these diagrams which we include also sum to zero. 

For instance, consider the top-left diagram in Fig. \ref{3-1seeds}, and suppose we have dressed the seed diagram with propagators which only stay within the Wilson loop on which \((j-1 \, j)\) lives. This one-loop diagram has vanishing colour factor, and yet its cut is included among the tree diagrams which contribute to 
\be
 \langle\mathcal{L}(\delta_1,\gamma_1,i_3,i_3+1,...,i_1-1,\alpha_1,\delta_1,j,j+1,...,j-1) \rangle.
\ee

\emph{However}, clearly from the other `seed diagrams' there are two further, equivalent diagrams which correspond to rotating round the other Wilson loop, which cyclically permutes the order of the three (plus any others which have been added) propagators running from that Wilson loop to the Lagrangian. The sum of these diagrams is then zero, which means that the sum of their cuts is also zero, and so they may be harmlessly included in our formula.

For the second Schubert solution, we replace \(C_1 \to C_2\), \(x \to \bar{x}\) and \((\alpha_1,\beta_1,\gamma_1,\delta_1) \to (\alpha_2,\beta_2,\gamma_2,\delta_2)\) as usual. In the case that spectator Wilson loops are present, just as for the earlier cases we sum over the possibilities for how we distribute the spectator Wilson loops across the three correlators in each of the three products to be summed over. 

\subsection{2-1-1 split}
Let us now turn to the case where the cut propagators are distributed across three Wilson loops, with two on one Wilson loop and one each on two others. Consider the cut 
\be
\langle A \, B \, i_1-1 \, i_1 \rangle = \langle A \, B \, i_2-1 \, i_2 \rangle = \langle A \, B \, j-1 \, j \rangle = \langle A \, B \, k-1 \, k \rangle = 0
\label{2-1-1split}
\ee
where \(i_1\) and \(i_2\) are on one Wilson loop, and  \(j\) and \(k\) are each on a different Wilson loop. We will identify \(\alpha_{1,2}\), \(\beta_{1,2}\), \(\gamma_{1,2}\) and \(\delta_{1,2}\) with the intersections of each Schubert solution with the lines \((i_1-1 \, i_1)\), \((i_2-1 \, i_2)\), \((j-1 \, j)\), and \((k-1 \, k)\) respectively.

\begin{figure}
\centering

\vspace{1cm}
\begin{tabular}{c@{\hspace{2cm}}c}
\begin{tikzpicture}[scale=0.6]
  \defineline{lin1m}{originx=1.25, originy=1.5, angle=270, pointcount=3}
  \defineline{lin4m}{originx=4.25, originy=1.5, angle=90, pointcount=3}
  \defineline{lin5m}{originx=-5, originy=0, angle=0, pointcount=3}
  \defineline{lin6m}{originx=-1.75, originy=-1, angle=0, pointcount=3}
  \defineline{linA}{originx=0, originy=-2.5, angle=180, pointcount=4,linelength=6}
 
\drawline[showtwistors=true,showvertex=true,wingorientation=outer,tolabel=$Z_B$,fromlabel=$Z_A$]{linA}
 
    \drawline[showwings=true, showtwistors=true, fromlabel={\tiny $\, \, \, \, \, \, Z_{i_2-1}$},tolabel={\tiny $Z_{i_2}$}]{lin1m}
  \drawline[showwings=true, showtwistors=true, fromlabel={\tiny $Z_{i_1-1} \, \, \,  \, \, \, $},tolabel={\tiny $Z_{i_1}$}]{lin4m}
  \drawline[showwings=true, showtwistors=true, fromlabel={\tiny \, $Z_{k-1}$},tolabel={\tiny $\,\,\,Z_{k}$}]{lin5m}
  \drawline[showwings=true, showtwistors=true, fromlabel={\tiny $Z_{j-1}$},tolabel={\tiny $\,\,Z_{j}$}]{lin6m}
   \drawpropagator{fromline=lin5m, toline=linA, fromip=2, toip=4}
   \drawpropagator{fromline=lin6m, toline=linA, fromip=2, toip=3}
   \drawpropagator{fromline=lin1m, toline=linA, fromip=2, toip=2}
   \drawpropagator{fromline=lin4m, toline=linA, fromip=2, toip=1}
\end{tikzpicture} &
\begin{tikzpicture}[scale=0.6]
  \defineline{lin1m}{originx=1.25, originy=1.5, angle=270, pointcount=3}
  \defineline{lin4m}{originx=4.25, originy=1.5, angle=90, pointcount=3}
  \defineline{lin5m}{originx=-5, originy=0, angle=0, pointcount=3}
  \defineline{lin6m}{originx=-1.75, originy=-1, angle=0, pointcount=3}
  \defineline{linA}{originx=0, originy=-2.5, angle=180, pointcount=4,linelength=6}
 
\drawline[showtwistors=true,showvertex=true,wingorientation=outer,tolabel=$Z_B$,fromlabel=$Z_A$]{linA}
 
    \drawline[showwings=true, showtwistors=true, fromlabel={\tiny $\, \, \, \, \, \, Z_{i_1-1}$},tolabel={\tiny $Z_{i_1}$}]{lin1m}
  \drawline[showwings=true, showtwistors=true, fromlabel={\tiny $Z_{i_2-1} \, \, \, \, \, \, $},tolabel={\tiny $Z_{i_2}$}]{lin4m}
  \drawline[showwings=true, showtwistors=true, fromlabel={\tiny \, $Z_{k-1}$},tolabel={\tiny $\,\,\,Z_{k}$}]{lin5m}
  \drawline[showwings=true, showtwistors=true, fromlabel={\tiny $Z_{j-1}$},tolabel={\tiny $\,\,Z_{j}$}]{lin6m}
   \drawpropagator{fromline=lin5m, toline=linA, fromip=2, toip=4}
   \drawpropagator{fromline=lin6m, toline=linA, fromip=2, toip=3}
   \drawpropagator{fromline=lin1m, toline=linA, fromip=2, toip=2}
   \drawpropagator{fromline=lin4m, toline=linA, fromip=2, toip=1}
\end{tikzpicture}\\
\begin{tikzpicture}[scale=0.6]
  \defineline{lin1m}{originx=1.25, originy=1.5, angle=270, pointcount=3}
  \defineline{lin4m}{originx=4.25, originy=1.5, angle=90, pointcount=3}
  \defineline{lin5m}{originx=-5, originy=0, angle=0, pointcount=3}
  \defineline{lin6m}{originx=-1.75, originy=-1, angle=0, pointcount=3}
  \defineline{linA}{originx=0, originy=-2.5, angle=180, pointcount=4,linelength=6}
 
\drawline[showtwistors=true,showvertex=true,wingorientation=outer,tolabel=$Z_B$,fromlabel=$Z_A$]{linA}
 
    \drawline[showwings=true, showtwistors=true, fromlabel={\tiny $\, \, \, \, \, \, Z_{i_2-1}$},tolabel={\tiny $Z_{i_2}$}]{lin1m}
  \drawline[showwings=true, showtwistors=true, fromlabel={\tiny $Z_{i_1-1} \, \, \, \, \, \, $},tolabel={\tiny $Z_{i_1}$}]{lin4m}
  \drawline[showwings=true, showtwistors=true, fromlabel={\tiny \, $Z_{j-1}$},tolabel={\tiny $\,\,\,Z_{j}$}]{lin5m}
  \drawline[showwings=true, showtwistors=true, fromlabel={\tiny $Z_{k-1}$},tolabel={\tiny $\,\,Z_{k}$}]{lin6m}
   \drawpropagator{fromline=lin5m, toline=linA, fromip=2, toip=4}
   \drawpropagator{fromline=lin6m, toline=linA, fromip=2, toip=3}
   \drawpropagator{fromline=lin1m, toline=linA, fromip=2, toip=2}
   \drawpropagator{fromline=lin4m, toline=linA, fromip=2, toip=1}
\end{tikzpicture} &
\begin{tikzpicture}[scale=0.6]
  \defineline{lin1m}{originx=1.25, originy=1.5, angle=270, pointcount=3}
  \defineline{lin4m}{originx=4.25, originy=1.5, angle=90, pointcount=3}
  \defineline{lin5m}{originx=-5, originy=0, angle=0, pointcount=3}
  \defineline{lin6m}{originx=-1.75, originy=-1, angle=0, pointcount=3}
  \defineline{linA}{originx=0, originy=-2.5, angle=180, pointcount=4,linelength=6}
 
\drawline[showtwistors=true,showvertex=true,wingorientation=outer,tolabel=$Z_B$,fromlabel=$Z_A$]{linA}
 
    \drawline[showwings=true, showtwistors=true, fromlabel={\tiny $\, \, \, \, \, \, Z_{i_1-1}$},tolabel={\tiny $Z_{i_1}$}]{lin1m}
  \drawline[showwings=true, showtwistors=true, fromlabel={\tiny $Z_{i_2-1} \, \, \, \, \, \, $},tolabel={\tiny $Z_{i_2}$}]{lin4m}
  \drawline[showwings=true, showtwistors=true, fromlabel={\tiny \, $Z_{j-1}$},tolabel={\tiny $\,\,\,Z_{j}$}]{lin5m}
  \drawline[showwings=true, showtwistors=true, fromlabel={\tiny $Z_{k-1}$},tolabel={\tiny $\,\,Z_{k}$}]{lin6m}
   \drawpropagator{fromline=lin5m, toline=linA, fromip=2, toip=4}
   \drawpropagator{fromline=lin6m, toline=linA, fromip=2, toip=3}
   \drawpropagator{fromline=lin1m, toline=linA, fromip=2, toip=2}
   \drawpropagator{fromline=lin4m, toline=linA, fromip=2, toip=1}
\end{tikzpicture}\\
\end{tabular}

\vspace{1.5cm}

\begin{tabular}{c@{\hspace{2cm}}c}
\begin{tikzpicture}[scale=0.6]
  \defineline{lin1m}{originx=-0.75, originy=1.5, angle=270, pointcount=3}
  \defineline{lin4m}{originx=3.75, originy=1.5, angle=90, pointcount=3}
  \defineline{lin5m}{originx=-4, originy=0, angle=0, pointcount=3}
  \defineline{lin6m}{originx=1.25, originy=-2, angle=0, pointcount=3}
  \defineline{linA}{originx=0, originy=-2.5, angle=180, pointcount=4,linelength=6}
 
\drawline[showtwistors=true,showvertex=true,wingorientation=outer,tolabel=$Z_B$,fromlabel=$Z_A$]{linA}
 
    \drawline[showwings=true, showtwistors=true, fromlabel={\tiny $\, \, \, \, \, \, Z_{i_2-1}$},tolabel={\tiny $Z_{i_2}$}]{lin1m}
  \drawline[showwings=true, showtwistors=true, fromlabel={\tiny $Z_{i_1-1} \, \, \, \, \, \, $},tolabel={\tiny $Z_{i_1}$}]{lin4m}
  \drawline[showwings=true, showtwistors=true, fromlabel={\tiny \, $Z_{k-1}$},tolabel={\tiny $\,\,\,Z_{k}$}]{lin5m}
  \drawline[showwings=true, showtwistors=true, fromlabel={\tiny $Z_{j-1}$},tolabel={\tiny $\,\,Z_{j}$}]{lin6m}
   \drawpropagator{fromline=lin5m, toline=linA, fromip=2, toip=4}
   \drawpropagator{fromline=lin6m, toline=linA, fromip=2, toip=2}
   \drawpropagator{fromline=lin1m, toline=linA, fromip=2, toip=3}
   \drawpropagator{fromline=lin4m, toline=linA, fromip=2, toip=1}
\end{tikzpicture} &
\begin{tikzpicture}[scale=0.6]
  \defineline{lin1m}{originx=-0.75, originy=1.5, angle=270, pointcount=3}
  \defineline{lin4m}{originx=3.75, originy=1.5, angle=90, pointcount=3}
  \defineline{lin5m}{originx=-4, originy=0, angle=0, pointcount=3}
  \defineline{lin6m}{originx=1.25, originy=-2, angle=0, pointcount=3}
  \defineline{linA}{originx=0, originy=-2.5, angle=180, pointcount=4,linelength=6}
 
\drawline[showtwistors=true,showvertex=true,wingorientation=outer,tolabel=$Z_B$,fromlabel=$Z_A$]{linA}
 
    \drawline[showwings=true, showtwistors=true, fromlabel={\tiny $\, \, \, \, \, \, Z_{i_2-1}$},tolabel={\tiny $Z_{i_2}$}]{lin1m}
  \drawline[showwings=true, showtwistors=true, fromlabel={\tiny $Z_{i_1-1} \, \, \, \, \, \, $},tolabel={\tiny $Z_{i_1}$}]{lin4m}
  \drawline[showwings=true, showtwistors=true, fromlabel={\tiny \, $Z_{j-1}$},tolabel={\tiny $\,\,\,Z_{j}$}]{lin5m}
  \drawline[showwings=true, showtwistors=true, fromlabel={\tiny $Z_{k-1}$},tolabel={\tiny $\,\,Z_{k}$}]{lin6m}
   \drawpropagator{fromline=lin5m, toline=linA, fromip=2, toip=4}
   \drawpropagator{fromline=lin6m, toline=linA, fromip=2, toip=2}
   \drawpropagator{fromline=lin1m, toline=linA, fromip=2, toip=3}
   \drawpropagator{fromline=lin4m, toline=linA, fromip=2, toip=1}
\end{tikzpicture}\\
\end{tabular}

\caption{The six N\(^2\)MHV 'seed' diagrams for the four-mass cut (\ref{2-1-1split}). Any other diagrams which one might draw are related to one of the shown diagrams by a cyclic permutation of the insertions on the Lagrangian, and as such are not really distinct.}
\label{2-1-1splitfig}
\end{figure}
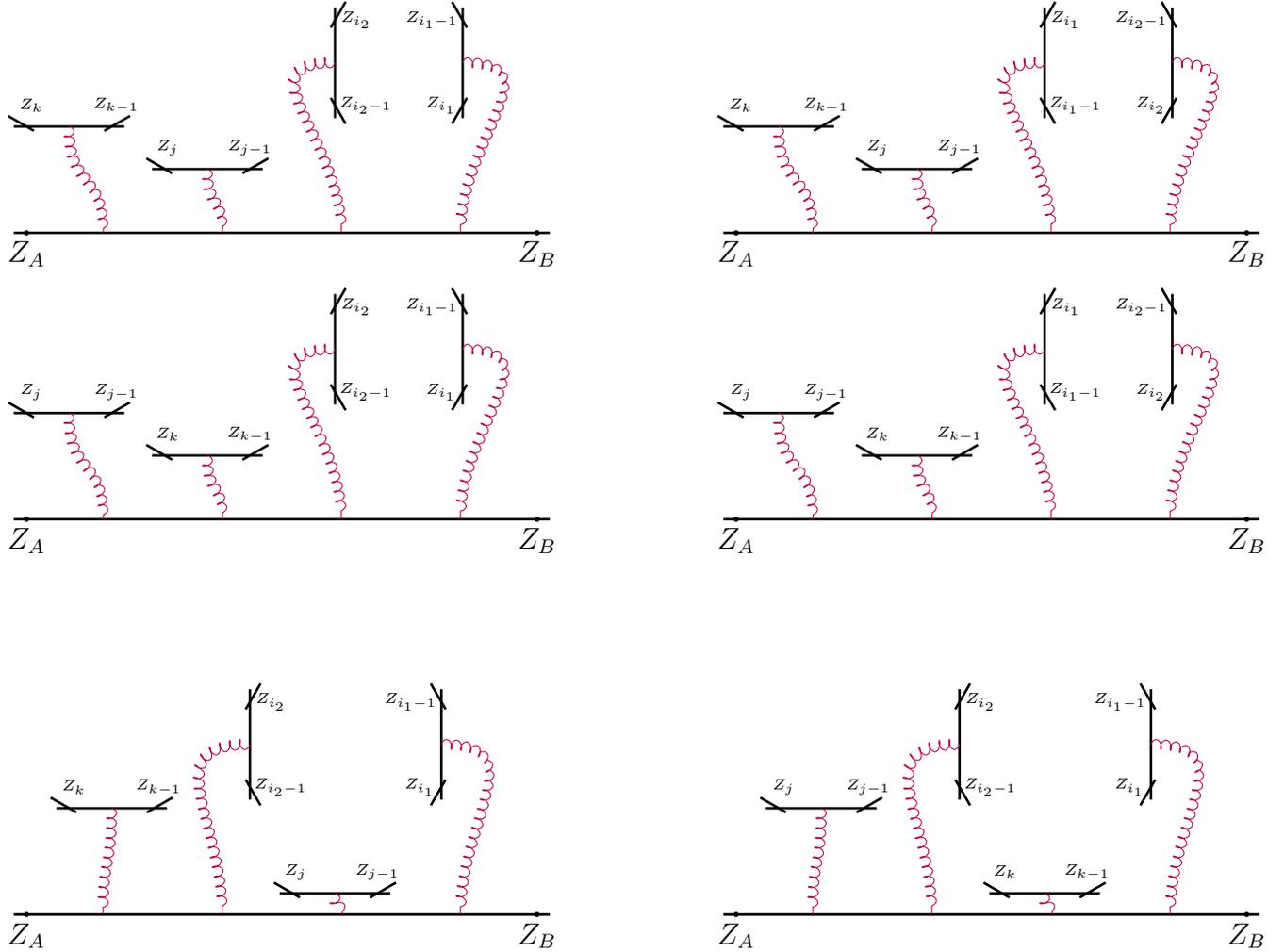

There are six `seed' diagrams up to cyclic permutations on the Lagrangian, all shown in Fig. \ref{2-1-1splitfig}; each of these is easily seen to decompose into a product of two correlators, and as usual these must be summed over, each dressed with the value of the cut of the original N\(^2\)MHV 'seed' diagram in the Abelian theory. 

As is by now familiar, we can read off the contribution from each of the six 'seed' diagrams very easily (including the pre-factors which are found by cutting the N$^2$MHV seed diagrams), and the end results for each diagram (clockwise from top left, to be summed over) are, for the first Schubert solution,

\begin{align}
\Bigl(-\frac{x(i_1,i_2,j,k)}{2}&-1\Bigr)C_1(i_1,i_2,j,k) \langle \mathcal{L}(\alpha_1,i_1,i_1+1,...,i_2-1,\beta_1)\rangle \notag \\
&\times \langle \mathcal{L}(\beta_1,i_2,i_2+1,...,i_1-1,\alpha_1,\delta_1,k,k+1,...,k-1,\delta_1,\gamma_1,j,j+1,...,j-1,\gamma_1)\rangle
\end{align}
\begin{align}
C_1(i_1,i_2,j,k) \langle & \mathcal{L}(\beta_1,i_2,i_2+1,...,i_1-1,\alpha_1)\rangle \notag \\
&\times \langle \mathcal{L}(\alpha_1,i_1,i_1+1,...,i_2-1,\beta_1,\delta_1,k,k+1,...,k-1,\delta_1,\gamma_1,j,j+1,...,j-1,\gamma_1)\rangle
\end{align}
\begin{align}
\Bigl(-\frac{x(i_1,i_2,j,k)}{2}&-1\Bigr)C_1(i_1,i_2,j,k) \langle  \mathcal{L}(\beta_1,i_2,i_2+1,...,i_1-1,\alpha_1)\rangle \notag \\
&\times \langle \mathcal{L}(\alpha_1,i_1,i_1+1,...,i_2-1,\beta_1,\gamma_1,j,j+1,...,j-1,\gamma_1,\delta_1,k,k+1,..,k-1,\delta_1)\rangle
\end{align}
\begin{align}
\frac{x(i_1,i_2,j,k)}{2}C_1(i_1,i_2,j,k)  \langle &\mathcal{L}(\gamma_1,j,j+1,...,j-1,\gamma_1,\beta_1,i_2,i_2+1,...,i_1-1,\alpha_1)\rangle\notag \\
&\times \langle  \mathcal{L}(\alpha_1,i_1,i_1+1,...,i_2-1,\beta_1,\delta_1,k,k+1,...,k-1,\delta_1)\rangle 
\end{align}
\begin{align}
\frac{x(i_1,i_2,j,k)}{2}C_1(i_1,i_2,j,k)  \langle &\mathcal{L}(\delta_1,k,k+1,...,k-1,\delta_1,\beta_1,i_2,i_2+1,...,i_1-1,\alpha_1)\rangle\notag \\
&\times \langle  \mathcal{L}(\alpha_1,i_1,i_1+1,...,i_2-1,\beta_1,\gamma_1,j,j+1,...,j-1,\gamma_1)\rangle 
\end{align}
\begin{align}
C_1(i_1,i_2,j,k)  \langle &\mathcal{L}(\alpha_1,i_1,i_1+1,...,i_2-1,\beta_1)\rangle\notag \\
&\times \langle  \mathcal{L}(\beta_1,i_2,i_2+1,...,i_1-1,\alpha_1,\gamma_1,j,j+1,...,j-1,\gamma_1,\delta_1,k,k+1,...,k-1,\delta_1)\rangle.
\end{align}

The cuts of one-loop diagrams which have zero colour trace due to at least one Wilson loop having fewer than two propagators running elsewhere will cancel via the same mechanism as spelled out in the 3-1 case.

As usual, for the second Schubert solution we replace \(C_1 \to C_2\), \(x \to \bar{x}\), and \((\alpha_1,\beta_1,\gamma_1,\delta_1) \to (\alpha_1,\beta_2,\gamma_2,\delta_2)\). In the case of spectator Wilson loops, for each of these six 'seed' diagrams we again replace the correlator part with the sum over all the possibilities for how to distribute the spectators across the two correlators in the product, to account for each region we could add the spectator to. 

\subsection{1-1-1-1 split}

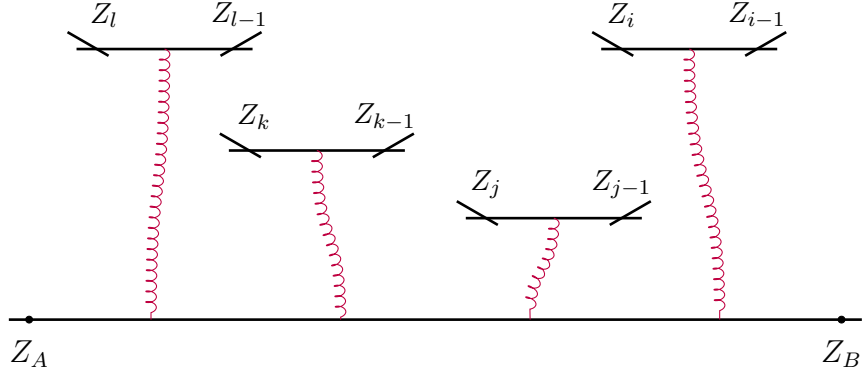
\begin{figure}
\begin{tikzpicture}[scale=0.9]

  \path[use as bounding box] (-3.5,-3) rectangle (3.5,2.5);

  \defineline{lin1m}{originx=7.75, originy=-1, angle=0, pointcount=3}
  \defineline{lin4m}{originx=9.75, originy=1.5, angle=0, pointcount=3}
  \defineline{lin5m}{originx=2, originy=1.5, angle=0, pointcount=3}
  \defineline{lin6m}{originx=4.25, originy=0, angle=0, pointcount=3}

  \defineline{linA}{originx=6, originy=-2.5, angle=180, pointcount=4,linelength=6}
  
\drawline[showtwistors=true,showvertex=true,wingorientation=outer,tolabel=$Z_B$,fromlabel=$Z_A$]{linA}
  
    \drawline[showwings=true, showtwistors=true, fromlabel={\small $Z_{j-1}$},tolabel={\small $Z_{j}$}]{lin1m}
  \drawline[showwings=true, showtwistors=true, fromlabel={\small $Z_{i-1}$},tolabel={\small $Z_{i}$}]{lin4m}
  \drawline[showwings=true, showtwistors=true, fromlabel={\small \,  $Z_{l-1}$},tolabel={\small $\,\,\,Z_{l}$}]{lin5m}
  \drawline[showwings=true, showtwistors=true, fromlabel={\small $Z_{k-1}$},tolabel={\small $\,\,Z_{k}$}]{lin6m}

   \drawpropagator{fromline=lin5m, toline=linA, fromip=2, toip=4}
   \drawpropagator{fromline=lin6m, toline=linA, fromip=2, toip=3}
   \drawpropagator{fromline=lin1m, toline=linA, fromip=2, toip=2}
   \drawpropagator{fromline=lin4m, toline=linA, fromip=2, toip=1}

\end{tikzpicture}
    \caption{One of the six N\(^2\)MHV 'seed' diagrams which contribute to a 1-1-1-1 four mass cut. The other five variations correspond to non-cyclic permutations of the order of the Wilson loops, e.g. keeping the \(i\)-label fixed and permuting the others.}
\label{1-1-1-1-diagram}
\end{figure}

Let us finally consider the case where one propagator is cut on each of four Wilson loops, initially without any spectator Wilson loops. In particular, let us take the cut
\be
\langle A \, B \, i-1 \, i \rangle =  \langle A \, B \, j-1 \, j \rangle =  \langle A \, B \, k-1 \, k \rangle =  \langle A \, B \, l-1 \, l \rangle = 0
\label{1-1-1-1cut}
\ee
Clearly there are six 'seed' diagrams at N\(^2\)MHV (vanishing for trace reasons in the $SU(N)$ theory) given by the one depicted in Fig. \ref{1-1-1-1-diagram} and the other five non-cyclic permutations of the order of the Wilson loops. Note that unlike all other cases, there is no longer any partioning into regions and as such we will have a \emph{single} Wilson loop rather than a product. 

Denoting the intersections of the Schubert solutions with \((i-1 \, i)\), \((j-1 \, j)\), \((k-1 \, k )\) and \((l-1 \, l)\) respectively as \(\alpha_{1,2}\), \(\beta_{1,2}\), \(\gamma_{1,2}\) and \(\delta_{1,2}\) respectively, we have that the contribution from the diagram depicted in Fig. \ref{1-1-1-1-diagram} is (as usual, the coefficients are determined by taking the cut of the N\(^2\)MHV 'seed' diagram, in the Abelian theory to avoid vanishing trace in the case of $SU(N)$)
\small{
\begin{align}
&C_1(i,j,k,l) \notag \\
&\langle\mathcal{L}(\alpha_1,i-1,i,i+1,...,i,\alpha_1,\beta_1,j-1,j,j+1,...,j,\beta_1,\gamma_1,k-1,k,k+1,...,k,\gamma_1,\delta_1,l-1,l,l+1,...,l,\delta_1) \rangle
\end{align}}
on the first Schubert solution, with the other five seed diagrams yielding
\small{
\begin{align}
&\bigl(-1-\frac{x(i,j,k,l)}{2}\bigr)C_1(i,j,k,l) \notag \\
&\langle\mathcal{L}(\alpha_1,i-1,i,i+1,...,i,\alpha_1,\beta_1,j-1,j,j+1,...,j,\beta_1,\delta_1,l-1,l,l+1,...,l,\delta_1,\gamma_1,k-1,k,k+1,...,k,\gamma_1) \rangle
\end{align}}
\small{
\begin{align}
&\frac{x(i,j,k,l)}{2}C_1(i,j,k,l) \notag \\
&\langle\mathcal{L}(\alpha_1,i-1,i,i+1,...,i,\alpha_1,\gamma_1,k-1,k,k+1,...,k,\gamma_1,\beta_1,j-1,j,j+1,...,j,\beta_1,\delta_1,l-1,l,l+1,...,l,\delta_1) \rangle
\end{align}}
\small{
\begin{align}
&\bigl(-1-\frac{x(i,j,k,l)}{2}\bigr)C_1(i,j,k,l) \notag \\
&\langle\mathcal{L}(\alpha_1,i-1,i,i+1,...,i,\alpha_1,\gamma_1,k-1,k,k+1,...,k,\gamma_1,\delta_1,l-1,l,l+1,...,l,\delta_1,\beta_1,j-1,j,j+1,...,j,\beta_1) \rangle
\end{align}}
\small{
\begin{align}
&C_1(i,j,k,l) \notag \\
&\langle\mathcal{L}(\alpha_1,i-1,i,i+1,...,i,\alpha_1,\delta_1,l-1,l,l+1,...,l,\delta_1,\gamma_1,k-1,k,k+1,...,k,\gamma_1,\beta_1,j-1,j,j+1,...,j,\beta_1) \rangle
\end{align}}
\small{
\begin{align}
&\frac{x(i,j,k,l)}{2}C_1(i,j,k,l) \notag \\
&\langle\mathcal{L}(\alpha_1,i-1,i,i+1,...,i,\alpha_1,\delta_1,l-1,l,l+1,...,l,\delta_1,\beta_1,j-1,j,j+1,...,j,\beta_1,\gamma_1,k-1,k,k+1,...,k,\gamma_1) \rangle
\end{align}}

Again, the cuts of one-loop diagrams which do not actually contribute to the planar limit because not all of the Wilson loops have at least two propagators running elsewhere will cancel via the same mechanism as spelled out in the 3-1 case.

As usual, the results on the second Schubert solution may be generated from the above by replacing \(C_1 \to C_2\), \(x \to \bar{x}\) and \((\alpha_1,\beta_1,\gamma_1,\delta_1) \to (\alpha_2,\beta_2,\gamma_2,\delta_2)\). In the case of spectator Wilson loops, each Wilson loop in the sum is simply replaced by the (connected part of the) correlator with all of those spectators; there is no longer a sum over how we distribute the spectators among the regions, because there is only one region! 

Note that, at first glance, the above formulae may appear to suggest that there is a contribution to this cut at N\({}^3\)MHV, since for each term we have a single Wilson loop correlator (which enters at NMHV) multiplying two $R$-invariants. \emph{However}, note that all six correlators are actually identical at NMHV, and the sum of the six bosonic factors that the seed diagrams come with is zero. It is therefore only once the correlator is taken to at least N\({}^2\)MHV (i.e. for the N\({}^4\)MHV contribution or above to the cut) that we receive a non-zero contribution.

\subsection{Lower mass example: 2-2 two-mass easy}
\label{lowermass}
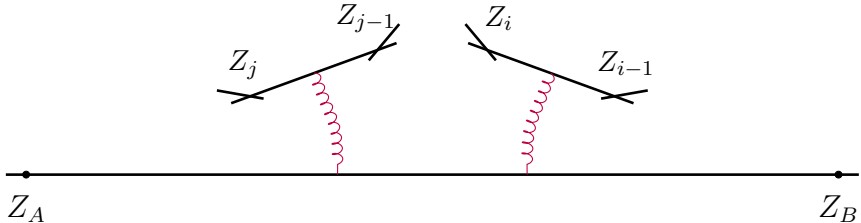
\begin{figure}
\begin{tikzpicture}[scale=0.9]

  \path[use as bounding box] (-3.5,-3) rectangle (3.5,2.5);

  \defineline{lin1m}{originx=7.75, originy=-1, angle=340, pointcount=3}
  \defineline{lin6m}{originx=4.25, originy=-1, angle=20, pointcount=3}

  \defineline{linA}{originx=6, originy=-2.5, angle=180, pointcount=4,linelength=6}
  
\drawline[showtwistors=true,showvertex=true,wingorientation=outer,tolabel=$Z_B$,fromlabel=$Z_A$]{linA}
  
    \drawline[showwings=true, showtwistors=true, fromlabel={\small $Z_{i-1}$},tolabel={\small $Z_{i}$}]{lin1m}
  \drawline[showwings=true, showtwistors=true, fromlabel={\small $Z_{j-1}$},tolabel={\small $\,\,Z_{j}$}]{lin6m}

   \drawpropagator{fromline=lin6m, toline=linA, fromip=2, toip=3}
   \drawpropagator{fromline=lin1m, toline=linA, fromip=2, toip=2}

\end{tikzpicture}
    \caption{A planar twistor diagram for the correlator of two Wilson loops at MHV which has non-zero residue on the quadruple pole in Eq. \ref{2-2_2mecut} for the first Schubert solution. The other three diagrams which contribute to this cut are related by \(i_1 \to i_1+1\) and/or \(j_1 \to j_1+1\).}
\label{22kermit}
\end{figure}

As in the case of a single Wilson loop, we would expect the lower mass leading singularities to follow smoothly from collinear limits on the appropriate four-mass version. However, it is also possible to derive them directly by using the same twistor diagrammatic arguments. As an example, let us consider the case of a 2-2 two-mass easy leading singularity on the first Schubert solution, e.g. the residue on 
\be
\langle A \, B \, i - 1 \, i \rangle = \langle A \, B \, i  \, i + 1 \rangle = \langle A \, B \, j -1 \, j \rangle = \langle A \, B \, j \, j+1 \rangle = 0
\label{2-2_2mecut}
\ee
when taking the Schubert solution \((AB) = (ij)\). As in the case of a single Wilson loop, there are four 'seed' diagrams with non-zero cut at MHV: the diagram depicted in Fig. \ref{22kermit}, and those related by \(i \to i+1\) and/or \(j \to j + 1\). Since all four seed diagrams clearly give the same correlator contributions, we may combine them into one, noting that the sum of the cuts of the four seed diagrams is \(-1\).

Recalling that for this Schubert solution we have \(\alpha_1 = \beta_1 = i\) and \(\gamma_1 = \delta_1 = j\) (where as usual \(\alpha_1,\beta_1,\gamma_1,\delta_1\) are the intersection of our Schubert solutions with the lines \((i-1 \, i)\), \((i \, i +1)\), \((j-1, \, j)\) and \((j \, j+1)\) respectively), we would naturally write down the correlator contribution 
\be
- \langle\mathcal{L}(\alpha_1,i,i+1,...,i-1,\alpha_1,\gamma_1,j,j+1,...,j-1,\gamma_1)\rangle
\label{incompleteCut1}
\ee
which we can write as 
\be
- \langle\mathcal{L}(\alpha_1,i,i+1,...,i-1,\beta_1,\gamma_1,j,j+1,...,j-1,\delta_1)\rangle
\label{incompleteCut2}
\ee
to make more manifest the descent of this correlator from the equivalent four-mass leading singularity. However, there is a subtlty: this includes contributions from diagrams where the additional propagators stay entirely within the first Wilson loop or the second Wilson loop, and these one-loop diagrams are zero due to the vanishing \(SU(N)\) generators. As such their cuts must be subtracted off from the correlator, which amounts to updating our expression to 
\be
- \langle\mathcal{L}(\alpha_1,i,i+1,...,i-1,\beta_1,\gamma_1,j,j+1,...,j-1,\delta_1)\rangle + \langle\mathcal{L}(\alpha_1,i,i+1,...,i-1) \rangle\langle\mathcal{L}(\beta_1,j,j+1,...,j-1) \rangle
\label{fixedCut}
\ee

Note that diagrams where one of the Wilson loops has a propagator running to the Lagrangian line and the other Wilson loop has propagators which only stay within it \emph{also} come with vanishing colour factor, and are not subtracted off in the above. However, these diagrams cancel out in cyclically related sums in the same way as seen for the traceless dressings of e.g. the 3-1 four-mass seed diagram, and so don't \emph{need} to be subtracted off.

This matches exactly with the formula obtained by taking collinear limits on the 2-2 four-mass leading singularity: the two terms are each collinear descendants of one of the four terms in the sum for the four-mass expression, and the other two terms have vanished since one of the Wilson loops in the product has degenerated to a two-cusp Wilson loop which necessarily vanishes by backtracking. 

\section{4-mass leading singularities of $\langle \mathcal{W}_1 \hdots \mathcal{W}_n\rangle^c$}
\label{Sec-allLeadingSing}

We have tested that, for lower-mass leading singularities, the expressions which arise by taking collinear limits of the appropriate four-mass case align with what is obtained by a direct, twistor-diagrammatic argument. Moreover, we have subjected the resulting formulae to robust checks by using them to generate \(O(g^2)\) correlators at N${^2}$MHV and N$^3$MHV in the chiral box expansion and testing that the resulting expressions are independent of the auxillary bitwistor \(X\). This is a highly non-trivial test which relies delicately on linear relations among the various leading singularities. 

We now collect the general formulae for the leading singularities of the connected part of a fully general correlator of multiple Wilson loops, which amounts (via e.g. the chiral box expansion) to the general solution to the \(O(g^2)\) problem.

For brevity, we present just the 4-mass leading singularities here, but we include the lower mass leading singularities in appendix \ref{lowermassleading}. The 4-mass leading singularities are the cuts corresponding to
\begin{equation}
    \langle A B I i \rangle =\langle A B J j\rangle =\langle A B K k \rangle = \langle A B L l \rangle=0
\end{equation}
where \(I=i-1\), \(J=j-1\), \(K=k-1\), and \(L=l-1\). As outlined in Section \ref{schubertproblem}, the singular configurations of the Lagrangian line are those which simultaneously intersects the 4 lines $(Ii),(Jj),(Kk)$ and $(L,l)$ simultaneously. This set of equations has two solutions. The intersection point of the line which solves this problem with the 4 lines $(I i),(J,j),(K,k),(L,l)$ are given explicitly by \\\\
Schubert Solution 1: 
\begin{equation}
    Z_{\alpha_1}=Z_i+X_1Z_{I},\hspace{2mm} Z_{\beta_1}=Z_{J}+X_2Z_j, \hspace{2mm} Z_{\gamma_1}=(k\hspace{0.5mm}K)\cap(L\hspace{0.5mm}l\hspace{0.5mm}\alpha_1), \hspace{2mm} Z_{\delta_1}=(L\hspace{0.5mm}l)\cap (k\hspace{0.5mm}K\hspace{0.5mm}\beta_1)
\end{equation}
Schubert Solution 2:
\begin{equation}
     Z_{\alpha_2}=Z_{I}+Y_1Z_i,\hspace{2mm} Z_{\beta_2}=Z_J+Y_2Z_{j}, \hspace{2mm} Z_{\gamma_2}=(K\hspace{0.5mm}k)\cap(j\hspace{0.5mm}J\hspace{0.5mm}\alpha_2), \hspace{2mm} Z_{\delta_2}=(l\hspace{0.5mm}L)\cap (I\hspace{0.5mm}i\hspace{0.5mm}\beta_2)
\end{equation}
where 
\begin{align}
    &X_1=\frac{\langle i\hspace{1mm} J \hspace{1mm}j\hspace{1mm} (k\hspace{0.5mm}K)\cap (L\hspace{0.5mm}l\hspace{0.5mm}I)\rangle+\langle I\hspace{1mm} J\hspace{1mm} j\hspace{1mm} (k\hspace{0.5mm}K)\cap (L\hspace{0.5mm}l\hspace{0.5mm}i)\rangle+\langle i\hspace{1mm} I\hspace{1mm} k\hspace{1mm} K\rangle \langle J\hspace{1mm} j\hspace{1mm} L\hspace{1mm} l \rangle \Delta}{2\langle J\hspace{1mm} j\hspace{1mm} (k\hspace{0.5mm} K)\cap (L\hspace{0.5mm} l\hspace{0.5mm} I) \hspace{1mm}I \rangle}\notag\\
    &X_2=\frac{\langle J\hspace{1mm} i\hspace{1mm} I\hspace{1mm} (L\hspace{0.5mm}l)\cap (k\hspace{0.5mm}K\hspace{0.5mm}j)\rangle+\langle j\hspace{1mm} i\hspace{1mm} I\hspace{1mm} (L\hspace{0.5mm}l)\cap (k\hspace{0.5mm}K\hspace{0.5mm}J)\rangle+\langle i\hspace{0.5mm} I\hspace{1mm} k\hspace{1mm} K\rangle \langle J\hspace{1mm} j \hspace{1mm}L \hspace{1mm}l \rangle \Delta}{2\langle i\hspace{1mm} I\hspace{1mm} (L\hspace{0.5mm} l)\cap (k\hspace{0.5mm} K\hspace{0.5mm} j) \hspace{1mm}j \rangle}\notag\\
    &Y_1=\frac{\langle I\hspace{1mm} l \hspace{1mm}L\hspace{1mm} (K\hspace{0.5mm}k)\cap (j\hspace{0.5mm}J\hspace{0.5mm}i)\rangle+\langle i\hspace{1mm} l\hspace{1mm} L\hspace{1mm} (K\hspace{0.5mm}k)\cap (j\hspace{0.5mm}J\hspace{0.5mm}I)\rangle+\langle I\hspace{1mm} i\hspace{1mm} K\hspace{1mm} k\rangle \langle j\hspace{1mm} J\hspace{1mm} l\hspace{1mm} L \rangle \Delta}{2\langle l\hspace{1mm} L\hspace{1mm} (K\hspace{0.5mm} k)\cap (j\hspace{0.5mm} J\hspace{0.5mm} i) \hspace{1mm}i \rangle}\notag\\
    &Y_2=\frac{\langle j\hspace{1mm} K\hspace{1mm} k\hspace{1mm} (l\hspace{0.5mm}L)\cap (I\hspace{0.5mm}i\hspace{0.5mm}J)\rangle+\langle J\hspace{1mm} K\hspace{1mm} k\hspace{1mm} (l\hspace{0.5mm}L)\cap (I\hspace{0.5mm}i\hspace{0.5mm}j)\rangle+\langle I\hspace{0.5mm} i\hspace{1mm} K\hspace{1mm} k\rangle \langle j\hspace{1mm} J \hspace{1mm}l \hspace{1mm}L \rangle \Delta}{2\langle K\hspace{1mm} k\hspace{1mm} (l\hspace{0.5mm} L)\cap (I\hspace{0.5mm} i\hspace{0.5mm} J) \hspace{1mm}J \rangle}\notag\\
    &\Delta=\sqrt{(1-u-v)^2-4uv},\hspace{2mm}u=\frac{\langle I \hspace{1mm}i \hspace{1mm} J \hspace{1mm}j \rangle\langle K \hspace{1mm} k \hspace{1mm} L \hspace{1mm} l\rangle}{\langle I \hspace{1mm} i \hspace{1mm} K \hspace{1mm} k \rangle \langle J \hspace{1mm} j \hspace{1mm} L \hspace{1mm} l\rangle}, \hspace{2mm}v=\frac{\langle J \hspace{1mm}j \hspace{1mm} K \hspace{1mm}k \rangle\langle L \hspace{1mm} l \hspace{1mm} I \hspace{1mm} i\rangle}{\langle J \hspace{1mm} j \hspace{1mm} L \hspace{1mm} l \rangle \langle K \hspace{1mm} k \hspace{1mm} I \hspace{1mm} i\rangle}
    \label{leadingsingparamdefs}
\end{align}
Let us label the cut for Schubert solution 1 as $\textrm{\emph{Cut}}_1$ and the cut for Schubert solution 2 as $\textrm{\emph{Cut}}_2$. The expression for the cuts will depend on the distribution of the lines $(Ii),(Jj),(Kk),(Ll)$ from the Wilson loops within the correlator and we will give each case. To present the distribution of spectator Wilson loops in the leading singularities, we introduce some notation.

Let $P^{n,(m_1,\hdots,m_k)}_{q}$ be the set of ordered partitions of $\{\mathcal{W}_1,\hdots,\mathcal{W}_n\}\backslash\{\mathcal{W}_{m_1},\hdots,\mathcal{W}_{m_k}\}$ into $q$ subsets for $n-k\geq 1$ (spectator Wilson loops appear) and $q\geq 1$ (we distinguish between the ordering of the subsets but not the ordering of the Wilson loops within a subset). Let $P^{n,(m_1,\hdots,m_k)}_{q}\equiv\{\{\{\},\{\},\{\},\{\}\}\}$ for $n-k=0$ (no spectator Wilson loops). Let $p\in P^{n,(m_1,\hdots,m_k)}_{q}$, $\tilde{\mathcal{W}}_{i}\equiv\prod_{j} p_{i,j}$ for $p_{i}\neq \{\}$ and $\tilde{\mathcal{W}}_{i}\equiv\mathbbm{1}$ for $p_{i}=\{\}$. As a simple example, $P^{4,(1,2)}_{2}=\{\{\{\mathcal{W}_3,\mathcal{W}_4\},\{\}\},\{\{\mathcal{W}_3\},\{\mathcal{W}_4\}\},\{\{\mathcal{W}_4\},\{\mathcal{W}_3\}\},\{\{\},\{\mathcal{W}_3,\mathcal{W}_4\}\}\}$. For $p=\{\{\mathcal{W}_3,\mathcal{W}_4\},\{\}\}$, $\tilde{\mathcal{W}}_1=\mathcal{W}_3\mathcal{W}_4$ and $\tilde{\mathcal{W}}_2=\mathbbm{1}$. In the leading singularities, $\{\mathcal{W}_{m_1},\hdots,\mathcal{W}_{m_k}\}$ are the Wilson loops which have propagators appearing in the cut. $\{\mathcal{W}_1,\hdots,\mathcal{W}_n\}\backslash\{\mathcal{W}_{m_1},\hdots,\mathcal{W}_{m_k}\}$ is the set of spectator Wilson loops. $P^{n,(m_1,\hdots,m_k)}_{q}$ are the different ways of distributing the $n-k$ spectator Wilson loops amongst the $q$ correlator factors in the expression for the leading singularity. The 4-mass leading singularities are then given by
\begin{flalign}
 &\textrm{\textbf{4-0 4-mass}}&\notag \\
 &\{I,i,J,j,K,k,L,l\}\in \mathcal{W}_m\notag\\
 &\textrm{\emph{Cut}}_1=-\phi_1 [\delta_1,I,i,J,j][\beta_1,K,k,L,l]\sum_{p \hspace{0.5mm}\in P^{n,(m)}_{4}}\langle \mathcal{L}[w_{\alpha_1\beta_1}]\hspace{1mm} \tilde{\mathcal{W}}_{1}\rangle\langle\mathcal{L}[w_{\beta_1\gamma_1}]\hspace{1mm} \tilde{\mathcal{W}}_{2}\rangle\langle\mathcal{L}[w_{\gamma_1\delta_1}]\hspace{1mm}\tilde{\mathcal{W}}_{3}\rangle \langle\mathcal{L}[w_{\delta_1\alpha_1}]\hspace{1mm}\tilde{\mathcal{W}}_{4}\rangle\notag\\
 &\textrm{\emph{Cut}}_2=\phi_2 [\alpha_2,J,j,K,k][\gamma_2,L,l,I,i]\sum_{p \hspace{0.5mm}\in P^{n,(m)}_{4}}\langle \mathcal{L}[w_{\alpha_2\beta_2}]\hspace{1mm}\tilde{\mathcal{W}}_{1}\rangle\langle\mathcal{L}[w_{\beta_2\gamma_2}]\hspace{1mm} \tilde{\mathcal{W}}_{2} \rangle\langle\mathcal{L}[w_{\gamma_2\delta_2}]\hspace{1mm}\tilde{\mathcal{W}}_{3}\rangle \langle\mathcal{L}[w_{\delta_2\alpha_2}]\hspace{1mm}\tilde{\mathcal{W}}_{4}\rangle\notag\\
  &w_{\alpha_a\beta_a}=(\alpha_a,i,\dots,J,\beta_a),\hspace{2mm} w_{\beta_a\gamma_a}=(\beta_a,j,\dots,K,\gamma_a),\hspace{2mm} w_{\gamma_a\delta_a}=(\gamma_a,k,\dots,L,\delta_a), \hspace{2mm} w_{\delta_a\alpha_a}= (\delta_a,l,\dots,I,\alpha_a)
\end{flalign}
\begin{flalign}
 &\textrm{\textbf{3-1 4-mass}}&\notag\\
 &\{I,i,J,j,K,k\}\in \mathcal{W}_{m_1}, \hspace{5mm} \{L,l\}\in \mathcal{W}_{m_2}\notag\\
 &\textrm{\emph{Cut}}_1=-\phi_1[\delta_1,I,i,J,j][\beta_1,K,k,L,l]\sum_{p \hspace{0.5mm}\in P^{n,(m_1,m_2)}_{3}}\notag\Bigl(\langle \mathcal{L}[w_{\alpha_1\beta_1}]\tilde{\mathcal{W}}_1\rangle\langle\mathcal{L}[w_{\beta_1\gamma_1}]\tilde{\mathcal{W}}_2\rangle\langle\mathcal{L}[w_{\gamma_1\alpha_1},w_{\delta_1\delta_1}]\tilde{\mathcal{W}}_3\rangle\notag\\
    &\hspace{45.5mm}+\frac{x}{2}\langle \mathcal{L}[w_{\alpha_1\beta_1},w_{\delta_1\delta_1}]\tilde{\mathcal{W}}_1\rangle\langle\mathcal{L}[w_{\beta_1\gamma_1}]\tilde{\mathcal{W}}_2\rangle\langle\mathcal{L}[w_{\gamma_1\alpha_1}]\tilde{\mathcal{W}}_3\rangle&\notag\\
    &\hspace{45.5mm}+\left(-\frac{x}{2}-1\right)\langle \mathcal{L}[w_{\alpha_1\beta_1}]\tilde{\mathcal{W}}_1\rangle\langle\mathcal{L}[w_{\beta_1\gamma_1},w_{\delta_1\delta_1}]\tilde{\mathcal{W}}_2\rangle\langle\mathcal{L}[w_{\gamma_1\alpha_1}]\tilde{\mathcal{W}}_1\rangle\Bigr)\notag\\
 &\textrm{\emph{Cut}}_2=\phi_2[\alpha_2,J,j,K,k][\gamma_2,L,l,I,i]\sum_{p \hspace{0.5mm}\in P^{n,(m_1,m_2)}_{3}} \Bigl(\langle \mathcal{L}[w_{\alpha_2\beta_2}]\tilde{\mathcal{W}}_1\rangle\langle\mathcal{L}[w_{\beta_2\gamma_2}]\tilde{\mathcal{W}}_2\rangle\langle\mathcal{L}[w_{\gamma_2,\alpha_2},w_{\delta_2\delta_2}]\tilde{\mathcal{W}}_3\rangle\notag\\
    &\hspace{45.5mm}+\frac{\overline{x}}{2}\langle \mathcal{L}[w_{\alpha_2\beta_2},w_{\delta_2\delta_2}]\tilde{\mathcal{W}}_1\rangle\langle\mathcal{L}[w_{\beta_2\gamma_2}]\tilde{\mathcal{W}}_2\rangle\langle\mathcal{L}[w_{\gamma_2\alpha_2}]\tilde{\mathcal{W}}_3\rangle&\notag\\
    &\hspace{45.5mm}+\left(-\frac{\overline{x}}{2}-1\right)\langle \mathcal{L}[w_{\alpha_2\beta_2}]\tilde{\mathcal{W}}_1\rangle\langle\mathcal{L}[w_{\beta_2\gamma_2},w_{\delta_2\delta_2}]\tilde{\mathcal{W}}_2\rangle\langle\mathcal{L}[w_{\gamma_2\alpha_2}]\tilde{\mathcal{W}}_3\rangle\Bigr)\notag\\
    &w_{\alpha_a\beta_a}=(\alpha_a,i,\dots,J,\beta_a),\hspace{2mm} w_{\beta_a\gamma_a}=(\beta_a,j,\dots,K,\gamma_a),\hspace{2mm} w_{\gamma_a\alpha_a}=(\gamma_a,k,\dots,I,\alpha_a), \hspace{2mm} w_{\delta_a\delta_a}= (\delta_a,l,\dots,L,\delta_a)
\end{flalign}
\begin{flalign}
 &\textrm{\textbf{2-2 4-mass}}&\notag\\
 &\{I,i,J,j\}\in \mathcal{W}_{m_1}, \hspace{5mm} \{K,k,L,l\}\in \mathcal{W}_{m_2}\notag\\
 &\textrm{\emph{Cut}}_1=-\phi_1[\delta_1,I,i,J,j][\beta_1,K,k,L,l]  \sum_{p \hspace{0.5mm}\in P^{n,(m_1,m_2)}_{3}} \Bigl(\langle \mathcal{L}[w_{\alpha_1\beta_1},w_{\gamma_1\delta_1}]\tilde{\mathcal{W}}_1\rangle\langle\mathcal{L}[w_{\beta_1\alpha_1}]\tilde{\mathcal{W}}_2\rangle \langle\mathcal{L}[w_{\delta_1\gamma_1}]\tilde{\mathcal{W}}_3\rangle&\notag\\
 &\hspace{50.25mm}+\langle\mathcal{L}[w_{\alpha_1\beta_1}]\tilde{\mathcal{W}}_1\rangle \langle\mathcal{L}[w_{\gamma_1\delta_1}]\tilde{\mathcal{W}}_2\rangle \langle \mathcal{L}[w_{\beta_1\alpha_1},w_{\delta_1\gamma_1}]\tilde{\mathcal{W}}_3\rangle\notag\\
 &\hspace{50.25mm}+\left(-\frac{x}{2}-1\right)\big(\langle \mathcal{L}[w_{\alpha_1\beta_1},w_{\delta_1\gamma_1}]\tilde{\mathcal{W}}_1\rangle\notag\langle\mathcal{L}[w_{\beta_1\alpha_1}]\tilde{\mathcal{W}}_2\rangle \langle\mathcal{L}[w_{\gamma_1\delta_1}]\tilde{\mathcal{W}}_3\rangle&\notag\\
 &\hspace{50.25mm}+\langle\mathcal{L}[w_{\alpha_1\beta_1}]\tilde{\mathcal{W}}_1\rangle \langle\mathcal{L}[w_{\delta_1\gamma_1}]\tilde{\mathcal{W}}_2\rangle \langle \mathcal{L}[w_{\beta_1\alpha_1},w_{\gamma_1\delta_1}]\tilde{\mathcal{W}}_3\rangle\bigr)\Bigr)\notag\\
 &\textrm{\emph{Cut}}_2=\phi_2[\alpha_2,J,j,K,k][\gamma_2,L,l,I,i]\sum_{p \hspace{0.5mm}\in P^{n,(m_1,m_2)}_{3}}\Bigl(\langle \mathcal{L}[w_{\alpha_2\beta_2},w_{\gamma_2\delta_2}]\tilde{\mathcal{W}}_1\rangle \langle\mathcal{L}[w_{\beta_2\alpha_2}]\tilde{\mathcal{W}}_2\rangle \langle\mathcal{L}[w_{\delta_2\gamma_2}]\tilde{\mathcal{W}}_3\rangle&\notag\\
 &\hspace{50.25mm}+\langle\mathcal{L}[w_{\alpha_2\beta_2}]\tilde{\mathcal{W}}_1\rangle \langle\mathcal{L}[w_{\gamma_2\delta_2}]\tilde{\mathcal{W}}_2\rangle \langle \mathcal{L}[w_{\beta_2\alpha_2},w_{\delta_2\gamma_2}]\tilde{\mathcal{W}}_3\rangle\notag\\
 &\hspace{50.25mm}+\left(-\frac{\overline{x}}{2}-1\right)\big(\langle \mathcal{L}[w_{\alpha_2\beta_2},w_{\delta_2\gamma_2}]\tilde{\mathcal{W}}_1\rangle\langle\mathcal{L}[w_{\beta_2\alpha_2}]\tilde{\mathcal{W}}_2\rangle \langle\mathcal{L}[w_{\gamma_2\delta_2}]\tilde{\mathcal{W}}_3\rangle&\notag\\
 &\hspace{50.25mm}+\langle\mathcal{L}[w_{\alpha_2\beta_2}]\tilde{\mathcal{W}}_1\rangle \langle\mathcal{L}[w_{\delta_2\gamma_2}]\tilde{\mathcal{W}}_2\rangle \langle \mathcal{L}[w_{\beta_2\alpha_2},w_{\gamma_2\delta_2}]\tilde{\mathcal{W}}_3\rangle\bigr)\Bigr)\notag\\
  &w_{\alpha_a\beta_a}=(\alpha_a,i,\dots,J,\beta_a),\hspace{2mm} w_{\beta_a\alpha_a}=(\beta_a,j,\dots,I,\alpha_a),\hspace{2mm} w_{\gamma_a\delta_a}=(\gamma_a,k,\dots,L,\delta_a), \hspace{2mm} w_{\delta_a\gamma_a}= (\delta_a,l,\dots,K,\gamma_a)
\end{flalign}
\begin{flalign}
 &\textrm{\textbf{2-1-1 4-mass}}&\notag\\
 &\{I,i,J,j\}\in \mathcal{W}_{m_1}, \hspace{2mm} \{K,k\}\in \mathcal{W}_{m_2}, \hspace{2mm} \{L,l\}\in \mathcal{W}_{m_3}\notag\\
 &\textrm{\emph{Cut}}_1=-\phi_1[\delta_1,I,i,J,j][\beta_1,K,k,L,l] \notag\\
 &\hspace{5mm}\times\sum_{p \hspace{0.5mm}\in P^{n,(m_1,m_2,m_3)}_{2}}\Bigl(\langle \mathcal{L}[w_{\alpha_1\beta_1}]\tilde{\mathcal{W}}_1\rangle \langle \mathcal{L}[w_{\beta_1\alpha_1},w_{\gamma_1\gamma_1},w_{\delta_1\delta_1}]\tilde{\mathcal{W}}_2\rangle+\langle \mathcal{L}[w_{\beta_1\alpha_1}]\tilde{\mathcal{W}}_1\rangle\langle \mathcal{L}[w_{\alpha_1\beta_1},w_{\delta_1\delta_1},w_{\gamma_1\gamma_1}]\tilde{\mathcal{W}}_2\rangle \notag\\
   &\hspace{5mm}+\frac{x}{2}\bigl(\langle \mathcal{L}[w_{\alpha_1\beta_1},w_{\gamma_1\gamma_1}]\tilde{\mathcal{W}}_1\rangle \langle \mathcal{L}[w_{\beta_1\alpha_1},w_{\delta_1\delta_1}]\tilde{\mathcal{W}}_2\rangle+\langle \mathcal{L}[w_{\beta_1\alpha_1},w_{\gamma_1\gamma_1}]\tilde{\mathcal{W}}_1\rangle\langle \mathcal{L}[w_{\alpha_1\beta_1},w_{\delta_1\delta_1}]\tilde{\mathcal{W}}_2\rangle\bigr)\notag\\
   &\hspace{5mm}+\left(-\frac{x}{2}-1\right)\bigl(\langle \mathcal{L}[w_{\alpha_1\beta_1},w_{\gamma_1\gamma_1},w_{\delta_1\delta_1}]\tilde{\mathcal{W}}_1\rangle\langle \mathcal{L}[w_{\beta_1\alpha_1}]\tilde{\mathcal{W}}_2\rangle +\langle \mathcal{L}[w_{\beta_1\alpha_1},w_{\delta_1\delta_1},w_{\gamma_1\gamma_1}]\tilde{\mathcal{W}}_1\rangle\langle \mathcal{L}[w_{\alpha_1\beta_1}]\tilde{\mathcal{W}}_2\rangle\bigr)
 \Bigr)\notag\\
 &\textrm{\emph{Cut}}_2=\phi_2[\alpha_2,J,j,K,k][\gamma_2,L,l,I,i]\notag\\
&\hspace{5mm}\times \sum_{p \hspace{0.5mm}\in P^{n,(m_1,m_2,m_3)}_{2}}\Bigl(\langle \mathcal{L}[w_{\alpha_2\beta_2}]\tilde{\mathcal{W}}_1\rangle \langle \mathcal{L}[w_{\beta_2\alpha_2},w_{\gamma_2\gamma_2},w_{\delta_2\delta_2}]\tilde{\mathcal{W}}_2\rangle+\langle \mathcal{L}[w_{\beta_2\alpha_2}]\tilde{\mathcal{W}}_1\rangle\langle \mathcal{L}[w_{\alpha_2\beta_2},w_{\delta_2\delta_2},w_{\gamma_2\gamma_2}]\tilde{\mathcal{W}}_2\rangle \notag\\
   &\hspace{5mm}+\frac{\overline{x}}{2}\bigl(\langle \mathcal{L}[w_{\alpha_2\beta_2},w_{\gamma_2\gamma_2}]\tilde{\mathcal{W}}_1\rangle \langle \mathcal{L}[w_{\beta_2\alpha_2},w_{\delta_2\delta_2}]\tilde{\mathcal{W}}_2\rangle+\langle \mathcal{L}[w_{\beta_2\alpha_2},w_{\gamma_2\gamma_2}]\tilde{\mathcal{W}}_1\rangle\langle \mathcal{L}[w_{\alpha_2\beta_2},w_{\delta_2\delta_2}]\tilde{\mathcal{W}}_2\rangle\bigr)\notag\\
   &\hspace{5mm}+\left(-\frac{\overline{x}}{2}-1\right)\bigl(\langle \mathcal{L}[w_{\alpha_2\beta_2},w_{\gamma_2\gamma_2},w_{\delta_2\delta_2}]\tilde{\mathcal{W}}_1\rangle\langle \mathcal{L}[w_{\beta_2\alpha_2}]\tilde{\mathcal{W}}_2\rangle+\langle \mathcal{L}[w_{\beta_2\alpha_2},w_{\delta_2\delta_2},w_{\gamma_2\gamma_2}]\tilde{\mathcal{W}}_1\rangle\langle \mathcal{L}[w_{\alpha_2\beta_2}]\tilde{\mathcal{W}}_2\rangle\bigr) \Bigr)\notag\\
   &w_{\alpha_a\beta_a}=(\alpha_a,i,\dots,J,\beta_a),\hspace{2mm} w_{\beta_a\alpha_a}=(\beta_a,j,\dots,I,\alpha_a),\hspace{2mm} w_{\gamma_a\gamma_a}=(\gamma_a,k,\dots,K,\gamma_a), \hspace{2mm} w_{\delta_a\delta_a}= (\delta_a,l,\dots,L,\delta_a)
\end{flalign}
\begin{flalign}
 &\textrm{\textbf{1-1-1-1 4-mass}}&\notag\\
 &\{I,i\}\in \mathcal{W}_{m_1}, \hspace{2mm} \{J,j\}\in \mathcal{W}_{m_2}, \hspace{2mm} \{K,k\}\in \mathcal{W}_{m_3}, \hspace{2mm} \{L,l\}\in \mathcal{W}_{m_4}, \hspace{2mm}  p \in P_{1}^{n,(m_1,\hdots,m_4)}\\
 &\textrm{\emph{Cut}}_1=-\phi_1[\delta_1,I,i,J,j][\beta_1,K,k,L,l]\notag\\
 &\hspace{16mm}\times\Bigl(\langle \mathcal{L}[w_{\alpha_1\alpha_1},w_{\beta_1\beta_1},w_{\gamma_1\gamma_1},w_{\delta_1\delta_1}]\tilde{\mathcal{W}}_1 \rangle+\langle \mathcal{L}[w_{\alpha_1\alpha_1},w_{\delta_1\delta_1},w_{\gamma_1\gamma_1},w_{\beta_1\beta_1}]\tilde{\mathcal{W}}_1 \rangle\notag\\
 &\hspace{16mm}+\frac{x}{2}\bigl(\langle \mathcal{L}[w_{\alpha_1\alpha_1},w_{\gamma_1\gamma_1},w_{\beta_1\beta_1},w_{\delta_1\delta_1}]\tilde{\mathcal{W}}_1 \rangle+\langle \mathcal{L}[w_{\alpha_1\alpha_1},w_{\delta_1\delta_1},w_{\beta_1\beta_1},w_{\gamma_1\gamma_1}]\tilde{\mathcal{W}}_1 \rangle\bigr)\notag\\
 &\hspace{16mm}+\left(-\frac{x}{2}-1\right)\bigl(\langle \mathcal{L}[w_{\alpha_1\alpha_1},w_{\gamma_1\gamma_1},w_{\delta_1\delta_1},w_{\beta_1\beta_1}]\tilde{\mathcal{W}}_1 \rangle+\langle \mathcal{L}[w_{\alpha_1\alpha_1},w_{\beta_1\beta_1},w_{\delta_1\delta_1},w_{\gamma_1\gamma_1}]\tilde{\mathcal{W}}_1 \rangle\bigr)\Bigr)\notag\\
 &\textrm{\emph{Cut}}_2=\phi_2[\alpha_2,J,j,K,k][\gamma_2,L,l,I,i]&\notag\\
 &\hspace{16mm}\times \Bigl(\langle \mathcal{L}[w_{\alpha_2\alpha_2},w_{\beta_2\beta_2},w_{\gamma_2\gamma_2},w_{\delta_2\delta_2}]\tilde{\mathcal{W}}_1 \rangle+\langle \mathcal{L}[w_{\alpha_2\alpha_2},w_{\delta_2\delta_2},w_{\gamma_2\gamma_2},w_{\beta_2\beta_2}]\tilde{\mathcal{W}}_1 \rangle\notag\\
 &\hspace{16mm}+\frac{\bar{x}}{2}\bigl(\langle \mathcal{L}[w_{\alpha_2\alpha_2},w_{\gamma_2\gamma_2},w_{\beta_2\beta_2},w_{\delta_2\delta_2}]\tilde{\mathcal{W}}_1 \rangle+\langle \mathcal{L}[w_{\alpha_2\alpha_2},w_{\delta_2\delta_2},w_{\beta_2\beta_2},w_{\gamma_2\gamma_2}]\tilde{\mathcal{W}}_1 \rangle\bigr)\notag\\
 &\hspace{16mm}+\left(-\frac{\bar{x}}{2}-1\right)\bigl(\langle \mathcal{L}[w_{\alpha_2\alpha_2},w_{\gamma_2\gamma_2},w_{\delta_2\delta_2},w_{\beta_2\beta_2}]\tilde{\mathcal{W}}_1 \rangle+\langle \mathcal{L}[w_{\alpha_2\alpha_2},w_{\beta_2\beta_2},w_{\delta_2\delta_2},w_{\gamma_2\gamma_2}]\tilde{\mathcal{W}}_1 \rangle\bigr)\Bigr)\notag\\
  &w_{\alpha_a\alpha_a}=(\alpha_a,i,\dots,I,\alpha_a),\hspace{2mm} w_{\beta_a\beta_a}=(\beta_a,j,\dots,J,\beta_a),\hspace{2mm} w_{\gamma_a\gamma_a}=(\gamma_a,k,\dots,K,\gamma_a), \hspace{2mm} w_{\delta_a\delta_a}= (\delta_a,l,\dots,L,\delta_a)
\end{flalign}
All the correlators denote the tree-level, connected part. The parameters appearing in the coefficients are
\begin{align}
    &\phi_1=\left(1-\frac{\langle \beta_1\hspace{1mm} l\hspace{1mm} I\hspace{1mm} i \rangle \langle \delta_1\hspace{1mm} j\hspace{1mm} K\hspace{1mm} k \rangle}{\langle \beta_1\hspace{1mm} l\hspace{1mm} K\hspace{1mm} k \rangle \langle \delta_1\hspace{1mm} j \hspace{1mm}I\hspace{1mm} i \rangle}\right)^{-1},\hspace{2mm}\phi_2=\left(1-\frac{\langle \alpha_2 \hspace{1mm} k \hspace{1mm} L \hspace{1mm} l \rangle\langle \gamma_2 \hspace{1mm} i \hspace{1mm} J \hspace{1mm} j\rangle}{\langle \alpha_2 \hspace{1mm} k \hspace{1mm} J \hspace{1mm} j \rangle \langle \gamma_2 \hspace{1mm} i \hspace{1mm} L \hspace{1mm} l \rangle}\right)^{-1},\notag\\
    &x=\Delta-u+v-1,\hspace{2mm} \overline{x}=-\Delta-u+v-1
    \label{leadingsingparamdefs2}
\end{align}

\section{Conclusion}
We have presented a novel derivation of the leading singularities of the expectation value of planar \(\mathcal{N}=4\) scattering amplitudes which makes no direct reference to e.g. the unitary properties of scattering amplitudes, and which generalises straightforwardly to the case of multiple Wilson loop correlators. Via the chiral or scalar box expansion of local one-loop integrals, this amounts to the general solution to the \(O(g^2)\) problem for light-like Wilson loop correlators, for any number of Wilson loops, any combination of multiplicities, and any MHV degrees. At higher order in the coupling, this data should provide the basic ingredient for a computation using the generalised \(\bar{Q}\)-equation presented in \cite{Drummond:2026lvq} to obtain NMHV \(O(g^4)\) and even \(O(g^6)\) MHV contributions. This would bring the data available for multiple Wilson loop correlators to the equivalent cutting edge of the amplitude/single Wilson loop case, although we defer such computations to future work.

One obvious avenue for further investigation is the possibility that a similar procedure may allow a derivation for the full, non-planar leading singularities of Wilson loop correlators. On one hand, the arguments presented here rely strongly on the notion of planarity in order to partition our Wilson loop diagrams into regions. On the other hand, the basic observation that each one-loop cut is equivalent to a single tree-level diagram does not rely on planarity, and it seems natural that the collection of tree-level diagrams which give the non-planar leading singularity should organise themselves into an expression in terms of simple, tree-level objects, for instance involving correlators between the different regions. 

It would be interesting to make more explicit the connection between the procedure we present here and the more familiar derivation of the leading singularities of a single Wilson loop using on-shell diagrams. Another perhaps related direction is to extend our analysis here to higher-loop leading singularities. It would also be interesting to investigate whether the formulae which we present in this paper for the leading singularities of multiple Wilson loops can be generated using a Grassmannian integral, as is the case for a single Wilson loop \cite{Arkani-Hamed:2009ljj,Mason:2009qx,Arkani-Hamed:2009nll}.

\section*{Acknowledgements}
The authors thank \"Omer G\"urdo\u gan and Matteo Parisi for helpful discussions on related topics. All authors are supported by the STFC consolidated grant ST/X000583/1.

\newpage

\appendix

\section{Lower-mass leading singularities of $\langle W_1 \hdots W_n \rangle^c$}
\label{lowermassleading}

In Section \ref{Sec-allLeadingSing}, the general expressions were given for all of the 4-mass cuts of $\langle W_1 \hdots W_n \rangle^c$. Here, we give the lower mass leading singularities, which can be obtained in an analagous fashion to the lower mass example given in \ref{lowermass}. 
\begin{flalign}
 &\textrm{\textbf{4-0 3-mass}}&\notag \\
 &\{I,i=J,j,K,k,L,l\}\in \mathcal{W}_m\notag\\
 &\textrm{\emph{Cut}}_1=-[\beta_1,K,k,L,l]\sum_{p\hspace{0.5mm}\in P^{n,(m)}_3}\langle\mathcal{L}[w_{\beta_1\gamma_1}]\hspace{1mm} \tilde{\mathcal{W}}_1\rangle\langle\mathcal{L}[w_{\gamma_1\delta_1}]\hspace{1mm}\tilde{\mathcal{W}}_2\rangle \langle\mathcal{L}[w_{\delta_1\alpha_1}]\hspace{1mm}\tilde{\mathcal{W}}_3\rangle\notag\\
 &\textrm{\emph{Cut}}_2=[\alpha_2,J,j,K,k][\gamma_2,L,l,I,i]\sum_{p\hspace{0.5mm}\in P^{n,(m)}_4}\langle \mathcal{L}[w_{\alpha_2\beta_2}]\hspace{1mm}\tilde{\mathcal{W}}_1\rangle\langle\mathcal{L}[w_{\beta_2\gamma_2}]\hspace{1mm} \tilde{\mathcal{W}}_2\rangle\langle\mathcal{L}[w_{\gamma_2\delta_2}]\hspace{1mm}\tilde{\mathcal{W}}_3\rangle \langle\mathcal{L}[w_{\delta_2\alpha_2}]\hspace{1mm}\tilde{\mathcal{W}}_4\rangle\notag\\
  &w_{\alpha_a\beta_a}=(\alpha_a,i=J,\beta_a),\hspace{2mm} w_{\beta_a\gamma_a}=(\beta_a,j,\dots,K,\gamma_a),\hspace{2mm} w_{\gamma_a\delta_a}=(\gamma_a,k,\dots,L,\delta_a), \hspace{2mm} w_{\delta_a\alpha_a}= (\delta_a,l,\dots,I,\alpha_a)\notag\\
  &\alpha_1=i,\hspace{2mm}\beta_1=J,\hspace{2mm}\gamma_1=(j J)\cap(L l i),\hspace{2mm}\delta_1=(L l)\cap(k K J)\notag\\
  &\alpha_2=(I i) \cap (l L \gamma_2),\hspace{2mm}\beta_2=(j J) \cap(K k \delta_2),\hspace{2mm}\gamma_2=(K k) \cap (j J I), \hspace{2mm}\delta_2=(l L)\cap(I i j)
\end{flalign}
\begin{flalign}
 &\textrm{\textbf{4-0 2-mass hard}}&\notag \\
 &\{I,i=J,j=K,k,L,l\}\in \mathcal{W}_m\notag\\
 &\textrm{\emph{Cut}}_1=-[\beta_1,K,k,L,l]\sum_{p\hspace{0.5mm}\in P^{n,(m)}_3}\langle\mathcal{L}[w_{\beta_1\gamma_1}]\hspace{1mm} \tilde{\mathcal{W}}_1\rangle\langle\mathcal{L}[w_{\gamma_1\delta_1}]\hspace{1mm}\tilde{\mathcal{W}}_2\rangle \langle\mathcal{L}[w_{\delta_1\alpha_1}]\hspace{1mm}\tilde{\mathcal{W}}_3\rangle\notag\\
 &\textrm{\emph{Cut}}_2=[\gamma_2,L,l,I,i]\sum_{p\hspace{0.5mm}\in P^{n,(m)}_3}\langle \mathcal{L}[w_{\alpha_2\beta_2}]\hspace{1mm}\tilde{\mathcal{W}}_1\rangle\langle\mathcal{L}[w_{\gamma_2\delta_2}]\hspace{1mm}\tilde{\mathcal{W}}_2\rangle \langle\mathcal{L}[w_{\delta_2\alpha_2}]\hspace{1mm}\tilde{\mathcal{W}}_3\rangle\notag\\
  &w_{\alpha_a\beta_a}=(\alpha_a,i=J,\beta_a),\hspace{2mm} w_{\beta_a\gamma_a}=(\beta_a,j=K,\gamma_a),\hspace{2mm} w_{\gamma_a\delta_a}=(\gamma_a,k,\dots,L,\delta_a), \hspace{2mm} w_{\delta_a\alpha_a}= (\delta_a,l,\dots,I,\alpha_a)\notag\\
  &\alpha_1=i,\hspace{2mm}\beta_1=J,\hspace{2mm}\gamma_1=(j J)\cap(L l i),\hspace{2mm}\delta_1=(L l)\cap(k K J)\notag\\
  &\alpha_2=(I i) \cap (l L K),\hspace{2mm}\beta_2=j,\hspace{2mm}\gamma_2=K, \hspace{2mm}\delta_2=(l L)\cap(I i j)
\end{flalign}
\begin{flalign}
 &\textrm{\textbf{4-0 2-mass easy}}&\notag \\
 &\{I,i=J,j,K,k=L,l\}\in \mathcal{W}_m\notag\\
 &\textrm{\emph{Cut}}_1=-\sum_{p\hspace{0.5mm}\in P^{n,(m)}_2}\langle\mathcal{L}[w_{\beta_1\gamma_1}]\hspace{1mm} \tilde{\mathcal{W}}_1\rangle\langle\mathcal{L}[w_{\delta_1\alpha_1}]\hspace{1mm}\tilde{\mathcal{W}}_2\rangle\notag\\
 &\textrm{\emph{Cut}}_2= [\alpha_2,J,j,K,k][\gamma_2,L,l,I,i]\sum_{p\hspace{0.5mm}\in P^{n,(m)}_4}\langle \mathcal{L}[w_{\alpha_2\beta_2}]\hspace{1mm}\tilde{\mathcal{W}}_1\rangle\langle\mathcal{L}[w_{\beta_2\gamma_2}]\hspace{1mm} \tilde{\mathcal{W}}_2\rangle\langle\mathcal{L}[w_{\gamma_2\delta_2}]\hspace{1mm}\tilde{\mathcal{W}}_3\rangle \langle\mathcal{L}[w_{\delta_2\alpha_2}]\hspace{1mm}\tilde{\mathcal{W}}_4\rangle\notag\\
  &w_{\alpha_a\beta_a}=(\alpha_a,i=J,\beta_a),\hspace{2mm} w_{\beta_a\gamma_a}=(\beta_a,j,\dots,K,\gamma_a),\hspace{2mm} w_{\gamma_a\delta_a}=(\gamma_a,k=L,\delta_a), \hspace{2mm} w_{\delta_a\alpha_a}= (\delta_a,l,\dots,I,\alpha_a)\notag\\
 &\alpha_1=i,\hspace{2mm}\beta_1=J,\hspace{2mm}\gamma_1=k,\hspace{2mm}\delta_1=L\notag\\
  &\alpha_2=(I i) \cap (l L K),\hspace{2mm}\beta_2=(j J) \cap(K k l),\hspace{2mm}\gamma_2=(K k) \cap (j J I), \hspace{2mm}\delta_2=(l L)\cap(I i j)
\end{flalign}
\begin{flalign}
 &\textrm{\textbf{4-0 1-mass}}&\notag \\
 &\{I,i=J,j=K,k=L,l\}\in \mathcal{W}_m\notag\\
 &\textrm{\emph{Cut}}_1=-\sum_{p\hspace{0.5mm}\in P^{n,(m)}_2}\langle\mathcal{L}[w_{\beta_1\gamma_1}]\hspace{1mm} \tilde{\mathcal{W}}_1\rangle\langle\mathcal{L}[w_{\delta_1\alpha_1}]\hspace{1mm}\tilde{\mathcal{W}}_2\rangle\notag\\
 &\textrm{\emph{Cut}}_2=[\gamma_2,L,l,I,i]\sum_{p\hspace{0.5mm}\in P^{n,(m)}_3}\langle \mathcal{L}[w_{\alpha_2\beta_2}]\hspace{1mm}\tilde{\mathcal{W}}_1\rangle\langle\mathcal{L}[w_{\gamma_2\delta_2}]\hspace{1mm}\tilde{\mathcal{W}}_2\rangle \langle\mathcal{L}[w_{\delta_2\alpha_2}]\hspace{1mm}\tilde{\mathcal{W}}_3\rangle\notag\\
  &w_{\alpha_a\beta_a}=(\alpha_a,i=J,\beta_a),\hspace{2mm} w_{\beta_a\gamma_a}=(\beta_a,j=K,\gamma_a),\hspace{2mm} w_{\gamma_a\delta_a}=(\gamma_a,k=L,\delta_a), \hspace{2mm} w_{\delta_a\alpha_a}= (\delta_a,l,\dots,I,\alpha_a)\notag\\
 &\alpha_1=i,\hspace{2mm}\beta_1=J,\hspace{2mm}\gamma_1=k,\hspace{2mm}\delta_1=L\notag\\
  &\alpha_2=(I i) \cap (l L K),\hspace{2mm}\beta_2=j,\hspace{2mm}\gamma_2=K, \hspace{2mm}\delta_2=(l L)\cap(I i j)
\end{flalign}
\begin{flalign}
 &\textrm{\textbf{4-0 0-mass}}&\notag \\
 &\{l=I,i=J,j=K,k=L\}\in \mathcal{W}_m\notag\\
 &\textrm{\emph{Cut}}_1=-\sum_{p\hspace{0.5mm}\in P^{n,(m)}_2}\langle\mathcal{L}[w_{\beta_1\gamma_1}]\hspace{1mm} \tilde{\mathcal{W}}_1\rangle\langle\mathcal{L}[w_{\delta_1\alpha_1}]\hspace{1mm}\tilde{\mathcal{W}}_2\rangle\notag\\
 &\textrm{\emph{Cut}}_2=\sum_{p\hspace{0.5mm}\in P^{n,(m)}_2}\langle \mathcal{L}[w_{\alpha_2\beta_2}]\hspace{1mm}\tilde{\mathcal{W}}_1\rangle\langle\mathcal{L}[w_{\gamma_2\delta_2}]\hspace{1mm}\tilde{\mathcal{W}}_2\rangle \notag\\
  &w_{\alpha_a\beta_a}=(\alpha_a,i=J,\beta_a),\hspace{2mm} w_{\beta_a\gamma_a}=(\beta_a,j=K,\gamma_a),\hspace{2mm} w_{\gamma_a\delta_a}=(\gamma_a,k=L,\delta_a), \hspace{2mm} w_{\delta_a\alpha_a}= (\delta_a,l=I,\alpha_a),\notag\\
 &\alpha_1=i,\hspace{2mm}\beta_1=J,\hspace{2mm}\gamma_1=k,\hspace{2mm}\delta_1=L\notag\\
  &\alpha_2=I,\hspace{2mm}\beta_2=j,\hspace{2mm}\gamma_2=K, \hspace{2mm}\delta_2=l
\end{flalign}
\begin{flalign}
 &\textrm{\textbf{3-1 3-mass}}&\notag\\
 &\{I,i=J,j,K,k\}\in \mathcal{W}_{m_1}, \hspace{5mm} \{L,l\}\in \mathcal{W}_{m_2}\notag\\
 &\textrm{\emph{Cut}}_1=-[\beta_1,K,k,L,l]\sum_{p\hspace{0.5mm}\in P^{n,(m_1,m_2)}_2}\notag\Bigl(\langle\mathcal{L}[w_{\beta_1\gamma_1}]\tilde{\mathcal{W}}_1\rangle\langle\mathcal{L}[w_{\gamma_1\alpha_1},w_{\delta_1\delta_1}]\tilde{\mathcal{W}}_2\rangle\notag\\
    &\hspace{35mm}+\left(-1\right)\langle\mathcal{L}[w_{\beta_1\gamma_1},w_{\delta_1\delta_1}]\tilde{\mathcal{W}}_1\rangle\langle\mathcal{L}[w_{\gamma_1\alpha_1}]\tilde{\mathcal{W}}_2\rangle\Bigr)\notag\\
 &\textrm{\emph{Cut}}_2=[\alpha_2,J,j,K,k][\gamma_2,L,l,I,i]\sum_{p\hspace{0.5mm}\in P^{n,(m_1,m_2)}_3} \Bigl(\langle \mathcal{L}[w_{\alpha_2\beta_2}]\tilde{\mathcal{W}}_1\rangle\langle\mathcal{L}[w_{\beta_2\gamma_2}]\tilde{\mathcal{W}}_2\rangle\langle\mathcal{L}[w_{\gamma_2,\alpha_2},w_{\delta_2\delta_2}]\tilde{\mathcal{W}}_3\rangle\notag\\
    &\hspace{45.5mm}+\left(v-1\right)\langle \mathcal{L}[w_{\alpha_2\beta_2},w_{\delta_2\delta_2}]\tilde{\mathcal{W}}_1\rangle\langle\mathcal{L}[w_{\beta_2\gamma_2}]\tilde{\mathcal{W}}_2\rangle\langle\mathcal{L}[w_{\gamma_2\alpha_2}]\tilde{\mathcal{W}}_3\rangle&\notag\\
    &\hspace{45.5mm}+\left(-v\right)\langle \mathcal{L}[w_{\alpha_2\beta_2}]\tilde{\mathcal{W}}_1\rangle\langle\mathcal{L}[w_{\beta_2\gamma_2},w_{\delta_2\delta_2}]\tilde{\mathcal{W}}_2\rangle\langle\mathcal{L}[w_{\gamma_2\alpha_2}]\tilde{\mathcal{W}}_3\rangle\Bigr)\notag\\
    &w_{\alpha_a\beta_a}=(\alpha_a,i=J,\beta_a),\hspace{2mm} w_{\beta_a\gamma_a}=(\beta_a,j,\dots,K,\gamma_a),\hspace{2mm} w_{\gamma_a\alpha_a}=(\gamma_a,k,\dots,I,\alpha_a), \hspace{2mm} w_{\delta_a\delta_a}= (\delta_a,l,\dots,L,\delta_a),\notag\\
     &\alpha_1=i,\hspace{2mm}\beta_1=J,\hspace{2mm}\gamma_1=(j J)\cap(L l i),\hspace{2mm}\delta_1=(L l)\cap(k K J)\notag\\
    &\alpha_2=(I i) \cap (l L \gamma_2),\hspace{2mm}\beta_2=(j J) \cap(K k \delta_2),\hspace{2mm}\gamma_2=(K k) \cap (j J I), \hspace{2mm}\delta_2=(l L)\cap(I i j)
\end{flalign}
\begin{flalign}
 &\textrm{\textbf{3-1 2-mass hard}}&\notag\\
 &\{I,i=J,j=K,k\}\in \mathcal{W}_{m_1}, \hspace{5mm} \{L,l\}\in \mathcal{W}_{m_2}\notag\\
 &\textrm{\emph{Cut}}_1=-[\beta_1,K,k,L,l]\sum_{p\hspace{0.5mm}\in P^{n,(m_1,m_2)}_2} \notag\Bigl(\langle\mathcal{L}[w_{\beta_1\gamma_1}]\tilde{\mathcal{W}}_1\rangle\langle\mathcal{L}[w_{\gamma_1\alpha_1},w_{\delta_1\delta_1}]\tilde{\mathcal{W}}_2\rangle\notag\\
    &\hspace{35mm}+\left(-1\right)\langle\mathcal{L}[w_{\beta_1\gamma_1},w_{\delta_1\delta_1}]\tilde{\mathcal{W}}_1\rangle\langle\mathcal{L}[w_{\gamma_1\alpha_1}]\tilde{\mathcal{W}}_2\rangle\Bigr)\notag\\
 &\textrm{\emph{Cut}}_2=[\gamma_2,L,l,I,i]\sum_{p\hspace{0.5mm}\in P^{n,(m_1,m_2)}_2}\Bigl(\langle \mathcal{L}[w_{\alpha_2\beta_2}]\tilde{\mathcal{W}}_1\rangle\langle\mathcal{L}[w_{\gamma_2,\alpha_2},w_{\delta_2\delta_2}]\tilde{\mathcal{W}}_2\rangle\notag\\
    &\hspace{32.5mm}+\left(-1\right)\langle \mathcal{L}[w_{\alpha_2\beta_2},w_{\delta_2\delta_2}]\tilde{\mathcal{W}}_1\rangle\langle\mathcal{L}[w_{\gamma_2\alpha_2}]\tilde{\mathcal{W}}_2\rangle\Bigr)\notag\\
    &w_{\alpha_a\beta_a}=(\alpha_a,i=J,\beta_a),\hspace{2mm} w_{\beta_a\gamma_a}=(\beta_a,j=K,\gamma_a),\hspace{2mm} w_{\gamma_a\alpha_a}=(\gamma_a,k,\dots,I,\alpha_a), \hspace{2mm} w_{\delta_a\delta_a}= (\delta_a,l,\dots,L,\delta_a)\notag\\
    &\alpha_1=i,\hspace{2mm}\beta_1=J,\hspace{2mm}\gamma_1=(k K)\cap (L l i),\hspace{2mm}\delta_1=(L l)\cap(k K J)\notag\\
  &\alpha_2=(I i) \cap (l L K),\hspace{2mm}\beta_2=j,\hspace{2mm}\gamma_2=K, \hspace{2mm}\delta_2=(l L)\cap(I i j)
\end{flalign}
\begin{flalign}
 &\textrm{\textbf{2-2 3-mass}}&\notag\\
 &\{I,i=J,j\}\in \mathcal{W}_{m_1}, \hspace{5mm} \{K,k,L,l\}\in \mathcal{W}_{m_2}\notag\\
 &\textrm{\emph{Cut}}_1=-[\beta_1,K,k,L,l]  \sum_{p\hspace{0.5mm}\in P^{n,(m_1,m_2)}_2}\Bigl( \langle\mathcal{L}[w_{\gamma_1\delta_1}]\tilde{\mathcal{W}}_1\rangle \langle \mathcal{L}[w_{\beta_1\alpha_1},w_{\delta_1\gamma_1}]\tilde{\mathcal{W}}_2\rangle\notag\\
 &\hspace{34mm}+\left(-1\right) \langle\mathcal{L}[w_{\delta_1\gamma_1}]\tilde{\mathcal{W}}_1\rangle \langle \mathcal{L}[w_{\beta_1\alpha_1},w_{\gamma_1\delta_1}]\tilde{\mathcal{W}}_2\rangle\Bigr)\notag\\
 &\textrm{\emph{Cut}}_2=[\alpha_2,J,j,K,k][\gamma_2,L,l,I,i]\sum_{p\hspace{0.5mm}\in P^{n,(m_1,m_2)}_3}\Bigl(\langle \mathcal{L}[w_{\alpha_2\beta_2},w_{\gamma_2\delta_2}]\tilde{\mathcal{W}}_1\rangle \langle\mathcal{L}[w_{\beta_2\alpha_2}]\tilde{\mathcal{W}}_2\rangle \langle\mathcal{L}[w_{\delta_2\gamma_2}]\tilde{\mathcal{W}}_3\rangle&\notag\\
 &\hspace{50.25mm}+\langle\mathcal{L}[w_{\alpha_2\beta_2}]\tilde{\mathcal{W}}_1\rangle \langle\mathcal{L}[w_{\gamma_2\delta_2}]\tilde{\mathcal{W}}_2\rangle \langle \mathcal{L}[w_{\beta_2\alpha_2},w_{\delta_2\gamma_2}]\tilde{\mathcal{W}}_3\rangle\notag\\
 &\hspace{50.25mm}+\left(-v\right)\big(\langle \mathcal{L}[w_{\alpha_2\beta_2},w_{\delta_2\gamma_2}]\tilde{\mathcal{W}}_1\rangle\langle\mathcal{L}[w_{\beta_2\alpha_2}]\tilde{\mathcal{W}}_2\rangle \langle\mathcal{L}[w_{\gamma_2\delta_2}]\tilde{\mathcal{W}}_3\rangle&\notag\\
 &\hspace{50.25mm}+\langle\mathcal{L}[w_{\alpha_2\beta_2}]\tilde{\mathcal{W}}_1\rangle \langle\mathcal{L}[w_{\delta_2\gamma_2}]\tilde{\mathcal{W}}_2\rangle \langle \mathcal{L}[w_{\beta_2\alpha_2},w_{\gamma_2\delta_2}]\tilde{\mathcal{W}}_3\rangle\bigr)\Bigr)\notag\\
  &w_{\alpha_a\beta_a}=(\alpha_a,i=J,\beta_a),\hspace{2mm} w_{\beta_a\alpha_a}=(\beta_a,j,\dots,I,\alpha_a),\hspace{2mm} w_{\gamma_a\delta_a}=(\gamma_a,k,\dots,L,\delta_a), \hspace{2mm} w_{\delta_a\gamma_a}= (\delta_a,l,\dots,K,\gamma_a)\notag\\
      &\alpha_1=i,\hspace{2mm}\beta_1=J,\hspace{2mm}\gamma_1=(k K)\cap (L l i),\hspace{2mm}\delta_1=(L l)\cap(k K J)\notag\\
  &\alpha_2=(I i) \cap (l L \gamma_2),\hspace{2mm}\beta_2=(j J)\cap(K k \delta_2),\hspace{2mm}\gamma_2=(K k) \cap (j J I), \hspace{2mm}\delta_2=(l L)\cap(I i j)
\end{flalign}
\begin{flalign}
 &\textrm{\textbf{2-2 2-mass easy}}&\notag\\
 &\{I,i=J,j\}\in \mathcal{W}_{m_1}, \hspace{5mm} \{K,k=L,l\}\in \mathcal{W}_{m_2}\notag\\
 &\textrm{\emph{Cut}}_1=-\bigl(\langle \mathcal{L}[w_{\beta_1\alpha_1},w_{\delta_1\gamma_1}]\tilde{\mathcal{W}}_1\rangle\big|_{p\hspace{0.5mm}\in P^{n,(m_1,m_2)}_1}+\sum_{p\hspace{0.5mm}\in P^{n,(m_1,m_2)}_2}\left(-1\right) \langle\mathcal{L}[w_{\delta_1\gamma_1}]\tilde{\mathcal{W}}_1\rangle \langle \mathcal{L}[w_{\beta_1\alpha_1}]\tilde{\mathcal{W}}_2\rangle\bigr)\notag\\
 &\textrm{\emph{Cut}}_2=[\alpha_2,J,j,K,k][\gamma_2,L,l,I,i]\sum_{p\hspace{0.5mm}\in P^{n,(m_1,m_2)}_3}\Bigl(\langle \mathcal{L}[w_{\alpha_2\beta_2},w_{\gamma_2\delta_2}]\tilde{\mathcal{W}}_1\rangle \langle\mathcal{L}[w_{\beta_2\alpha_2}]\tilde{\mathcal{W}}_2\rangle \langle\mathcal{L}[w_{\delta_2\gamma_2}]\tilde{\mathcal{W}}_3\rangle&\notag\\
 &\hspace{50.25mm}+\langle\mathcal{L}[w_{\alpha_2\beta_2}]\tilde{\mathcal{W}}_1\rangle \langle\mathcal{L}[w_{\gamma_2\delta_2}]\tilde{\mathcal{W}}_2\rangle \langle \mathcal{L}[w_{\beta_2\alpha_2},w_{\delta_2\gamma_2}]\tilde{\mathcal{W}}_3\rangle\notag\\
 &\hspace{50.25mm}+\left(-v\right)\big(\langle \mathcal{L}[w_{\alpha_2\beta_2},w_{\delta_2\gamma_2}]\tilde{\mathcal{W}}_1\rangle\langle\mathcal{L}[w_{\beta_2\alpha_2}]\tilde{\mathcal{W}}_2\rangle \langle\mathcal{L}[w_{\gamma_2\delta_2}]\tilde{\mathcal{W}}_3\rangle&\notag\\
 &\hspace{50.25mm}+\langle\mathcal{L}[w_{\alpha_2\beta_2}]\tilde{\mathcal{W}}_1\rangle \langle\mathcal{L}[w_{\delta_2\gamma_2}]\tilde{\mathcal{W}}_2\rangle \langle \mathcal{L}[w_{\beta_2\alpha_2}w_{\gamma_2\delta_2}]\tilde{\mathcal{W}}_3\rangle\bigr)\Bigr)\notag\\
  &w_{\alpha_a\beta_a}=(\alpha_a,i=J,\beta_a),\hspace{2mm} w_{\beta_a\alpha_a}=(\beta_a,j,\dots,I,\alpha_a),\hspace{2mm} w_{\gamma_a\delta_a}=(\gamma_a,k=L,\delta_a), \hspace{2mm} w_{\delta_a\gamma_a}= (\delta_a,l,\dots,K,\gamma_a)\notag\\
      &\alpha_1=i,\hspace{2mm}\beta_1=J,\hspace{2mm}\gamma_1=(k K)\cap (L l i),\hspace{2mm}\delta_1=(L l)\cap(k K J)\notag\\
  &\alpha_2=(I i) \cap (l L K),\hspace{2mm}\beta_2=(j J)\cap(K k l),\hspace{2mm}\gamma_2=(K k) \cap (j J I), \hspace{2mm}\delta_2=(l L)\cap(I i j)
\end{flalign}
\begin{flalign}
  &\textrm{\textbf{2-1-1 3-mass}}\notag\\
  &\{I,i=J,j\}\in \mathcal{W}_{m_1}, \hspace{2mm} \{K,k\}\in \mathcal{W}_{m_2}, \hspace{2mm} \{L,l\}\in \mathcal{W}_{m_3}\notag\\
  &\textrm{\emph{Cut}}_1=-[\beta_1,K,k,L,l]\Bigl(\langle \mathcal{L}[w_{\beta_1\alpha_1},w_{\gamma_1\gamma_1},w_{\delta_1\delta_1}]\tilde{\mathcal{W}}_1\rangle+\left(-1\right)\langle \mathcal{L}[w_{\beta_1\alpha_1},w_{\delta_1\delta_1},w_{\gamma_1\gamma_1}]\tilde{\mathcal{W}}_{1}\rangle
 \Bigr)\big|_{p\hspace{0.5mm}\in P^{n,(m_1,m_2,m_3)}_1}\notag\\
  &\textrm{\emph{Cut}}_2=[\alpha_2,J,j,K,k][\gamma_2,L,l,I,i]\notag\\
  &\hspace{5mm}\times \sum_{\vec{l}\hspace{0.5mm}=\hspace{0.5mm}1}^2\Bigl(\langle \mathcal{L}[w_{\alpha_2\beta_2}]\tilde{\mathcal{W}}_1\rangle \langle \mathcal{L}[w_{\beta_2\alpha_2},w_{\gamma_2\gamma_2},w_{\delta_2\delta_2}]\tilde{\mathcal{W}}_2\rangle+\langle \mathcal{L}[w_{\beta_2\alpha_2}]\tilde{\mathcal{W}}_1\rangle\langle \mathcal{L}[w_{\alpha_2\beta_2},w_{\delta_2\delta_2},w_{\gamma_2\gamma_2}]\tilde{\mathcal{W}}_2\rangle \notag\\
  &\hspace{5mm}+(v-1)\bigl(\langle \mathcal{L}[w_{\alpha_2\beta_2},w_{\gamma_2\gamma_2}]\tilde{\mathcal{W}}_1\rangle \langle \mathcal{L}[w_{\beta_2\alpha_2},w_{\delta_2\delta_2}]\tilde{\mathcal{W}}_2\rangle+\langle \mathcal{L}[w_{\beta_2\alpha_2},w_{\gamma_2\gamma_2}]\tilde{\mathcal{W}}_1\rangle\langle \mathcal{L}[w_{\alpha_2\beta_2},w_{\delta_2\delta_2}]\tilde{\mathcal{W}}_2\rangle\bigr)\notag\\
  &\hspace{5mm}+\left(-v\right)\bigl(\langle \mathcal{L}[w_{\alpha_2\beta_2},w_{\gamma_2\gamma_2},w_{\delta_2\delta_2}]\tilde{\mathcal{W}}_1\rangle\langle \mathcal{L}[w_{\beta_2\alpha_2}]\tilde{\mathcal{W}}_2\rangle+\langle \mathcal{L}[w_{\beta_2\alpha_2},w_{\delta_2\delta_2},w_{\gamma_2\gamma_2}]\tilde{\mathcal{W}}_1\rangle\langle \mathcal{L}[w_{\alpha_2\beta_2}]\tilde{\mathcal{W}}_2\rangle\bigr) \Bigr)\notag\\
  &w_{\alpha_a\beta_a}=(\alpha_a,i=J,\beta_a),\hspace{2mm} w_{\beta_a\alpha_a}=(\beta_a,j,\dots,I,\alpha_a),\hspace{2mm} w_{\gamma_a\gamma_a}=(\gamma_a,k,\dots,K,\gamma_a), \hspace{2mm} w_{\delta_a\delta_a}= (\delta_a,l,\dots,L,\delta_a)\notag\\
  &\alpha_1=i,\hspace{2mm}\beta_1=J,\hspace{2mm}\gamma_1=(j J)\cap(L l i),\hspace{2mm}\delta_1=(L l)\cap(k K J)\notag\\
  &\alpha_2=(I i) \cap (l L \gamma_2),\hspace{2mm}\beta_2=(j J) \cap(K k \delta_2),\hspace{2mm}\gamma_2=(K k) \cap (j J I), \hspace{2mm}\delta_2=(l L)\cap(I i j)
\end{flalign}
where all correlators appearing above denote the tree-level, connected parts and $v$ and $P^{n,(m_1,\hdots,m_k)}_q$ are as defined in equation \ref{leadingsingparamdefs}.

\end{document}